\documentclass[aps,prl,rmp,floatfix,twocolumn]{revtex4-1}
\usepackage[usenames]{color}
\usepackage{graphicx}
\usepackage{multirow}
\usepackage{subfigure}
\usepackage{epstopdf}
\usepackage{bm}
\usepackage{amssymb}
\usepackage{amsmath}
\usepackage{times}
\include{macros}

\def\be{\begin{equation}} \def\ee{\end{equation}}
\def\bea{\begin{eqnarray}} \def\eea{\end{eqnarray}}

\begin{document}
\title{The quantum anomalous Hall effect}
\author{Chao-Xing Liu, Shou-Cheng Zhang and Xiao-Liang Qi}
\address{Department of Physics, The Penn State University, University Park, PA, }
\address{Department of Physics, Stanford University, Stanford, CA 94305}

\begin{abstract}
The quantum anomalous Hall effect is defined as a quantized Hall effect realized in a system without external magnetic field. Quantum anomalous Hall effect is a novel manifestation of topological structure in many-electron systems, and may have potential applications in future electronic devices. In recent years, quantum anomalous Hall effect has been proposed theoretically and realized experimentally. In this review article, we provide a systematic overview of the theoretical and experimental developments in this field.
\end{abstract}

\maketitle

\tableofcontents

\section{Introduction}
\label{sec:intro}

\begin{figure} [h]
\includegraphics[width=8cm]{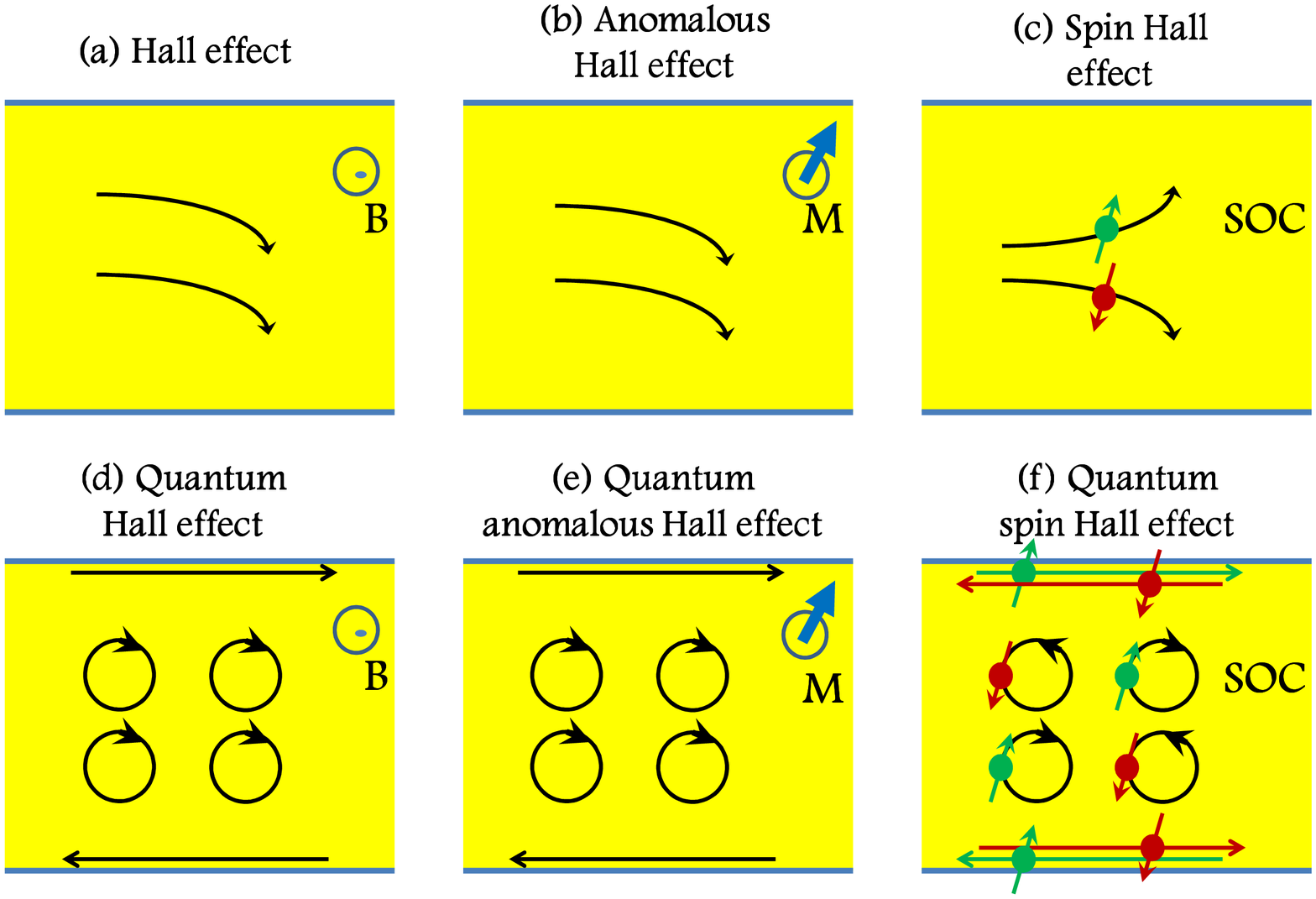}
\caption{(Color online) Six members in the family of Hall effect. (a) Hall effect; (b) Anomalous Hall effect; (c) Spin Hall effect; (d) Quantum Hall effect; (e) Quantum anomalous Hall effect; and (f) Quantum spin Hall effect.  } \label{Hall}
\end{figure}

More than a century ago, in 1879 Edwin Hall discovered the Hall effect \cite{hall1879}, which occurs in a sheet of metal in a perpendicular magnetic field. An electric voltage perpendicular to the charge current is induced by the magnetic field, due to the Lorentz force acting on electrons, as is illustrated in Fig. \ref{Hall}(a). 
In a two dimensional electron gas (2DEG) under low temperature and strong magnetic field, the Hall conductance becomes quantized in the unit of $\frac{e^2}{h}$, while longitudinal resistance drops to zero. This is known as the integer quantum Hall (IQH) effect \cite{klitzing1980}. The IQH effect occurs when electrons enter an incompressible state, with a robust uni-directional, or ``chiral", edge state propagating along the boundary, and all other states localized, as shown in Fig. \ref{Hall}(d). At a lower temperature, fractional quantum Hall (FQH) effect \cite{tsui1982} can occur, which are realized by strongly correlated electron states with fractionally quantized Hall conductivity (in the same unit of $\frac{e^2}{h}$).

One year after his discovery of the Hall effect, Edwin Hall also noticed that the Hall voltage was ten times larger in ferromagnetic materials than that in non-magnetic conductors, which is later known as the "anomalous Hall effect" \cite{hall1881} (Fig. \ref{Hall}(b)). The physical origin of anomalous Hall effect has been studied extensively in the past 50 years \cite{nagaosa2010}. Although the general mechanism is complicated with multiple contributions, the key ingredients responsible for the anomalous Hall effect that have been studied are magnetism, spin-orbit coupling and disorder effects. With the discovery of the quantum Hall effect, it is natural to ask whether there is a similar quantum version of the anomalous Hall effect, or the quantum anomalous Hall (QAH) effect. More precisely, the question is whether a quantized Hall conductance can be induced by mechanisms different from an external magnetic field, such as magnetism, spin-orbit coupling and/or disorder. An extreme form of this question is whether a state with a quantized Hall conductance can exist without external magnetic field at all. The first step of answering this question came from understanding the general topological reason of the Hall conductance quantization \cite{thouless1982,avron1985,niu1985}. A geometrical gauge field can be defined in the space of momentum or twist-boundary conditions, which has a quantized total flux that determines the Hall conductance quantum. Therefore, the IQH effect can exist in more general systems as long as the geometrical gauge field is nontrivial. In 1988, F. D. M. Haldane proposed the first explicit model realizing quantized Hall conductance without orbital magnetic field \cite{haldane1988}, which is known as the Haldane model. The Haldane model is a tight-binding model defined on a honeycomb lattice with real nearest neighbor hopping and complex next-nearest-neighbor hopping.

Although the Haldane model demonstrated the theoretical possibility of realizing a QAH state, how to realize such a state in realistic materials remained unknown for another two decades. The solution was eventually discovered through a ``detour" into another cousin of Hall effect---the spin Hall effect and the quantum spin Hall (QSH) effect. In a system with spin-orbit coupling, electrons with different spin directions can experience different forces. In particular, the motion of electrons in an external electric field can be biased to different directions depending on their spin direction, as is illustrated in Fig. \ref{Hall}(c). This is named as the spin Hall effect \cite{sinova2004,murakami2003,hirsch1999}, since in the direction perpendicular to the electric field there is a net spin current though no net charge current. Different from the charge Hall effect, the spin Hall effect does not require time-reversal symmetry breaking, and thus can occur in a semiconductor without magnetic field. Since 2005, the quantum version of the spin Hall effect (Fig. \ref{Hall}(f)) was proposed theoretically \cite{kane2005a,bernevig2006a,bernevig2006c} and realized experimentally in HgTe/CdTe quantum wells \cite{koenig2007} soon after the theoretical proposal \cite{bernevig2006c}. The discovery of the QSH effect has lead to the latter discovery of topological insulators and topological superconductors \cite{hasan2010,qi2011}, which greatly expanded our understanding of topological states of matter. The QSH effect is deeply related to the QAH effect, since the simplest models realizing QSH effect describe electrons with opposite spin carrying opposite quantized Hall conductance, as is illustrated in Fig. \ref{Hall}(f). The QH state of each spin component can be either a IQH state with Landau levels \cite{bernevig2006a} or a QAH state \cite{kane2005a,qi2005}. Although the QSH state is only protected by time-reversal symmetry and does not require the two spin components to be decoupled \cite{kane2005b}, the simple QSH states with opposite QAH of each spin component pointed out a new route of realizing the QAH effect. Intuitively one can say that for electrons in a QSH state, each spin component is already in a QAH state, and the only problem is the Hall conductance of two spin components cancels each other due to time-reversal symmetry. Therefore the only remaining task is to remove the QAH effect of one spin component, so that the net Hall conductance becomes nonvanishing. This goal was achieved in Ref. \cite{liu2009C}, in which HgTe/CdTe quantum wells doped with Mn impurities, in proper thickness range, was proposed to realize the QAH state, given that the Mn spins are polarized. However, Mn spins are paramagnetic instead of ferromagnetic in this material, so that a small external magnetic field is still required to polarize them in order to realize the QAH phase. Indeed, the early transport experiment \cite{buhmann2002} has observed a Hall conductance $-\frac{e^2}h$ plateau starting at a quite small magnetic field and persisting for an extremely wide range of magnetic field, agreeing with the theoretical prediction.

More recently, a new QAH state was proposed in magnetically doped topological insulator films in the family of ${\rm Bi_2Se_3}$\cite{yu2010}, which is ferromagnetic and thus finally leads to the realization of the QAH effect completely without external magnetic field. ${\rm Bi_2Se_3}$ and related materials ${\rm Bi_2Te_3}$ and ${\rm Sb_2Te_3}$ are three-dimensional topological insulators \cite{hjzhang2009,xia2009}. Similar to the QSH state, the three-dimensional TI turns out to be also deeply related to the QAH state. A universal feature of three-dimensional TI is that its two-dimensional surface state always obtains a half quantized Hall conductance $\sigma_H=\frac{e^2}{h}\left(n+\frac12\right)$ (with $n$ an integer) once time reversal is broken \cite{qi2008b,fu2007b}. This becomes the key ingredient of realizing QAH state in the thin film of TI doped with magnetic impurities such as Cr or V. The first experimental realization of this proposal was achieved in Cr doped ${\rm Bi(Sb)_2Te_3}$ thin films \cite{chang2013a}. The QAH state occurs when the bulk carrier density is tuned to be closed to zero by applying a gate voltage. The magnetic field dependence of the Hall conductivity shows a hysteresis loop that agrees with the anomalous nature of the QH state. In particular, the Hall conductance is quantized to $\pm \frac{e^2}h$ at zero external magnetic field.  Since then this QAH state has been firmly verified in more and more works \cite{checkelsky2014,kou2014,bestwick2014,kou2015,feng2015,chang2014,kandala2015}, in both Cr doped and V doped ${\rm (Bi,Sb)_2Te_3}$ films. An overview on the QAH effect is provided in Ref. \cite{wang2014c}.

This review article is organized as follows. In Sec. II, we will first present a theoretical toy model, and then discuss the general conditions for the occurence of the QAH effect. Sec. III is devoted to the theoretical prediction of different classes of realistic QAH materials and the recent experimental realization. Sec. IV contains the conclusion and further discussions.

\section{Theory of the quantum anomalous Hall effect}
\label{sec:theory}

\subsection{General setup and history}

\begin{figure} [h]
\includegraphics[width=7cm]{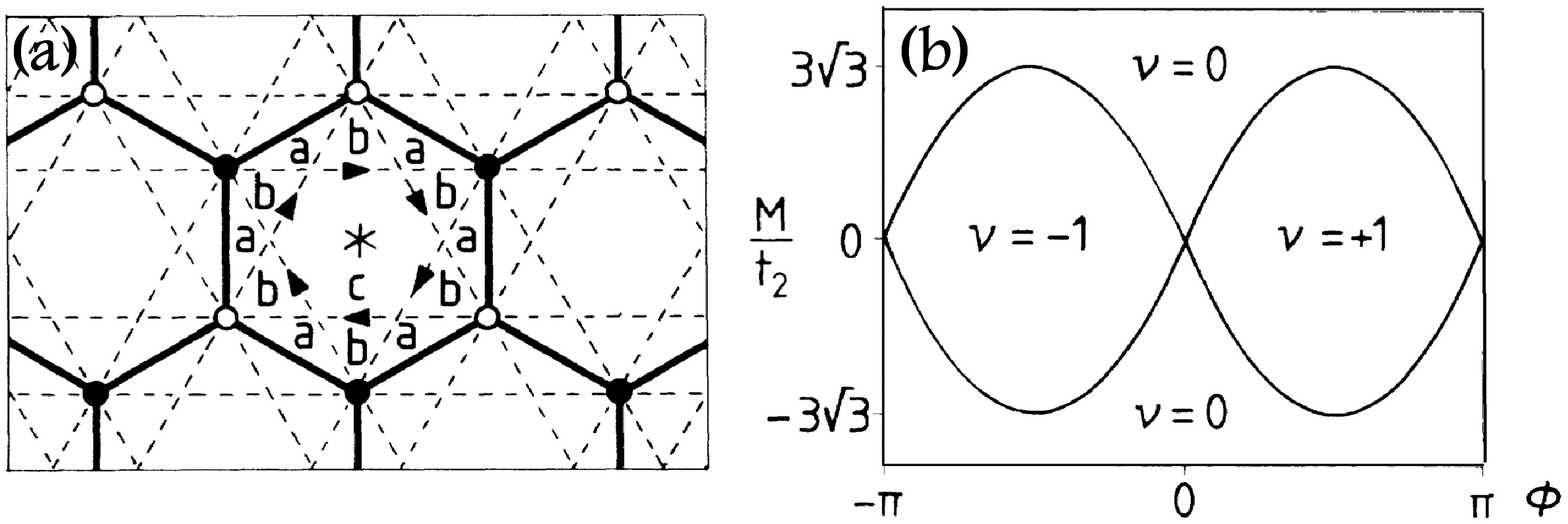}
\caption{(Color online) (a) The honeycomb lattice with the complex next nearest neighbor hopping. (b) The phase diagram of Haldane model where the Hall conductance is given by $\sigma_{xy}=\nu e^2/h$. From Ref. \cite{haldane1988}.  } \label{Haldane}
\end{figure}

In the QH effect, the Hall conductance is quantized to the integer value of $e^2/h$. Consider a general non-interacting electron Hamiltonian $H=\sum_{\bf k}\sum_{\alpha,\beta=1}^Nc_{{\bf k}\alpha}^\dagger h^{\alpha\beta}({\bf k})c_{{\bf k}\beta}$, with $\alpha,\beta=1,2,...,N$ labeling $N$ orbitals in each unit cell. Diagonalization of single particle Hamiltonian matrix $h({\bf k})$ gives the eigenstates $\left|n{\bf k}\right\rangle$. The components $\langle\alpha|n{\bf k}\rangle=u_{n{\bf k}}^\alpha$ describe the periodic part of Bloch wavefunctions for the $n$-th band. For a band insulator with $M$ of the $N$ bands fully occupied, the Hall conductance can be calculated directly using Kubo formula, which at zero temperature is reduced to the following formula:
\begin{eqnarray}
	\sigma_{H}&=&\frac{e^2}{h}\frac{2\pi}\int d^2{\bf k}\left(\partial_{k_x}a_y-\partial_{k_y}a_x\right)\label{eq:TKNN}
\end{eqnarray}
where
\begin{eqnarray}
a_{x(y)}&=&-i\sum_{n=1}^M\langle n{\bf k}|\frac{\partial}{\partial k_{x(y)}}|n{\bf k}\rangle.
\end{eqnarray}
${\bf a}=(a_x,a_y)$ is a gauge field in the momentum space which describes the Berry's phase obtained by adiabatically changing momentum ${\bf k}$ in the Brillouin zone. The Hall conductance is the net flux of the Berry's phase gauge field ${\bf a}$ times a universal constant $e^2/2\pi h$. This formula was first studied for the IQH state and is known as the Thouless-Kohmoto-Nightingale-Nijs (TKNN) formula \cite{thouless1982}. Mathematically, the flux of a gauge field on a compact manifold, such as the Brillouin zone torus, is always quantized, which is called the first Chern number. 

In the QH effect, non-zero Chern number comes from strong magnetic fields and the associated Landau levels. A natural question arises: can non-zero Chern number exist an electronic band without Landau levels? It has been known \cite{nagaosa2010} for a while that Berry curvature in the momentum space can emerge in an electronic band of a metallic system and leads to the so-called intrinsic anomalous Hall effect. Therefore, there is no constraint to forbid the appearance of non-zero Chern number for a band in an insulating system. The first explicit model with quantized Hall conductance in absent of magnetic fields was constructed by F. D. M. Haldane \cite{haldane1988} in the honeycomb lattice. Besides the nearest neighbor hopping between A and B sublattices in the honeycomb lattice, this tight-binding model takes into account the complex second-neighbour hopping induced by a periodic local magnetic flux. The directions of positive phase hopping is indicated by the arrows in Fig. \ref{Haldane}(a). It should be emphasized that the periodic local magnetic flux is chosen to keep zero total flux through one hexagonal unit cell so that the full lattice translation symmetry is respected. The Hamiltonian of this model is given by
\begin{eqnarray}
	&&H({\bf k})=2t_2\cos\phi (\sum_i\cos({\bf k}\cdot{\bf b}_i)){\bf I}\nonumber\\
	&&+t_1\left( \sum_i[\cos({\bf k}\cdot{\bf a}_i)\sigma_1+\sin({\bf k}\cdot{\bf a}_i)\sigma_2 \right)\nonumber\\
	&&+\left[ M-2t_2\sin\phi \sum_i \sin({\bf k}\cdot{\bf b}_i) \right]\sigma_3,
	\label{eq:Ham_Haldane}
\end{eqnarray}
where $\sigma_i$ are the Pauli matrices for A and B sublattices, $t_1$ is for the nearest neighbour hopping, $t_2$ is for the complex second-neighbour hopping, $\phi$ is the phase accumulated by the $t_2$ hopping and $M$ is for on-site energy between A and B sublattices. ${\bf a}_i$ are the displacements from a B site to its three nearest-neighbor A site and ${\bf b}_i$ are displacements for nearest neighbor sites in the same sublattices. When both $M$ and $t_2$ are zero, two Dirac cones appear at the corner of the Brillouin zone (usually denoted as K and K'), which have been well studied ever since the discovery of graphene. As shown in Fig. \ref{Haldane}(b), when the phase $\phi$ of the next nearest neighbor hopping ($t_2$ term) is included, mass terms appear in the Dirac equation and gap the gapless Dirac cones. This gaped system exhibits a non-zero Hall conductivity ($\sigma_{xy}=\nu e^2/h$), and thus realizes the QAH phase.


\subsection{A minimal two-band model}
\label{sec:minimalmodel}

Although Haldane's remarkable model demonstrates the realization of quantized Hall conductance in the absence of magnetic field, it remains theoretical and has not been realized in real materials with honeycomb lattice. In 2005, Ref. \cite{qi2005} discussed the possibility of the QAH effect in semiconductors, and proposed a different two-band tight-binding model with the QAH phase, which is defined on the square lattice and has low energy excitations around momentum ${\bf k}=0$. Compared to the Haldane model, this model turns out to be more closely related to the realistic material proposals of the QAH states\cite{liu2009C,yu2010}. Furthermore, this model is also closely related to the so-called BHZ model of B. A. Bernevig, T. L. Hughes and S. C. Zhang\cite{bernevig2006c}, which describes the first QSH insulator, the HgTe quantum well. This model can be viewed as a lattice regularized Hamiltonian of Dirac fermions, which can be generalized straightfowardly to lattice Dirac models in general dimensions. It turns out that the lattice Dirac models can be used as ``prototype models" which describes all free-fermion topological insulators in different dimensions and symmetry classes\cite{qi2008b}. In the following, we will introduce this simple two-band model and analyze it's physical properties. Based on the explicit understanding gained in this model, we will then discuss the necessary conditions for realizing the QAH phase in realistic systems.

\begin{figure} [h]
\includegraphics[width=7.5cm]{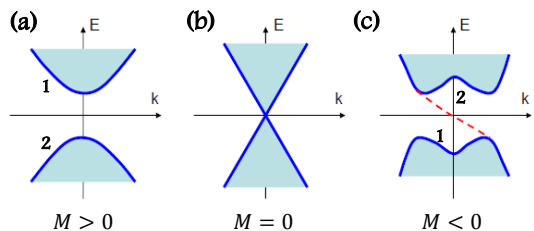}
\caption{(Color online) (a) Normal band structure; (b) Band structure at the critical point; and (c) Inverted band structure. Here we always keep the parameter $B>0$. } \label{bandinversion}
\end{figure}

The minimal two-band model was first introduced by Qi \cite{qi2005}, given by
\begin{eqnarray}
h({\bf k})=\sum_a d_a({\bf k})\sigma^a.\label{QAH}
\end{eqnarray}
where
\begin{eqnarray}
{\bf d} (\mathbf{k})&=&\left(A \sin k_x,-A \sin k_y,\mathbb{M}(\mathbf{k})\right),\nonumber\\
\mathbb{M}(\mathbf{k})&=& M+2B(2-\cos k_x-\cos k_y).\label{eq:twoband}
\end{eqnarray} \noindent
$A,B,M$ are some material related parameters and their physical meaning depends on detailed properties of materials.

The topological property of the model (\ref{QAH}) is determined by the unit-vector $\hat{\bf d}({\bf k})={\bf d}({\bf k})/|{\bf d}({\bf k})|$ in the momentum space.
For generic models with general form (\ref{QAH}), without specifying the form of vector $d_a({\bf k})$ one can directly simplify the TKNN formula (\ref{eq:TKNN}) of the zero temperature Hall conductivity as
\begin{eqnarray}
\sigma_H=\frac{e^2}{h}\frac1{4\pi}\int dk_x\int dk_y\,\hat{\bf d}\cdot
\left(\frac{\partial
\hat{\bf d}}{\partial k_x}\times\frac{\partial \hat{\bf d}}{\partial
k_y}\right).\label{winding2band}
\end{eqnarray}
The expression (\ref{winding2band}) has an explicit geometric meaning. The function $\hat{\bf d}({\bf k})$ defines a mapping from the momentum ${\bf k}$ to the unit vector ($|\hat{\bf d}|=1$) in a sphere, denoted as $\hat{\bf d}({\bf k}): K \rightarrow S^2$. Here $K$ denote the momentum space and in a lattice, it corresponds to the Brillouin zone, belonging to the torus $T^2$. One notices that $\hat{\bf d}\cdot\left(\frac{\partial\hat{\bf d}}{\partial k_x}\times\frac{\partial \hat{\bf d}}{\partial k_y}\right)$ is exactly the Jacobian of this mapping. Thus, the integration over the whole Brillouin zone gives the area which the vector $\hat{\bf d}$ covers, which equals to the integer number times the area of a unit-sphere surface, $4\pi n$. This integer number $n$ is just the winding number for this mapping\cite{semenoff1984,khveshchenko1989,volovik2003,qi2005}.
Consequently, the Hall conductivity is given by $\sigma_H=n\frac{e^2}{h}$.

A more intuitive picture can be obtained by taking the continuum limit of this model around ${\bf k}=0$, given by
\begin{eqnarray}
d_x=A k_x, d_y=-A k_y, d_z=M+B(k^2_x+k^2_y). \label{eq:twoband2}
\end{eqnarray} \noindent
Here we expand the Hamiltonian up to the second order in ${\bf k}$. Without the parabolic term ($B=0$), this Hamiltonian is nothing but massive 2+1D Dirac Hamiltonian with the Dirac mass $M$, which was studied by Jackiw \cite{jackiw1984}. For our purpose, it turns out that the parabolic term is essential, as discussed below. This continuous model corresponds to the case where the gap for large ${\bf k}$ is much larger than that near ${\bf k}=0$, so the corresponding topological property is determined by low energy physics around ${\bf k}=0$. In this case, the Brillouin zone is approximated by a 2D infinite plane. For most general function ${\bf d}({\bf k})$, the quantization of the integral in Eq. (\ref{winding2band}) may fail because ${\bf k}$ is not defined on a compact manifold anymore. However, this is not a problem in the model in Eq. (\ref{eq:twoband}) because ${\bf d}({\bf k})$ has a simple behavior at large ${\bf k}$. For large enough ${\bf k}$, the parabolic term will dominate over other terms, leading to $\hat{\bf d}=sign(B)\hat{\bf z}$. Therefore, there is no contribution to the winding number (\ref{winding2band}) from large momentum, and the quantization of Hall conductivity still holds.
For the Hamiltonian (\ref{eq:twoband}) together with (\ref{eq:twoband2}), one can explicitly show that $\hat{\bf d}=sign(M)\hat{\bf z}$ for $k=0$ and $\hat{\bf d}=sign(B)\hat{\bf z}$ for $k\rightarrow\infty$. Therefore, when $M/B<0$, the $\hat{\bf d}$ vector is perpendicular to the plane and points in opposite directions at $k=0$ and $k\rightarrow\infty$. For other momenta, the $\hat{\bf d}$ vector tilts to the plane and thus forms a skyrmion configuration in the momentum space with a winding number 1, leading to the Hall conductance $e^2/h$. In contrast, for $M/B>0$, the $\hat{\bf d}$ vector points the same direction for $k=0$ and $k\rightarrow\infty$ and gives rise to zero winding number, as well as zero Hall conductance. Thus, the Hall conductivity is quantized to be $e^2/h$ for the regime $M/B<0$ and zero for $M/B>0$. The energy band dispersions for $M/B>0$ and $M/B<0$ are shown in Fig. \ref{bandinversion}. We called the case with $M/B>0$ as normal band structure and that with $M/B<0$ as the ``inverted'' band structure. The ``inverted'' band structure usually possesses topologically non-trivial property. Between these two distinct phases, a topological phase transition occurs at the gapless critical point with $M=0$ and the corresponding system is described by a massless Dirac fermion. The ``inverted'' band structure provides us an intuitive picture to understand how non-trivial topological property occurs in realistic materials and also a helpful guidance to search for realistic topoloigical materials.



Just like the QH effect, one expects chiral edge modes to appear at the boundary of the above two-band model in the inverted regime as a result of non-zero Hall conductance. To see edge states explicitly, let us consider the eigen equation $h(-i\partial_x,k_y)\psi(x)=E\psi(x)$ of the two-band model in a semi-infinite sample ($x>0$) with the boundary condition $\psi(x=0)=0$ and $\psi(x\rightarrow\infty)=0$. We set $k_y=0$ and search for zero energy state ($E=0$). We take the wave function ansatz $\psi\sim e^{\lambda x}\xi$ with two component spinor $\xi$. With this ansatz, the eigen equation is simplified as $(M-B\lambda^2)\xi+A\lambda \sigma_y\xi=0$. Thus, the spinor $\xi$ should be the eigen state of $\sigma_y$, $\sigma_y\xi_s=s\xi_s$ ($s=\pm$), and $\lambda$ satisfies the equation $B\lambda_s^2-sA\lambda_s-M=0$. This suggests that the general solution of the wave function should be given by
\begin{eqnarray}
	\psi(x)&=&(c_{+1}e^{\lambda_{1}x}+c_{+2}e^{\lambda_{2}x})\xi_{+}\nonumber\\
	&+&(c_{-1}e^{-\lambda_{1}x}+c_{-2}e^{-\lambda_{2}x})\xi_{-},
	\label{eq:sol_edge}
\end{eqnarray}
where $\lambda_{i}$ ($i=1,2$) is given by $\lambda_{1,2}=\frac{1}{2B}(A\pm\sqrt{A^2+2BM})$. The boundary condition $\psi(x=0)=0$ leads to $c_{s1}+c_{s2}=0$ for $s=\pm$ because $\xi_+$ and $\xi_-$ are orthogonal to each other. As a consequence, in order to satisfy the condition $\psi(x\rightarrow\infty)=0$, one needs $\lambda_{1}\lambda_{2}=-M/B>0$, which is exactly the condition for band inversion discussed above. Under this condition, one can further show that for $A/B<0$, $\psi(x)=c_{+1}(e^{\lambda_{1}x}-e^{\lambda_{2}x})\xi_{+}$ due to $Re(\lambda_{1,2})<0$, while $A/B>0$, $\psi(x)=c_{-1}(e^{-\lambda_{1}x}-e^{-\lambda_{2}x})\xi_{-}$. Therefore, only one kind of spin state is normalizable for the zero mode. Furthermore, we can consider the dispersion along the $k_y$ direction for the zero mode. To the leading order, if we assume the wavefunction of the zero mode does not depend on $k_y$, i.e. if the edge state wavefunction with momentum $k_y$ is given by $\phi_{k_y}(x,y)=\psi(x)e^{ik_yy}$, the energy of this state is simply
\begin{eqnarray}
	E\left(k_y\right)=\left\{
	\begin{array}{c}
		-Ak_y \qquad A/B<0\\
		Ak_y \qquad A/B>0
	\end{array}
	\right.
	\label{eq:Heff_edge}
\end{eqnarray}
The linear dispersion shows that the edge state is chiral, with a velocity determined by $A$ and $B$. 
The existence of chiral edge modes directly demonstrate the non-zero quanitzed Hall conductivity in the 2D bulk system.

\subsection{Two conditions for realistic materials}
\label{sec:conditions}
In the above, we have presented a minimal model for the QAH effect and obtain the quantized Hall conductivity, as well as chiral edge modes, for this model. Based on this understanding, one may ask what are the physical criteria for the QAH effect that will guide our search for realistic QAH materials. One essential condition is inverted band structure (or $M/B<0$ for the minimal model). In realistic systems, it is known that time reversal invariant topological insulators possess this type of band structure\cite{bernevig2006c}. For topological insulators, inverted band structure is driven by strong spin-orbit coupling and time reversal symmetry is preserved. Thus, Hall conductance must be zero and its edge/surface state is of helical type, consisting of both spin. We may make an analog and require the band inversion also to occur only for one species of spin, accompanied by the breaking of time reversal symmetry. This can be achieved in ferromagnetic materials with a large spin splitting due to the so-called exchange coupling between magnetic moments and electron spin. Thus, to look for the QAH effect, we require ferromagnetic insulators with a strong exchange coupling.

This disucssion suggests two necessary conditions for the QAH effect: (1) inverted band structures and (2) ferromagnetic insulators. For the first condition, we may consider materials with electronic band structure in the inverted regime or close to the transition point when there is no magnetization. Only for these materials, it is possible for the exchange coupling to drive the system into the inverted regime. This indicates that we should consider TR invariant TIs, of which the band structure is already inverted, or other narrow gap (or zero gap) semiconductors. The second condition is not easy to satisfy since ferromagnetism usually coexists with metallic behaviors and ferromagnetic insulators are rare. The reason is that in most ferromagnetic systems, ferromagnetism originates from RKKY mechanism, of which ferromagnetic coupling is usually mediated by free carriers. For example, the QAH effect was first predicted in Mn doped HgTe quantum wells \cite{liu2009C}. However, due to the lack of free carriers, magnetic moments of Mn atoms in this system is coupled antiferromagnetically rather than ferromagnetically. This difficulty prevents a direct observation of the QAH effect in this system. There are several ferromagnetic insulators, such as EuO and GdN. However, inverted band structure is difficult to be realized in these materials due to the large band gap. This difficulty was overcome in magnetically doped Bi or Sb chalcogenides. A key insight \cite{yu2010} is that strong spin-orbit coupling can significantly enhance spin susceptibility, particularly in the inverted band regime, through the Van Vleck mechanism. With this mechanism for ferromagnetic insulators in an inverted material, the last missing piece for realizing QAH state was found, and this theoretical prediction was verified by experiments soon afterwards. In the next section, we will follow the history of this field and discuss Mn doped HgTe quantum wells first, which is also helpful for illustrating the general mechanism of relating the QAH effect with time-reversal invariant topological insulators. Then we will discuss the rapid development of the theoretical and experimental studies of this effect in magnetically doped Bi or Sb chalcogenides. Finally, we will comments on other possible materials.

\section{Realistic materials and experiments}
\label{sec:mat}

\subsection{Magnetically doped HgTe and InAs/GaSb quantum wells}
HgTe and CdTe are typical II-VI group compound semiconductors, which possess the zinc-blende lattice structure with both the anion and cation atoms forming two interpenetrating face-centered-cubic lattices. Similar to other II-VI and III-V semiconductors, the bands of HgTe and CdTe near Fermi energy are a s-type band ($\Gamma_6$), and a p-type band split by spin-orbit coupling into a $J=3/2$ band ($\Gamma_8$) and a $J=1/2$ band ($\Gamma_7$), where $J$ denotes the total angular momentum. CdTe has an energy gap ($\sim 1.6$eV) and its band ordering is normal with s-type ($\Gamma_6$) bands above p-type ($\Gamma_7$ and $\Gamma_8$) bands. In constrast, HgTe has an inverted band structure \cite{bernevig2006c}: $\Gamma_6$ bands are below $\Gamma_8$ bands with a negative energy gap of $\sim 0.3$eV. HgTe can form a quantum well (QW) structure with CdTe as a barrier, in which both the conduction and valence bands are splitted into sub-bands, denoted as $|E_n\rangle$ for electron $\Gamma_6$ bands and $|H_n\rangle$ for heavy-hole $\Gamma_8$ bands ($n=1,2,\cdots$), due to the quantum confinement. Because of the opposite band sequence between CdTe and HgTe, a topological phase transition exists in this system at a critical thickness of the quantum well, denoted as $d_c$ ($d_c\approx 6.3$nm), across which the band sequence between $|E_1\rangle$ and $|H_1\rangle$ sub-bands is reversed. For the thickness $d<d_c$, $|E_1\rangle$ state has higher energy than $|H_1\rangle$ state (normal band structure) while for $d>d_c$, $|E_1\rangle$ is below $|H_1\rangle$, leading to a similar inverted band structure as descibed in the last section. However, different from our simple model, we need to consider electron spin states for both $|E_1\rangle$ and $|H_1\rangle$ sub-bands. Thus, under the basis $|E_1,+\rangle$, $|H_1,+\rangle$, $|E_1,-\rangle$ and $|H_1,-\rangle$ (here $\pm$ labels spin states), the Hamiltonian describing the low energy physics in HgTe/CdTe quantum wells is written as
\begin{eqnarray}
      H=\left(
      \begin{array}{cc}
	      h(k)&0\\
	      0&h^*(-k)
      \end{array}
      \right),
	\label{eq:BHZ}
\end{eqnarray}
where $h(k)$ is given by the Hamiltonian (\ref{eq:twoband}). Therefore, this Hamiltonian is just two copies of the two band model with each copy described by the minimal two-band model (\ref{QAH}) and is first described by B. A. Bernevig, T. L. Hughes and S. C. Zhang\cite{bernevig2006c}, known as the BHZ Hamiltonian. These two copies are for spin up and spin down states and related to each other by TR symmetry. The BHZ Hamiltonian forms the basis of the QSH effect, for which edge states also exist and reveal a helical nature.

\begin{figure} [h]
\includegraphics[width=8cm]{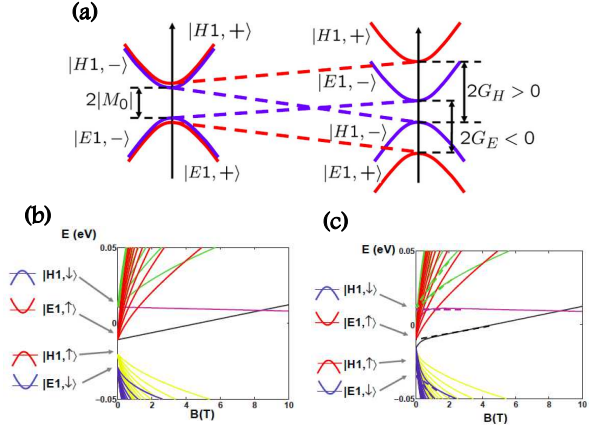}
\caption{(Color online) (a) Evolution of band structure and edge states upon increasing the spin splitting for Mn doped HgTe quantum wells. Band inversion occurs for one species of spin but not for the other. This figure and more details can be found in Ref. \cite{liu2009C}. The Landau levels are shown as a function of magnetic fields for the case (b) with initial magnetization and (c) without initial magnetization.  } \label{HgTe}
\end{figure}

The band inversion for electrons with both spins does not directly give QAH effect since time-reversal symmetry has not been broken. However, it does make the realization of QAH state easier, since there is actually two QAH effects (for opposite spins) canceling each other, and we simply need to break time reversal and turn off one of them. For this purpose, magnetization is introduced to this system\cite{liu2009C}, so that the two spin blocks are no longer related. Physically, magnetization is introduced by Mn doping. ${\rm Mn}^{2+}$ substitutes ${\rm Hg^{2+}}$, and the localized spin of Mn ion couples with that of itinerant electrons through exchange coupling 
\begin{eqnarray}
	H_s=\left(
	\begin{array}{cccc}
		G_E&0&0&0\\
		0&G_H&0&0\\
		0&0&-G_E&0\\
		0&0&0&-G_H
	\end{array}
	\right),
	\label{eq:Hex}
\end{eqnarray}
where spin splitting is $2G_E$ for the $|E_1,\pm\rangle$ band and $2G_H$ for the $|H_1,\pm\rangle$ band. The corresponding energy gap is $E_g^+=2M_0+G_E-G_H$ for spin up block while $E_g^-=2M_0-G_E+G_H$ for spin down block. To realized the QAH effect, we require that (1) the state with one kind of spin is in the inverted regime while the other goes to the normal regime; and (2) the entire system is in the insulating regime. Combining these two requirement leads to a simple condition $G_EG_H<0$, which means that the spin splittings for electron sub-bands and hole sub-bands are opposite. As shown in Fig. \ref{HgTe}(a), in this case, the energy gap for one copy of spin is increased while that for the other spin is decreased. Thus, the low energy physics is just described by the two-band model (\ref{eq:twoband}), which can result in the QAH phase.
In the other regime where $G_EG_H>0$, electrons with opposite spin still acquire different gaps, but the system becomes metallic before the two blocks develop an opposite Hall conductance.

The above discussion suggests that to realize inverted band structure for one spin but normal band dispersion for the other, the key condition is that spin splitting for electron and heavy hole sub-bands has opposite sign ($G_EG_H<0$). Fortunately, in HgTe QW doped with Mn, sp-d exchange coupling indeed gives the opposite signs for spin splitting of $|E_1\rangle$ and $|H_1\rangle$ bands. The underyling reason is different exchange mechanism for the coupling between electron spin and magnetic moments for $\Gamma_6$ and $\Gamma_8$ bands. For $\Gamma_8$ bands consisting of p orbitals, there is a large hybridization between p orbital from anions, such as Te, and d orbital from Mn which replace cation positions. As a result, the so-called p-d exchange mechanism gives rise to antiforromagnetic coupling between $\Gamma_8$ bands and magnetic moments. In contrast, the $\Gamma_6$ bands come from s orbitals of cations, such as Hg, and its hybridization to d orbitals of Mn atoms is forbidden by symmetry. Thus, the direct exchange coupling is dominant here, leading to ferromagnetic coupling. The direct calculation based on the eight band Kane model indeed confirms the opposite exchange coupling for electron and heavy hole sub-bands, suggesting that the required inverted band structure can be satisfied in this system.

However, the difficulty of this system lies in the fact that it is a paramagnetic rather than ferromagnetic material. The magnetization in Mn doped HgTe satisfies the Curie-Weiss rule
\begin{eqnarray}
	\langle\mathcal{S}\rangle=-S_0B_{5/2}\frac{5g_{Mn}\mu_BB}{2k_B(T+T_0)},
	\label{eq:CurieWeiss}
\end{eqnarray}
where $S_0=5/2$, $B_{5/2}$ is the Brillouin function and $T_0$ stands for a weak antiferromagnetic coupling between magnetic moments. Due to $T_0$, Mn doped HgTe reveals spin glass behaviors at low temperatures, which is the main obstacle to observe the QAH effect in this system experimentally. Therefore, a small magnetic field is required to magnetize the system and the orbital effect (Landau levels) is inevitable. However, it is still possible to distinguish the effect of magnetization from the orbital effect by checking the behaviors of Landau levels. We may first check the Landau level with initial magnetization at a zero magnetic field, as shown in Fig. \ref{HgTe}(b). One can see that the gap at a zero magnetic field corresponds to the regimes with $-e^2/h$ Hall conductance in the Landau level fan diagram. This indicates that the QAH effect and the corresponding QH effect are adiabatically connected. For the realistic situation, there is no magnetization at a zero magnetic field, so the zero field gap is connected to the zero Hall conductance regime. However, from Fig. \ref{HgTe}(c), one can see that the energy of the zeroth Landau level of the valence band (black line) first shows a rapid and nonlinear increase in energy at low magnetic field and then follow the linear behavior at higher field. This is because magnetic moments can be easily polarized by magnetic field and saturate at a magnetic field weaker than 1T. Therefore, if Fermi energy is tuned into the bulk gap, we expect the quantized Hall conductance can be observed at a small magnetic field. Indeed, transport experiments performed by Molenkamp's group \cite{buhmann2002} have indeed observed an extremely long quantized Hall plateau with the Hall conductivity $-e^2/h$ with magnetic fields ranging from 0.3T up to 30T. This provides the evidence of the quantized Hall conductance plateau induced by magnetization, rather than the magnetic field in this system. Moreover, different behaviors between exchange coupling of magnetic moments and the orbital effect from magnetic fields can also be unveiled by rotating magnetic fields \cite{hsu2013}.

%


More recently, it was proposed that the QAH effect can be realized in another diluted magnetic semiconductor heterostructure, Mn doped type II InAs/GaSb quantum wells \cite{wang2014}. Both InAs and GaSb are III-V group semiconductors. The unique feature of these two materials is that the conduction band minimum of InAs has lower energy than the valence band maximum of GaSb, due to the large band offset. As a consequence, for the InAs/GaSb/AlSb quantum wells, in which InAs and GaSb together serve as well layers and AlSb layers are for barriers, the first electron subband $|E_1\rangle$ of InAs layers lies below the first hole subband $|H_1\rangle$ of GaSb layers, when the well thickness is large enough. The effective model for this system is essentially the same as that for HgTe/CdTe QWs and this material has been theoretically predicted to be a QSH system \cite{liu2008}. Therefore, it is natural to expect that magnetic doping can drive this system into the QAH state. It turns out that there are two salient features making this system more attractive than Mn doped HgTe QWs. Firstly, Since the $|E_1\rangle$ band in the InAs layer is lower than the $|H_1\rangle$ band in the GaSb layer, there is a charge transfer between these two layers. Therefore, if one neglects the coupling between these two layers, InAs is electron-doped while GaSb is hole-doped. The coupling between these two layers can open a hybridization gap. However, since two bands are separated into two layers, this hybridization gap is quite small. Correspondingly, in the effective model (\ref{eq:BHZ}), the coefficient $A$ in the linear term is much smaller in this system than that in HgTe QWs. Due to the small $A$ term, energy dispersion reveals a Mexican hat shape for both the conduction and valence bands, resulting in a peak of the density of states around band edges. This band edge singularity is responsible for the enhancement of spin susceptibility discussed below. Secondly, both Mn doped InAs and GaSb are diluted magnetic semiconductors, and the exchange mechanism is consistent with free-hole-mediated ferromagnetism within a mean field Zener model \cite{dietl2000,jungwirth2006,sato2010a,dietl2010}. If one neglects the coupling between InAs and GaSb layers, there are free carriers in both layers due to charge transfer, leading to ferromagnetism in the system. However, when hybridization gap is taken into account, free-hole mediated mechanism does not work. Thanks to the band edge singularity, spin susceptibility is enhanced even for the Fermi energy in the hybridization gap. Thus, the ferromagnetic insulating phase is possible in this system. By combining the first principles calculations and the analytical Kane model, it was found that both conditions can be naturally satisfied for magnetically doped InAs/GaSb quantum wells, which provide good candidates for realizing the QAH insulator at a high temperature.

Mn doped HgTe QWs and InAs/GaSb QWs provide examples of the QAH states in II-VI and III-V compound semiconductor heterostructures. A more general scheme based on p-n junction structure has been proposed by H.J. Zhang {\it et al} \cite{zhang2014} for other II-VI, III-V and IV semiconductors. The authors have shown that for some narrow gap semiconductors, inverted band structure can naturally exist at the interface between p and n-type region. These structures are dubbed ``junction quantum wells'' in the limit where both the p-type and n-type regions are narrow, and comparable in size to the interface region, and can host the QSH state. This scheme can also be applied to conventional diluted magnetic semiconductors, leading to the QAH state.

\subsection{Magnetically doped (BiSb)$_2$(TeSe)$_3$ family of materials}

Tetradymite semiconductors Bi$_2$Te$_3$, Bi$_2$Se$_3$ and Sb$_2$Te$_3$ is a family of three dimensional TR invariant TIs with the helical surface state of a single Dirac cone at each surface\cite{hjzhang2009,xia2009}. It has been shown that a thin film of 3D TI crosses over to a 2D insulator which is either a TI (the QSH state) or a trivial insulator, depending on the thickness of the film in an oscillatory fashion\cite{liu2010a,lu2010}. Therefore, inverted band structures can be realized in this class of materials in a controllable manner. In addition, it is remarkable that the ferromagnetism can be realized in this class of materials by doping with proper transition metal elements, such as Cr, Fe and V, even in the insulating regime, which is predicted by Yu {\it et al} \cite{yu2010}. Based on the first principles calculations and the mean field Zener model, Yu {\it et al} showed that spin susceptibility is significantly enhanced due to the Van Vleck mechanism in the inverted regime. Therefore, inverted band structure can drive the system into a ferromagnetic insulating phase. This  prediction was soon confirmed by the experiment, in which by tuning the chemical composition of Te and Se in Bi$_2$(Te$_{1-x}$Se$_x$)$_3$, a topological phase transition between TIs and normal insulators is accompanied by a magnetic phase transition between ferromagnetism and paramagnetism \cite{zhang2013a}. Therefore, both conditions for the QAH effect are naturally satisfied in this class of materials.

\begin{figure} [h]
\includegraphics[width=8cm]{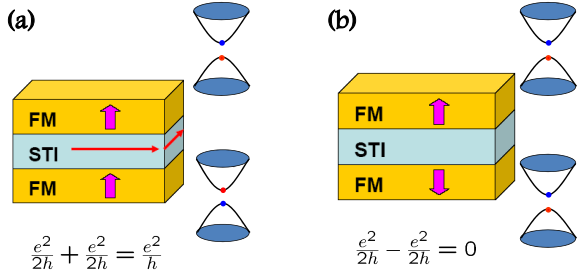}
\caption{(Color online)  Illustration of the Hall conductance in ferromagnet-topological insulator heterostructure. (a) When the magnetizations at the top and bottom surfaces are parallel, there is a quantized Hall conductance accompanied by a chiral edge state in topological insulator films. (b) When the magnetizations are anti-parallel, the Hall conductances from the top and bottom surfaces cancel each other. Here STI is for strong topoogical insulator and FM is for ferromagnetism \cite{qi2008b}.
} \label{TIsurface}
\end{figure}

To understand the mechanism of QAH effect in this family of materials, we do not need to start from the full 3D theory. Instead, it is sufficient to start from the two massless surface states at the top and bottom surfaces of a TI film.
For each surface Dirac cone, its effective model is similar to our two band model, but with the mass term vanishing $M({\bf k})=0$ due to time reversal symmetry. When magnetization is introduced in the surface and couples to surface electron spin, the mass $M$ can be non-zero and open a gap for surface Dirac cones. We emphasize that in this case, the quadratic term $B$ in Eq. (\ref{eq:twoband}) is still zero. For this case, we find the $\hat{\bf d}$ vector is perpendicular to the plane for $k=0$ but lies in the plane for $k\rightarrow \infty$. Therefore, it forms half of a skyrmion configuration, known as a ``meron''. For such a "meron" configuration, the $\hat{\bf d}$ vector only covers half of the unit sphere, leading to a winding number $\pm 1/2$ and corresponding to a Hall conductance $\pm \frac{e^2}{2h}$. The sign of the Hall conductance depends on the magnetization direction (or the sign of mass $M$). This half Hall conductance of 2D massive Dirac model has been studied for a long time in high energy physics, known as the ``parity anomaly''\cite{redlich1984,jackiw1984,haldane1988,fu2007b,qi2008b}. We emphasize that this half Hall conductance only exists at one surface of 3D TIs and distinguish topological surface states from any purely 2D states. In a realistic 2D system, this half Hall conductance cannot be observed since there are always even number of 2D Dirac fermions, known as the ``fermion doubling theorem'' \cite{nielsen1981,nielsen1981a}. For our interest, we may consider a TI thin film, for which there are two surfaces, top and bottom surfaces. Thus, the total number of Dirac cones is two in the whole film, as is required by the fermion doubling theorem. We may introduce magnetization to gap both surfaces. When magnetic moments on both surfaces are in parallel, we find both surface states contribute to the Hall conductance with the same sign, resulting in a total Hall conductance of $\pm e^2/h$, as shown in Fig. \ref{TIsurface}(a) \cite{qi2008b}. This exactly corresponds to the QAH effect that we are looking for. When magnetic moments of two surfaces are opposite, the Hall conductance for two surface states cancels each other, so there is no total Hall conductance (Fig. \ref{TIsurface}(b)). This discussion concludes that uniform magnetization in TI thin films can lead to the QAH effect.

In a thin film of TI, one cannot neglect the coupling between top and bottom surface states. A low energy effective model, taking into account this type of coupling, was proposed in Ref. \cite{yu2010} as
\begin{eqnarray}
	H_{sf}=\left(
	\begin{array}{cc}
		v_f(-k_x\sigma_y+k_y\sigma_x)&m_k\\
		m_k&v_f(k_x\sigma_y-k_y\sigma_x)
	\end{array}
	\right),
	\label{eq:Hsurf}
\end{eqnarray}
where $\sigma$ is the Pauli matrices for spin and the upper and lower blocks denote two surfaces respectively. Thus, the basis is given by $|t\uparrow\rangle$, $|t\downarrow\rangle$, $|b\uparrow\rangle$ and $|b\downarrow\rangle$ where $t,b$ are for top and bottom surfaces and $\uparrow,\downarrow$ are for spin. $v_f$ is Fermi velocity and $m_k=M+B(k_x^2+k_y^2)$ describes the coupling between top and bottom surfaces. The exchange coupling between magnetic moments and electron spin is given by
\begin{eqnarray}
	H_{ex}=\left(
	\begin{array}{cc}
		G_0\sigma_z&0\\
		0&G_0\sigma_z
	\end{array}
	\right).
	\label{eq:Hex2}
\end{eqnarray}
At the first sight, this Hamiltonian is different from the Hamiltonian (\ref{eq:BHZ}). One can consider the binding and antibinding states of two surface states, $|\pm\sigma\rangle=(|t\sigma\rangle\pm|b\sigma\rangle)/\sqrt{2}$, with $\sigma=\uparrow,\downarrow$. Under these basis, the Hamiltonian is transformed to
\begin{eqnarray}
	H_{sf}+H_{ex}=\left(
	\begin{array}{cc}
		h(k)+G_0\sigma_z&0\\
		0&h^*(k)-G_0\sigma_z
	\end{array}
	\right)
	\label{eq:Hsurf2}
\end{eqnarray}
under the basis $|+\uparrow\rangle$, $|-\downarrow\rangle$, $|+\downarrow\rangle$
and $|-\uparrow\rangle$. Here $h(k)=m_k\sigma_z+v_f(k_y\sigma_x-k_x\sigma_y)$, which is similar to that in the Hamiltonian (\ref{eq:BHZ}). For exchange coupling, compare to the case of Mn doped HgTe QWs, one can notice that the situation here indeed corresponds to the case with $G_EG_H<0$ in the exchange coupling (\ref{eq:Hex}). Thus, the conditions for the inverted band structure and an ferromagnetic insulating state are automatically satisfied and this analysis suggests the existence of the QAH effect in magnetically doped Bi$_2$Te$_3$ family of materials.

\begin{figure} [h]
\includegraphics[width=8cm]{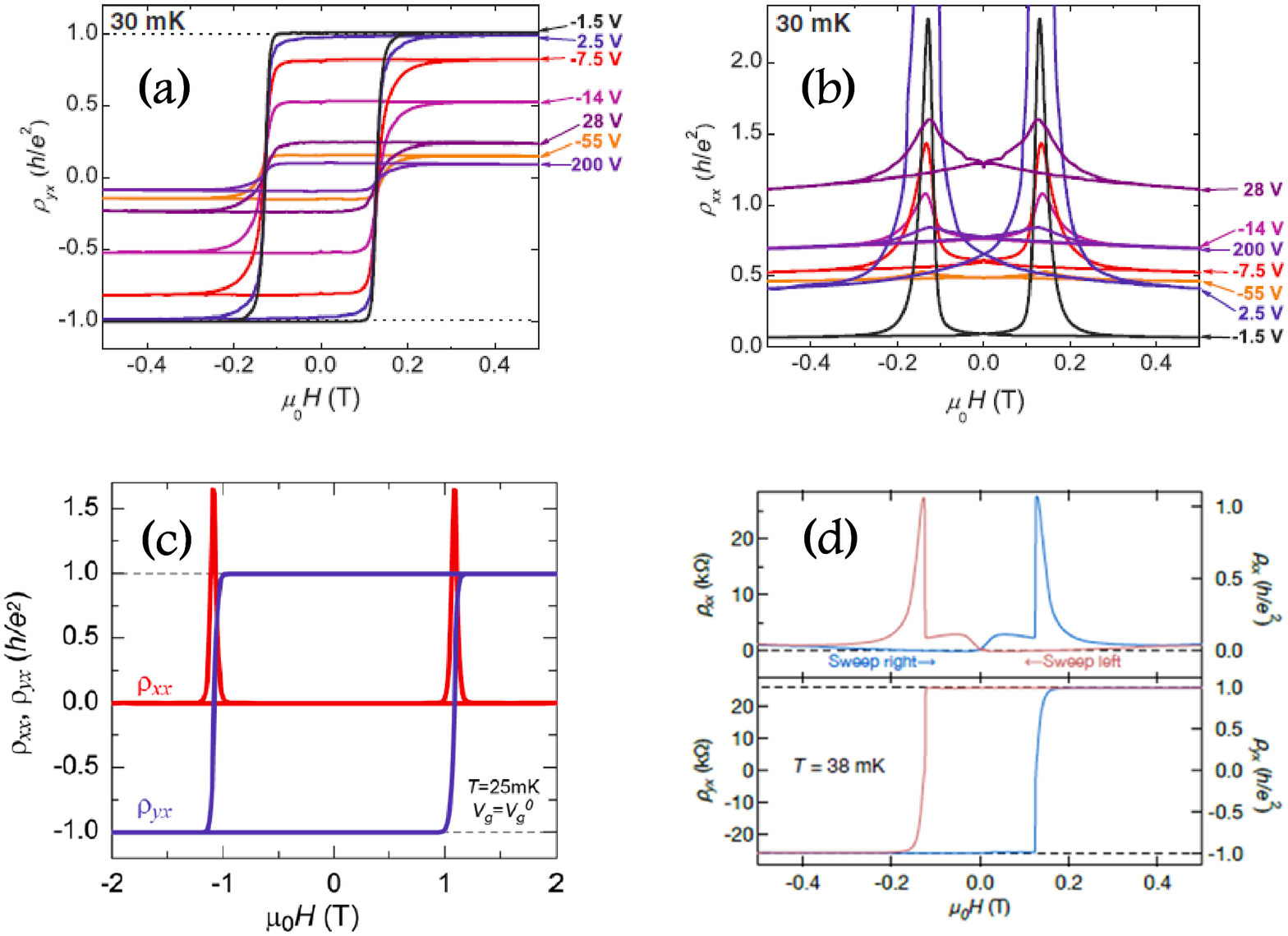}
\caption{(Color online) (a) The Hall resistance and (b) longitudinal resistance are measured as a function of magnetic fields for different gate voltages in Cr doped (Bi,Sb)$_2$Te$_3$ films \cite{chang2013a}. The quantized Hall resistance is found at the gate voltage $-1.5V$. Precise quantization of Hall resistance and negligible longitudinal resistance at a zero magnetic field were observed in both (c) V doped \cite{chang2014} and (d) Cr doped (Bi,Sb)$_2$Te$_3$ films \cite{bestwick2014}.   } \label{QAHexp}
\end{figure}

This theoretical prediction was first confirmed experimentally in the transport measurement of Cr doped (Bi,Sb)$_2$Te$_3$ thin films \cite{chang2013a}. The first experimental attempt was to achieve stable ferromagnetism in Bi or Sb chalcogenides. It turns out that magnetic property of this family of materials dramatically depends on both parent compounds and magnetic dopants. For example, a robust out-of-plane ferromagnetism was observed in Cr, V and Mn doped Bi$_2$Te$_3$ or Sb$_2$Te$_3$ \cite{chang2013b,hor2010b,kou2013b,kou2013a,li2015,dyck2005}, while spin glassy behavior was observed in Fe doped Bi$_2$Te$_2$ \cite{kim2013a}. For Mn doped Bi$_2$Se$_3$, ferromagnetism is only found on the surface and magnetization measurements reveal hysteresis for both the out-of-plane and in-plane magnetizations \cite{bardeleben2013,zhang2012a}. For the purpose of realizing the QAH effect, we require ferromagnetism out of the plane. Therefore, it is natural to consider Cr, V and Mn doped Bi$_2$Te$_3$ or Sb$_2$Te$_3$.

The first experimental observation of the QAH effect was reported in Cr doped (Bi,Sb)$_2$Te$_3$ thin film with five quintuple layers \cite{chang2013a}. As shown in Fig. \ref{QAHexp}(a) and (b), hysterisis loop in Hall and longitudinal resistivity was observed and Hall resistivity was almost $h/e^2$ at a zero magnetic field when the temperature is lowered to $30mK$. At the same time, the longitudinal resistivity drops significantly to $0.098 h/e^2$ ($\sim2.53k\Omega$). Furthermore, the observed plateau at a zero magnetic field is adiabatically connected to the quantum Hall plateau at high magnetic fields. All these observations confirm the existence of the QAH effect in this system. This experimental breakthrough has immediately aroused great interest of researchers in this field and was soon reproduced by several other experimental groups \cite{checkelsky2014,kou2014,bestwick2014,kou2015,feng2015,chang2014,kandala2015}.

\begin{figure} [h]
\includegraphics[width=8cm]{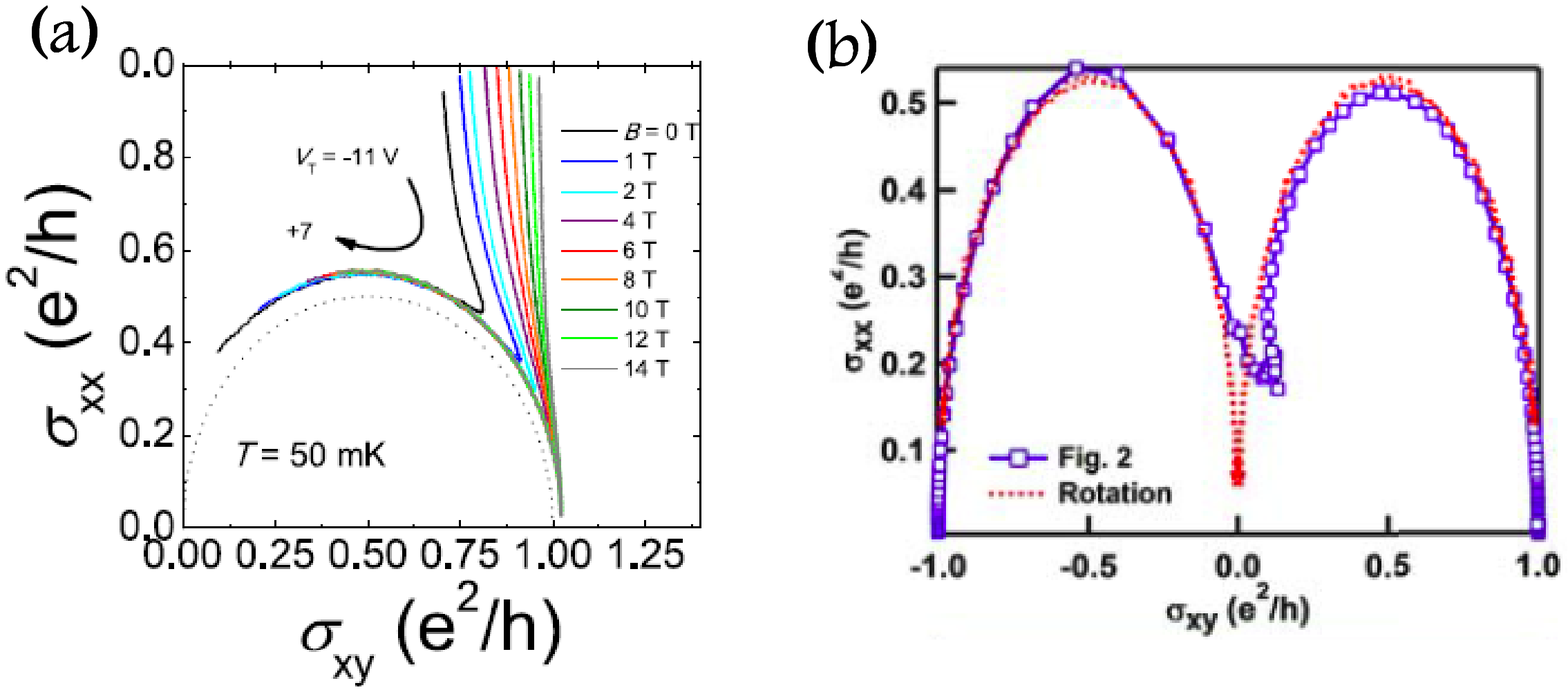}
\caption{(Color online) The global phase diagram of conductance tensor is constructed by (a) varying gate voltage for different magnetic fields \cite{checkelsky2014} and (b) tilting magnetic fields \cite{kou2015}.  } \label{phase}
\end{figure}

Nevertheless, for the first experiment of the QAH effect, a relatively large residual longitudinal resistance (around $2.5k\Omega$) always remains and the corresponding Hall conductance is around $0.987 e^2/h$ at zero magnetic field. Furthermore, this effect only occurs around 30mK and the longitudinal resistance will increase significantly once the temperature is increased to 1K. Therefore, an improvement of sample quality is required to demonstrate the precise quantization of this effect. In Cr doped (Bi,Sb)$_2$Te$_3$ films, the quantization of the Hall conductance was improved to the value within $0.01\%$ of $e^2/h$ and the longitudinal resistance reached as low as $15\Omega$ through adiabatic demagnetization cooling of the system
\cite{bestwick2014}. More recently, the QAH effect was observed in V doped (Bi,Sb)$_2$Te$_3$ \cite{chang2014} with the zero-field longitudinal resistance down to $0.00013\pm0.00007 h/e^2$ ($\sim3.35\pm1.76\Omega$), Hall conductance reaching $0.9998\pm0.0006e^2/h$ and the Hall angle becoming as high as $89.993\pm0.004^o$ at $T=25mK$. The advantage of this material lies in the fact that it is a hard ferromagnet with a large coercive field ($H_c>1.0T$) and a relative high Curie temperature.
These observations lay a firm foundation for the future experiments and applications of the QAH effect.

More recent experiments start exploring different aspects of the observed QAH effect. Ref. \cite{kou2014} observed the QAH effect in thicker Cr doped (Bi,Sb)$_2$Te$_3$ films up to 10 quintuple layers. Non-local transport measurements\cite{kou2014} have also been performed to reveal the coexistence of chiral edge modes and non-chiral ones, which has been predicted theoretically \cite{wang2013b}. Ref. \cite{checkelsky2014} first explored the global phase diagram of the QAH effect. By mapping the behavior of the conductivity tensor as a function of temperature, magnetic field and chemical potential, they find the delocalization behvior of the QAH effect quantitatively agrees with the theoretical prediction based on renormalization group analysis for the QH state \cite{wang2014a,kivelson1992,khmel1983}. For example, in Fig. \ref{phase}(a), all lines of conductance tensor for different magnetic fields clearly collapse onto the same semi-circle when the contribution of bulk carriers is not significant. The global phase diagram is further clarified by Ref. \cite{kou2015} by tilting the magnetic field. In Fig. \ref{phase}(b), two semi-circles for conductance tensor are revealed with the longitudinal conductance minimum at the $\pm e^2/h$ Hall plateau, as well as the unique zero resistance plateau \cite{kou2015,feng2015}, as predicted by Wang {\it et al} \cite{wang2014a}. These observations establish the equivalence between the QAH effect and QH effect in term of their topological nature.

\subsection{Other classes of materials}

Besides the two classes of materials reviewed above, the QAH effect was also proposed in a large variety of materials, including graphene, silicene, germanene and stanene systems\cite{haldane1988,qiao2010,tse2011,qiao2014,zhang2012b,ezawa2012,pan2014,wu2014,xu2013}, transition metal oxides \cite{xiao2011,wang2014b,cook2014}, and others \cite{dong2014}. Hybrid structures combining magnetic insulators with TIs are also discussed \cite{garrity2013}.

Graphene of a honeycomb lattice structure was the first model system for the study of the QAH effect, as proposed by Haldane \cite{haldane1988}. In Ref. \cite{raghu2008} it was shown that a strong interaction can induce a spontaneous symmetry breaking and drive the system into the QAH state. More sophisticated theoretical calculations suggest that the QAH can also be realized in graphene once exchange coupling and spin-orbit coupling are introduced \cite{qiao2010,tse2011,qiao2014} by, for example, decoration with 5d transition-metal adatoms \cite{zhang2012b}. The next development along this line is about other monolayer materials with honeycomb lattice structure, such as silecene, germanene and stanene, which are formed by a single layer of buckled honeycomb lattice of silicon, germanium and tin, respectively. Since these elements are heavier than carbon, intrinsic spin-orbit coupling in these materials is much stronger than that in graphene. Thus, once exchange coupling is introduced to these systems, the QAH effect is naturally expected \cite{ezawa2012,pan2014,wu2014,xu2013}. It should be emphasized that for germanene and stanene, low energy physics is dominated by the band inversion between s orbitals and $p_x-p_y$ orbitals, rather than $p_z$ orbitals \cite{xu2013}. It has been shown that ferromagnetism can emerge in germanene and stanene by controlling surface functionality instead of magnetic doping \cite{wu2014}.

Transition metal oxides (TMO) also provide another potential family of materials to search for QAH effect since 4d and 5d elements of transition metals naturally carry large magnetic moments and strong spin-orbit coupling. It was first pointed out by D. Xiao \cite{xiao2011} that the (111) bilayer heterostructure of perovskite TMO can be viewed as a buckled honeycomb lattice and has natively inverted band structure. This leads to material proposals of bilayer LaAuO$_3$, LaNiO$_3$ and LaCoO$_3$ heterostructures \cite{wang2014b} and some TMO materials with double perovskite structure \cite{zhang2014b,cook2014}. 

Besides these families of materials, hybrid structures \cite{garrity2013,zhang2014a} have also been discussed for the QAH effect and these systems have the advantage of separating topological nature and ferromagnism into two different materials, which can possibly allow the effect to be realized in a stronger ferromagnet and at higher temperature. In particular, Ref. \cite{zhang2014a} proposed the QAH effect in materials with 5d elements.
Potential candidate materials for QAH effect with the Hall conductance higher than $e^2/h$ have also been considered theoretically\cite{wang2013a,zhang2013b,fang2014,dong2014}. The QAH effect has also been proposed in other magnetic structures \cite{liu2013a,liang2013}.




\section{Conclusion and discussions}


In this article, we have reviewed the theoretical prediction and experimental realization of the QAH effect. By making use of spin-orbit coupling and magnetism, a material without external magnetic field can have quantized Hall conductance, as a consequence of the Berry's phase gauge field in momentum space. This effect has been predicted and experimentally verified in magnetically doped topological insulators. The discovery of the QAH effect has significance in both fundamental science and potential applications. Different from conventional QH effect, the QAH effect is a direct demonstration of {\it intrinsic} topological properties of a material, rather than those imposed by an external magnetic field. Viewing from a different angle, the materials with QAH effect are deeply connected to topological properties of both two-dimensional and three-dimensional topological insulators. The observation of QAH effect has provided a direct demonstration of the topological response properties of topological insulators in electrical transport measurements. In addition, realizing QAH state also opens the door for future experiments to verify other topological properties of TI related to time-reversal breaking at the surface, such as the topological magneto-electric (TME) effect\cite{qi2008b}, the image monopole effect\cite{qi2009} and topological Faraday rotation\cite{qi2008b,maciejko2010b}.

There are various generalizations of QAH effects in other systems and physical situations. Similar to the QH effect, the QAH effect also has a fractional QH generalization in strongly interacting systems, which has been recently proposed and studied theoretically, known as the fractional Chern insulators (FCI) \cite{tang2011,neupert2011a,sun2011,sheng2011,qi2011a,regnault2011,parameswaran2012,barkeshli2012,wu2013,bergholtz2013}. 
The realization of FCI phases requires strong electron interaction and a narrow topologically nontrivial band that is partially filled. 
QAH effect has also been generalized to systems with periodic time-dependent Hamiltonian, known as the Floquet Chern insulators\cite{lindner2011,kitagawa2011,gu2011,wang2013d}.  
Effective topological band structure can be induced by the time-dependent coupling, such as optical transition between different bands for an electron in external photon field. Beyond electron systems, the topological band structure leading to QAH effect has also been proposed and even realized in cold atom systems \cite{wu2008,zhai2014,liu2010b,duca2015,goldman2013,jotzu2014}, photonic crystals \cite{haldane2008,wang2008,wang2009,rechtsman2013} and magnonics\cite{zhang2013c,shindou2013,mook2014}.


QAH effect also has potential applications in future electronic devices. A key signature of the QH and QAH states is the chiral edge state, which can carry electric current without dissipation, since all electrons have the same direction of current, and backscattering is impossible. Dissipationless transport is always very important since Joule heating becomes a more and more significant problem when the size of electronic devices is reduced\cite{kim2003}. If QAH edge states are used to transport electric current, the only source of resistance is the contact resistance which is independent from the length of the edge. Consequently, using such a ``chiral interconnect" will be more efficient than an ordinary metal wire as long as the length exceeds a critical value \cite{zhang2012c}. It has been shown that chiral edge states can also carry spin polarization \cite{wu2014a}, thus enabling the potential applications in spintronics. Another mechanism to realize dissipationless transport is, of course, using superconductors. Currently both QAH state and superconductivity requires very low temperature, which limits their practical application. However, one may suspect that it is in general easier to realize a room-temperature QAH state than to realize a room temperature superconductor, because magnetism can easily occur at room temperature in many materials. More experimental and theoretical efforts in seeking QAH materials are certainly required for finding the route towards room temperature QAH effect.

\section*{ACKNOWLEDGMENTS}
We would like to acknowledge the helpful discussions and collaborations with C. Z. Chang, M. Chan, X. Dai, Z. Fang, C. Felser, T. L. Hughes, X. Liu, L. Molenkamp, N. Samarth, J. Wang, K. L. Wang, Y. Y. Wang, Y. S. Wu, G. Xu, Q. K. Xue, H. J. Zhang and all other collaborators of us in related works. SCZ is supported by the US Department of Energy, Office of Basic Energy Sciences under contract DE-AC02-76SF00515 and the NSF under grant numbers DMR-1305677. XLQ is supported by the National Science Foundation through the grant No. DMR-1151786. 

\bibliography{QAH_ref}

\begin{thebibliography}{120}%
\makeatletter
\providecommand \@ifxundefined [1]{%
 \@ifx{#1\undefined}
}%
\providecommand \@ifnum [1]{%
 \ifnum #1\expandafter \@firstoftwo
 \else \expandafter \@secondoftwo
 \fi
}%
\providecommand \@ifx [1]{%
 \ifx #1\expandafter \@firstoftwo
 \else \expandafter \@secondoftwo
 \fi
}%
\providecommand \natexlab [1]{#1}%
\providecommand \enquote  [1]{``#1''}%
\providecommand \bibnamefont  [1]{#1}%
\providecommand \bibfnamefont [1]{#1}%
\providecommand \citenamefont [1]{#1}%
\providecommand \href@noop [0]{\@secondoftwo}%
\providecommand \href [0]{\begingroup \@sanitize@url \@href}%
\providecommand \@href[1]{\@@startlink{#1}\@@href}%
\providecommand \@@href[1]{\endgroup#1\@@endlink}%
\providecommand \@sanitize@url [0]{\catcode `\\12\catcode `\$12\catcode
  `\&12\catcode `\#12\catcode `\^12\catcode `\_12\catcode `\%12\relax}%
\providecommand \@@startlink[1]{}%
\providecommand \@@endlink[0]{}%
\providecommand \url  [0]{\begingroup\@sanitize@url \@url }%
\providecommand \@url [1]{\endgroup\@href {#1}{\urlprefix }}%
\providecommand \urlprefix  [0]{URL }%
\providecommand \Eprint [0]{\href }%
\providecommand \doibase [0]{http://dx.doi.org/}%
\providecommand \selectlanguage [0]{\@gobble}%
\providecommand \bibinfo  [0]{\@secondoftwo}%
\providecommand \bibfield  [0]{\@secondoftwo}%
\providecommand \translation [1]{[#1]}%
\providecommand \BibitemOpen [0]{}%
\providecommand \bibitemStop [0]{}%
\providecommand \bibitemNoStop [0]{.\EOS\space}%
\providecommand \EOS [0]{\spacefactor3000\relax}%
\providecommand \BibitemShut  [1]{\csname bibitem#1\endcsname}%
\let\auto@bib@innerbib\@empty
\bibitem [{\citenamefont {Hall}(1879)}]{hall1879}%
  \BibitemOpen
  \bibfield  {author} {\bibinfo {author} {\bibfnamefont {E.~H.}\ \bibnamefont
  {Hall}},\ }\href@noop {} {\bibfield  {journal} {\bibinfo  {journal} {American
  Journal of Mathematics}\ }\textbf {\bibinfo {volume} {2}},\ \bibinfo {pages}
  {287} (\bibinfo {year} {1879})}\BibitemShut {NoStop}%
\bibitem [{\citenamefont {Klitzing}\ \emph {et~al.}(1980)\citenamefont
  {Klitzing}, \citenamefont {Dorda},\ and\ \citenamefont
  {Pepper}}]{klitzing1980}%
  \BibitemOpen
  \bibfield  {author} {\bibinfo {author} {\bibfnamefont {K.~v.}\ \bibnamefont
  {Klitzing}}, \bibinfo {author} {\bibfnamefont {G.}~\bibnamefont {Dorda}}, \
  and\ \bibinfo {author} {\bibfnamefont {M.}~\bibnamefont {Pepper}},\ }\href
  {\doibase 10.1103/PhysRevLett.45.494} {\bibfield  {journal} {\bibinfo
  {journal} {Phys. Rev. Lett.}\ }\textbf {\bibinfo {volume} {45}},\ \bibinfo
  {pages} {494} (\bibinfo {year} {1980})}\BibitemShut {NoStop}%
\bibitem [{\citenamefont {Tsui}\ \emph {et~al.}(1982)\citenamefont {Tsui},
  \citenamefont {Stormer},\ and\ \citenamefont {Gossard}}]{tsui1982}%
  \BibitemOpen
  \bibfield  {author} {\bibinfo {author} {\bibfnamefont {D.~C.}\ \bibnamefont
  {Tsui}}, \bibinfo {author} {\bibfnamefont {H.~L.}\ \bibnamefont {Stormer}}, \
  and\ \bibinfo {author} {\bibfnamefont {A.~C.}\ \bibnamefont {Gossard}},\
  }\href {\doibase 10.1103/PhysRevLett.48.1559} {\bibfield  {journal} {\bibinfo
   {journal} {Phys. Rev. Lett.}\ }\textbf {\bibinfo {volume} {48}},\ \bibinfo
  {pages} {1559} (\bibinfo {year} {1982})}\BibitemShut {NoStop}%
\bibitem [{\citenamefont {Hall}(1881)}]{hall1881}%
  \BibitemOpen
  \bibfield  {author} {\bibinfo {author} {\bibfnamefont {E.~H.}\ \bibnamefont
  {Hall}},\ }\href@noop {} {\bibfield  {journal} {\bibinfo  {journal} {The
  London, Edinburgh, and Dublin Philosophical Magazine and Journal of Science}\
  }\textbf {\bibinfo {volume} {12}},\ \bibinfo {pages} {157} (\bibinfo {year}
  {1881})}\BibitemShut {NoStop}%
\bibitem [{\citenamefont {Nagaosa}\ \emph {et~al.}(2010)\citenamefont
  {Nagaosa}, \citenamefont {Sinova}, \citenamefont {Onoda}, \citenamefont
  {MacDonald},\ and\ \citenamefont {Ong}}]{nagaosa2010}%
  \BibitemOpen
  \bibfield  {author} {\bibinfo {author} {\bibfnamefont {N.}~\bibnamefont
  {Nagaosa}}, \bibinfo {author} {\bibfnamefont {J.}~\bibnamefont {Sinova}},
  \bibinfo {author} {\bibfnamefont {S.}~\bibnamefont {Onoda}}, \bibinfo
  {author} {\bibfnamefont {A.~H.}\ \bibnamefont {MacDonald}}, \ and\ \bibinfo
  {author} {\bibfnamefont {N.~P.}\ \bibnamefont {Ong}},\ }\href@noop {}
  {\bibfield  {journal} {\bibinfo  {journal} {Rev. Mod. Phys.}\ }\textbf
  {\bibinfo {volume} {82}},\ \bibinfo {pages} {1539} (\bibinfo {year}
  {2010})}\BibitemShut {NoStop}%
\bibitem [{\citenamefont {Thouless}\ \emph {et~al.}(1982)\citenamefont
  {Thouless}, \citenamefont {Kohmoto}, \citenamefont {Nightingale},\ and\
  \citenamefont {den Nijs}}]{thouless1982}%
  \BibitemOpen
  \bibfield  {author} {\bibinfo {author} {\bibfnamefont {D.~J.}\ \bibnamefont
  {Thouless}}, \bibinfo {author} {\bibfnamefont {M.}~\bibnamefont {Kohmoto}},
  \bibinfo {author} {\bibfnamefont {M.~P.}\ \bibnamefont {Nightingale}}, \ and\
  \bibinfo {author} {\bibfnamefont {M.}~\bibnamefont {den Nijs}},\ }\href@noop
  {} {\bibfield  {journal} {\bibinfo  {journal} {Phys. Rev. Lett.}\ }\textbf
  {\bibinfo {volume} {49}},\ \bibinfo {pages} {405} (\bibinfo {year}
  {1982})}\BibitemShut {NoStop}%
\bibitem [{\citenamefont {Avron}\ and\ \citenamefont
  {Seiler}(1985)}]{avron1985}%
  \BibitemOpen
  \bibfield  {author} {\bibinfo {author} {\bibfnamefont {J.~E.}\ \bibnamefont
  {Avron}}\ and\ \bibinfo {author} {\bibfnamefont {R.}~\bibnamefont {Seiler}},\
  }\href {\doibase 10.1103/PhysRevLett.54.259} {\bibfield  {journal} {\bibinfo
  {journal} {Phys. Rev. Lett.}\ }\textbf {\bibinfo {volume} {54}},\ \bibinfo
  {pages} {259} (\bibinfo {year} {1985})}\BibitemShut {NoStop}%
\bibitem [{\citenamefont {Niu}\ \emph {et~al.}(1985)\citenamefont {Niu},
  \citenamefont {Thouless},\ and\ \citenamefont {Wu}}]{niu1985}%
  \BibitemOpen
  \bibfield  {author} {\bibinfo {author} {\bibfnamefont {Q.}~\bibnamefont
  {Niu}}, \bibinfo {author} {\bibfnamefont {D.~J.}\ \bibnamefont {Thouless}}, \
  and\ \bibinfo {author} {\bibfnamefont {Y.-S.}\ \bibnamefont {Wu}},\
  }\href@noop {} {\bibfield  {journal} {\bibinfo  {journal} {Phys. Rev. B}\
  }\textbf {\bibinfo {volume} {31}},\ \bibinfo {pages} {3372} (\bibinfo {year}
  {1985})}\BibitemShut {NoStop}%
\bibitem [{\citenamefont {Haldane}(1988)}]{haldane1988}%
  \BibitemOpen
  \bibfield  {author} {\bibinfo {author} {\bibfnamefont {F.~D.~M.}\
  \bibnamefont {Haldane}},\ }\href@noop {} {\bibfield  {journal} {\bibinfo
  {journal} {Phys. Rev. Lett.}\ }\textbf {\bibinfo {volume} {61}},\ \bibinfo
  {pages} {2015} (\bibinfo {year} {1988})}\BibitemShut {NoStop}%
\bibitem [{\citenamefont {Sinova}\ \emph {et~al.}(2004)\citenamefont {Sinova},
  \citenamefont {Culcer}, \citenamefont {Niu}, \citenamefont {Sinitsyn},
  \citenamefont {Jungwirth},\ and\ \citenamefont {MacDonald}}]{sinova2004}%
  \BibitemOpen
  \bibfield  {author} {\bibinfo {author} {\bibfnamefont {J.}~\bibnamefont
  {Sinova}}, \bibinfo {author} {\bibfnamefont {D.}~\bibnamefont {Culcer}},
  \bibinfo {author} {\bibfnamefont {Q.}~\bibnamefont {Niu}}, \bibinfo {author}
  {\bibfnamefont {N.~A.}\ \bibnamefont {Sinitsyn}}, \bibinfo {author}
  {\bibfnamefont {T.}~\bibnamefont {Jungwirth}}, \ and\ \bibinfo {author}
  {\bibfnamefont {A.~H.}\ \bibnamefont {MacDonald}},\ }\href@noop {} {\bibfield
   {journal} {\bibinfo  {journal} {Phys. Rev. Lett.}\ }\textbf {\bibinfo
  {volume} {92}},\ \bibinfo {pages} {126603} (\bibinfo {year}
  {2004})}\BibitemShut {NoStop}%
\bibitem [{\citenamefont {Murakami}\ \emph {et~al.}(2003)\citenamefont
  {Murakami}, \citenamefont {Nagaosa},\ and\ \citenamefont
  {Zhang}}]{murakami2003}%
  \BibitemOpen
  \bibfield  {author} {\bibinfo {author} {\bibfnamefont {S.}~\bibnamefont
  {Murakami}}, \bibinfo {author} {\bibfnamefont {N.}~\bibnamefont {Nagaosa}}, \
  and\ \bibinfo {author} {\bibfnamefont {S.~C.}\ \bibnamefont {Zhang}},\
  }\href@noop {} {\bibfield  {journal} {\bibinfo  {journal} {Science}\ }\textbf
  {\bibinfo {volume} {301}},\ \bibinfo {pages} {1348} (\bibinfo {year}
  {2003})}\BibitemShut {NoStop}%
\bibitem [{\citenamefont {Hirsch}(1999)}]{hirsch1999}%
  \BibitemOpen
  \bibfield  {author} {\bibinfo {author} {\bibfnamefont {J.}~\bibnamefont
  {Hirsch}},\ }\href@noop {} {\bibfield  {journal} {\bibinfo  {journal}
  {Physical Review Letters}\ }\textbf {\bibinfo {volume} {83}},\ \bibinfo
  {pages} {1834} (\bibinfo {year} {1999})}\BibitemShut {NoStop}%
\bibitem [{\citenamefont {Kane}\ and\ \citenamefont
  {Mele}(2005{\natexlab{a}})}]{kane2005a}%
  \BibitemOpen
  \bibfield  {author} {\bibinfo {author} {\bibfnamefont {C.~L.}\ \bibnamefont
  {Kane}}\ and\ \bibinfo {author} {\bibfnamefont {E.~J.}\ \bibnamefont
  {Mele}},\ }\href@noop {} {\bibfield  {journal} {\bibinfo  {journal} {Phys.
  Rev. Lett.}\ }\textbf {\bibinfo {volume} {95}},\ \bibinfo {pages} {226801}
  (\bibinfo {year} {2005}{\natexlab{a}})}\BibitemShut {NoStop}%
\bibitem [{\citenamefont {Bernevig}\ \emph
  {et~al.}(2006{\natexlab{a}})\citenamefont {Bernevig}, \citenamefont
  {Orenstein},\ and\ \citenamefont {Zhang}}]{bernevig2006a}%
  \BibitemOpen
  \bibfield  {author} {\bibinfo {author} {\bibfnamefont {B.~A.}\ \bibnamefont
  {Bernevig}}, \bibinfo {author} {\bibfnamefont {J.}~\bibnamefont {Orenstein}},
  \ and\ \bibinfo {author} {\bibfnamefont {S.-C.}\ \bibnamefont {Zhang}},\
  }\href@noop {} {\bibfield  {journal} {\bibinfo  {journal} {Phys. Rev. Lett.}\
  }\textbf {\bibinfo {volume} {97}},\ \bibinfo {pages} {236601} (\bibinfo
  {year} {2006}{\natexlab{a}})}\BibitemShut {NoStop}%
\bibitem [{\citenamefont {Bernevig}\ \emph
  {et~al.}(2006{\natexlab{b}})\citenamefont {Bernevig}, \citenamefont
  {Hughes},\ and\ \citenamefont {Zhang}}]{bernevig2006c}%
  \BibitemOpen
  \bibfield  {author} {\bibinfo {author} {\bibfnamefont {B.~A.}\ \bibnamefont
  {Bernevig}}, \bibinfo {author} {\bibfnamefont {T.~L.}\ \bibnamefont
  {Hughes}}, \ and\ \bibinfo {author} {\bibfnamefont {S.~C.}\ \bibnamefont
  {Zhang}},\ }\href@noop {} {\bibfield  {journal} {\bibinfo  {journal}
  {Science}\ }\textbf {\bibinfo {volume} {314}},\ \bibinfo {pages} {1757}
  (\bibinfo {year} {2006}{\natexlab{b}})}\BibitemShut {NoStop}%
\bibitem [{\citenamefont {K\"onig}\ \emph {et~al.}(2007)\citenamefont
  {K\"onig}, \citenamefont {Wiedmann}, \citenamefont {Br\"une}, \citenamefont
  {Roth}, \citenamefont {Buhmann}, \citenamefont {Molenkamp}, \citenamefont
  {Qi},\ and\ \citenamefont {Zhang}}]{koenig2007}%
  \BibitemOpen
  \bibfield  {author} {\bibinfo {author} {\bibfnamefont {M.}~\bibnamefont
  {K\"onig}}, \bibinfo {author} {\bibfnamefont {S.}~\bibnamefont {Wiedmann}},
  \bibinfo {author} {\bibfnamefont {C.}~\bibnamefont {Br\"une}}, \bibinfo
  {author} {\bibfnamefont {A.}~\bibnamefont {Roth}}, \bibinfo {author}
  {\bibfnamefont {H.}~\bibnamefont {Buhmann}}, \bibinfo {author} {\bibfnamefont
  {L.}~\bibnamefont {Molenkamp}}, \bibinfo {author} {\bibfnamefont {X.-L.}\
  \bibnamefont {Qi}}, \ and\ \bibinfo {author} {\bibfnamefont {S.-C.}\
  \bibnamefont {Zhang}},\ }\href@noop {} {\bibfield  {journal} {\bibinfo
  {journal} {Science}\ }\textbf {\bibinfo {volume} {318}},\ \bibinfo {pages}
  {766} (\bibinfo {year} {2007})}\BibitemShut {NoStop}%
\bibitem [{\citenamefont {Hasan}\ and\ \citenamefont {Kane}(2010)}]{hasan2010}%
  \BibitemOpen
  \bibfield  {author} {\bibinfo {author} {\bibfnamefont {M.~Z.}\ \bibnamefont
  {Hasan}}\ and\ \bibinfo {author} {\bibfnamefont {C.~L.}\ \bibnamefont
  {Kane}},\ }\href {\doibase 10.1103/RevModPhys.82.3045} {\bibfield  {journal}
  {\bibinfo  {journal} {Rev. Mod. Phys.}\ }\textbf {\bibinfo {volume} {82}},\
  \bibinfo {pages} {3045} (\bibinfo {year} {2010})}\BibitemShut {NoStop}%
\bibitem [{\citenamefont {Qi}\ and\ \citenamefont {Zhang}(2011)}]{qi2011}%
  \BibitemOpen
  \bibfield  {author} {\bibinfo {author} {\bibfnamefont {X.-L.}\ \bibnamefont
  {Qi}}\ and\ \bibinfo {author} {\bibfnamefont {S.-C.}\ \bibnamefont {Zhang}},\
  }\href@noop {} {\bibfield  {journal} {\bibinfo  {journal} {Rev. Mod. Phys.}\
  }\textbf {\bibinfo {volume} {83}},\ \bibinfo {pages} {1057} (\bibinfo {year}
  {2011})}\BibitemShut {NoStop}%
\bibitem [{\citenamefont {Qi}\ \emph {et~al.}(2006)\citenamefont {Qi},
  \citenamefont {Wu},\ and\ \citenamefont {Zhang}}]{qi2005}%
  \BibitemOpen
  \bibfield  {author} {\bibinfo {author} {\bibfnamefont {X.~L.}\ \bibnamefont
  {Qi}}, \bibinfo {author} {\bibfnamefont {Y.~S.}\ \bibnamefont {Wu}}, \ and\
  \bibinfo {author} {\bibfnamefont {S.~C.}\ \bibnamefont {Zhang}},\ }\href@noop
  {} {\bibfield  {journal} {\bibinfo  {journal} {Phys. Rev. B}\ }\textbf
  {\bibinfo {volume} {74}},\ \bibinfo {pages} {085308} (\bibinfo {year}
  {2006})}\BibitemShut {NoStop}%
\bibitem [{\citenamefont {Kane}\ and\ \citenamefont
  {Mele}(2005{\natexlab{b}})}]{kane2005b}%
  \BibitemOpen
  \bibfield  {author} {\bibinfo {author} {\bibfnamefont {C.~L.}\ \bibnamefont
  {Kane}}\ and\ \bibinfo {author} {\bibfnamefont {E.~J.}\ \bibnamefont
  {Mele}},\ }\href@noop {} {\bibfield  {journal} {\bibinfo  {journal} {Phys.
  Rev. Lett.}\ }\textbf {\bibinfo {volume} {95}},\ \bibinfo {pages} {146802}
  (\bibinfo {year} {2005}{\natexlab{b}})}\BibitemShut {NoStop}%
\bibitem [{\citenamefont {Liu}\ \emph {et~al.}(2008{\natexlab{a}})\citenamefont
  {Liu}, \citenamefont {Qi}, \citenamefont {Dai}, \citenamefont {Fang},\ and\
  \citenamefont {Zhang}}]{liu2009C}%
  \BibitemOpen
  \bibfield  {author} {\bibinfo {author} {\bibfnamefont {C.-X.}\ \bibnamefont
  {Liu}}, \bibinfo {author} {\bibfnamefont {X.-L.}\ \bibnamefont {Qi}},
  \bibinfo {author} {\bibfnamefont {X.}~\bibnamefont {Dai}}, \bibinfo {author}
  {\bibfnamefont {Z.}~\bibnamefont {Fang}}, \ and\ \bibinfo {author}
  {\bibfnamefont {S.-C.}\ \bibnamefont {Zhang}},\ }\href@noop {} {\bibfield
  {journal} {\bibinfo  {journal} {Phys. Rev. Lett.}\ }\textbf {\bibinfo
  {volume} {101}},\ \bibinfo {pages} {146802} (\bibinfo {year}
  {2008}{\natexlab{a}})}\BibitemShut {NoStop}%
\bibitem [{\citenamefont {Buhmann}\ \emph {et~al.}(2002)\citenamefont
  {Buhmann}, \citenamefont {Liu}, \citenamefont {Gui}, \citenamefont {Daumer},
  \citenamefont {Koenig}, \citenamefont {Becker},\ and\ \citenamefont
  {Molenkamp}}]{buhmann2002}%
  \BibitemOpen
  \bibfield  {author} {\bibinfo {author} {\bibfnamefont {H.}~\bibnamefont
  {Buhmann}}, \bibinfo {author} {\bibfnamefont {J.}~\bibnamefont {Liu}},
  \bibinfo {author} {\bibfnamefont {Y.~S.}\ \bibnamefont {Gui}}, \bibinfo
  {author} {\bibfnamefont {V.}~\bibnamefont {Daumer}}, \bibinfo {author}
  {\bibfnamefont {M.}~\bibnamefont {Koenig}}, \bibinfo {author} {\bibfnamefont
  {C.~R.}\ \bibnamefont {Becker}}, \ and\ \bibinfo {author} {\bibfnamefont
  {L.~W.}\ \bibnamefont {Molenkamp}},\ }\href@noop {} {\enquote {\bibinfo
  {title} {Anomalous hall effect in magnetic quantum wells},}\ }\bibinfo
  {howpublished} {Proc. 15th Int. Conf. on High Magnetic Fields in
  Semiconductor Physics, Oxford, 5-9 August 2002.} (\bibinfo {year}
  {2002})\BibitemShut {NoStop}%
\bibitem [{\citenamefont {Yu}\ \emph {et~al.}(2010)\citenamefont {Yu},
  \citenamefont {Zhang}, \citenamefont {Zhang}, \citenamefont {Zhang},
  \citenamefont {Dai},\ and\ \citenamefont {Fang}}]{yu2010}%
  \BibitemOpen
  \bibfield  {author} {\bibinfo {author} {\bibfnamefont {R.}~\bibnamefont
  {Yu}}, \bibinfo {author} {\bibfnamefont {W.}~\bibnamefont {Zhang}}, \bibinfo
  {author} {\bibfnamefont {H.~J.}\ \bibnamefont {Zhang}}, \bibinfo {author}
  {\bibfnamefont {S.~C.}\ \bibnamefont {Zhang}}, \bibinfo {author}
  {\bibfnamefont {X.}~\bibnamefont {Dai}}, \ and\ \bibinfo {author}
  {\bibfnamefont {Z.}~\bibnamefont {Fang}},\ }\href@noop {} {\bibfield
  {journal} {\bibinfo  {journal} {Science}\ }\textbf {\bibinfo {volume}
  {329}},\ \bibinfo {pages} {61} (\bibinfo {year} {2010})}\BibitemShut
  {NoStop}%
\bibitem [{\citenamefont {Zhang}\ \emph {et~al.}(2009)\citenamefont {Zhang},
  \citenamefont {Liu}, \citenamefont {Qi}, \citenamefont {Dai}, \citenamefont
  {Fang},\ and\ \citenamefont {Zhang}}]{hjzhang2009}%
  \BibitemOpen
  \bibfield  {author} {\bibinfo {author} {\bibfnamefont {H.}~\bibnamefont
  {Zhang}}, \bibinfo {author} {\bibfnamefont {C.-X.}\ \bibnamefont {Liu}},
  \bibinfo {author} {\bibfnamefont {X.-L.}\ \bibnamefont {Qi}}, \bibinfo
  {author} {\bibfnamefont {X.}~\bibnamefont {Dai}}, \bibinfo {author}
  {\bibfnamefont {Z.}~\bibnamefont {Fang}}, \ and\ \bibinfo {author}
  {\bibfnamefont {S.-C.}\ \bibnamefont {Zhang}},\ }\href@noop {} {\bibfield
  {journal} {\bibinfo  {journal} {Nature Phys.}\ }\textbf {\bibinfo {volume}
  {5}},\ \bibinfo {pages} {438} (\bibinfo {year} {2009})}\BibitemShut {NoStop}%
\bibitem [{\citenamefont {Xia}\ \emph {et~al.}(2009)\citenamefont {Xia},
  \citenamefont {Wray}, \citenamefont {Qian}, \citenamefont {Hsieh},
  \citenamefont {Pal}, \citenamefont {Lin}, \citenamefont {Bansil},
  \citenamefont {Grauer}, \citenamefont {Hor}, \citenamefont {Cava},\ and\
  \citenamefont {Hasan}}]{xia2009}%
  \BibitemOpen
  \bibfield  {author} {\bibinfo {author} {\bibfnamefont {Y.}~\bibnamefont
  {Xia}}, \bibinfo {author} {\bibfnamefont {L.}~\bibnamefont {Wray}}, \bibinfo
  {author} {\bibfnamefont {D.}~\bibnamefont {Qian}}, \bibinfo {author}
  {\bibfnamefont {D.}~\bibnamefont {Hsieh}}, \bibinfo {author} {\bibfnamefont
  {A.}~\bibnamefont {Pal}}, \bibinfo {author} {\bibfnamefont {H.}~\bibnamefont
  {Lin}}, \bibinfo {author} {\bibfnamefont {A.}~\bibnamefont {Bansil}},
  \bibinfo {author} {\bibfnamefont {D.}~\bibnamefont {Grauer}}, \bibinfo
  {author} {\bibfnamefont {Y.}~\bibnamefont {Hor}}, \bibinfo {author}
  {\bibfnamefont {R.}~\bibnamefont {Cava}}, \ and\ \bibinfo {author}
  {\bibfnamefont {M.}~\bibnamefont {Hasan}},\ }\href@noop {} {\bibfield
  {journal} {\bibinfo  {journal} {Nature Phys.}\ }\textbf {\bibinfo {volume}
  {5}},\ \bibinfo {pages} {398} (\bibinfo {year} {2009})}\BibitemShut {NoStop}%
\bibitem [{\citenamefont {Qi}\ \emph {et~al.}(2008)\citenamefont {Qi},
  \citenamefont {Hughes},\ and\ \citenamefont {Zhang}}]{qi2008b}%
  \BibitemOpen
  \bibfield  {author} {\bibinfo {author} {\bibfnamefont {X.-L.}\ \bibnamefont
  {Qi}}, \bibinfo {author} {\bibfnamefont {T.}~\bibnamefont {Hughes}}, \ and\
  \bibinfo {author} {\bibfnamefont {S.-C.}\ \bibnamefont {Zhang}},\ }\href@noop
  {} {\bibfield  {journal} {\bibinfo  {journal} {Phys. Rev. B}\ }\textbf
  {\bibinfo {volume} {78}},\ \bibinfo {pages} {195424} (\bibinfo {year}
  {2008})}\BibitemShut {NoStop}%
\bibitem [{\citenamefont {Fu}\ \emph {et~al.}(2007)\citenamefont {Fu},
  \citenamefont {Kane},\ and\ \citenamefont {Mele}}]{fu2007b}%
  \BibitemOpen
  \bibfield  {author} {\bibinfo {author} {\bibfnamefont {L.}~\bibnamefont
  {Fu}}, \bibinfo {author} {\bibfnamefont {C.~L.}\ \bibnamefont {Kane}}, \ and\
  \bibinfo {author} {\bibfnamefont {E.~J.}\ \bibnamefont {Mele}},\ }\href@noop
  {} {\bibfield  {journal} {\bibinfo  {journal} {Phys. Rev. Lett.}\ }\textbf
  {\bibinfo {volume} {98}},\ \bibinfo {pages} {106803} (\bibinfo {year}
  {2007})}\BibitemShut {NoStop}%
\bibitem [{\citenamefont {Chang}\ \emph
  {et~al.}(2013{\natexlab{a}})\citenamefont {Chang}, \citenamefont {Zhang},
  \citenamefont {Feng}, \citenamefont {Shen}, \citenamefont {Zhang},
  \citenamefont {Guo}, \citenamefont {Li}, \citenamefont {Ou}, \citenamefont
  {Wei}, \citenamefont {Wang}, \citenamefont {Ji}, \citenamefont {Feng},
  \citenamefont {Ji}, \citenamefont {Chen}, \citenamefont {Jia}, \citenamefont
  {Dai}, \citenamefont {Fang}, \citenamefont {Zhang}, \citenamefont {He},
  \citenamefont {Wang}, \citenamefont {Lu}, \citenamefont {Ma},\ and\
  \citenamefont {Xue}}]{chang2013a}%
  \BibitemOpen
  \bibfield  {author} {\bibinfo {author} {\bibfnamefont {C.-Z.}\ \bibnamefont
  {Chang}}, \bibinfo {author} {\bibfnamefont {J.}~\bibnamefont {Zhang}},
  \bibinfo {author} {\bibfnamefont {X.}~\bibnamefont {Feng}}, \bibinfo {author}
  {\bibfnamefont {J.}~\bibnamefont {Shen}}, \bibinfo {author} {\bibfnamefont
  {Z.}~\bibnamefont {Zhang}}, \bibinfo {author} {\bibfnamefont
  {M.}~\bibnamefont {Guo}}, \bibinfo {author} {\bibfnamefont {K.}~\bibnamefont
  {Li}}, \bibinfo {author} {\bibfnamefont {Y.}~\bibnamefont {Ou}}, \bibinfo
  {author} {\bibfnamefont {P.}~\bibnamefont {Wei}}, \bibinfo {author}
  {\bibfnamefont {L.-L.}\ \bibnamefont {Wang}}, \bibinfo {author}
  {\bibfnamefont {Z.-Q.}\ \bibnamefont {Ji}}, \bibinfo {author} {\bibfnamefont
  {Y.}~\bibnamefont {Feng}}, \bibinfo {author} {\bibfnamefont {S.}~\bibnamefont
  {Ji}}, \bibinfo {author} {\bibfnamefont {X.}~\bibnamefont {Chen}}, \bibinfo
  {author} {\bibfnamefont {J.}~\bibnamefont {Jia}}, \bibinfo {author}
  {\bibfnamefont {X.}~\bibnamefont {Dai}}, \bibinfo {author} {\bibfnamefont
  {Z.}~\bibnamefont {Fang}}, \bibinfo {author} {\bibfnamefont {S.-C.}\
  \bibnamefont {Zhang}}, \bibinfo {author} {\bibfnamefont {K.}~\bibnamefont
  {He}}, \bibinfo {author} {\bibfnamefont {Y.}~\bibnamefont {Wang}}, \bibinfo
  {author} {\bibfnamefont {L.}~\bibnamefont {Lu}}, \bibinfo {author}
  {\bibfnamefont {X.-C.}\ \bibnamefont {Ma}}, \ and\ \bibinfo {author}
  {\bibfnamefont {Q.-K.}\ \bibnamefont {Xue}},\ }\href@noop {} {\bibfield
  {journal} {\bibinfo  {journal} {Science}\ }\textbf {\bibinfo {volume}
  {340}},\ \bibinfo {pages} {167} (\bibinfo {year}
  {2013}{\natexlab{a}})}\BibitemShut {NoStop}%
\bibitem [{\citenamefont {Checkelsky}\ \emph {et~al.}(2014)\citenamefont
  {Checkelsky}, \citenamefont {Yoshimi}, \citenamefont {Tsukazaki},
  \citenamefont {Takahashi}, \citenamefont {Kozuka}, \citenamefont {Falson},
  \citenamefont {Kawasaki},\ and\ \citenamefont {Tokura}}]{checkelsky2014}%
  \BibitemOpen
  \bibfield  {author} {\bibinfo {author} {\bibfnamefont {J.}~\bibnamefont
  {Checkelsky}}, \bibinfo {author} {\bibfnamefont {R.}~\bibnamefont {Yoshimi}},
  \bibinfo {author} {\bibfnamefont {A.}~\bibnamefont {Tsukazaki}}, \bibinfo
  {author} {\bibfnamefont {K.}~\bibnamefont {Takahashi}}, \bibinfo {author}
  {\bibfnamefont {Y.}~\bibnamefont {Kozuka}}, \bibinfo {author} {\bibfnamefont
  {J.}~\bibnamefont {Falson}}, \bibinfo {author} {\bibfnamefont
  {M.}~\bibnamefont {Kawasaki}}, \ and\ \bibinfo {author} {\bibfnamefont
  {Y.}~\bibnamefont {Tokura}},\ }\href@noop {} {\bibfield  {journal} {\bibinfo
  {journal} {Nature Physics}\ } (\bibinfo {year} {2014})}\BibitemShut {NoStop}%
\bibitem [{\citenamefont {Kou}\ \emph {et~al.}(2014)\citenamefont {Kou},
  \citenamefont {Guo}, \citenamefont {Fan}, \citenamefont {Pan}, \citenamefont
  {Lang}, \citenamefont {Jiang}, \citenamefont {Shao}, \citenamefont {Nie},
  \citenamefont {Murata}, \citenamefont {Tang} \emph {et~al.}}]{kou2014}%
  \BibitemOpen
  \bibfield  {author} {\bibinfo {author} {\bibfnamefont {X.}~\bibnamefont
  {Kou}}, \bibinfo {author} {\bibfnamefont {S.-T.}\ \bibnamefont {Guo}},
  \bibinfo {author} {\bibfnamefont {Y.}~\bibnamefont {Fan}}, \bibinfo {author}
  {\bibfnamefont {L.}~\bibnamefont {Pan}}, \bibinfo {author} {\bibfnamefont
  {M.}~\bibnamefont {Lang}}, \bibinfo {author} {\bibfnamefont {Y.}~\bibnamefont
  {Jiang}}, \bibinfo {author} {\bibfnamefont {Q.}~\bibnamefont {Shao}},
  \bibinfo {author} {\bibfnamefont {T.}~\bibnamefont {Nie}}, \bibinfo {author}
  {\bibfnamefont {K.}~\bibnamefont {Murata}}, \bibinfo {author} {\bibfnamefont
  {J.}~\bibnamefont {Tang}},  \emph {et~al.},\ }\href@noop {} {\bibfield
  {journal} {\bibinfo  {journal} {Physical review letters}\ }\textbf {\bibinfo
  {volume} {113}},\ \bibinfo {pages} {137201} (\bibinfo {year}
  {2014})}\BibitemShut {NoStop}%
\bibitem [{\citenamefont {Bestwick}\ \emph {et~al.}(2014)\citenamefont
  {Bestwick}, \citenamefont {Fox}, \citenamefont {Kou}, \citenamefont {Pan},
  \citenamefont {Wang},\ and\ \citenamefont {Goldhaber-Gordon}}]{bestwick2014}%
  \BibitemOpen
  \bibfield  {author} {\bibinfo {author} {\bibfnamefont {A.}~\bibnamefont
  {Bestwick}}, \bibinfo {author} {\bibfnamefont {E.}~\bibnamefont {Fox}},
  \bibinfo {author} {\bibfnamefont {X.}~\bibnamefont {Kou}}, \bibinfo {author}
  {\bibfnamefont {L.}~\bibnamefont {Pan}}, \bibinfo {author} {\bibfnamefont
  {K.~L.}\ \bibnamefont {Wang}}, \ and\ \bibinfo {author} {\bibfnamefont
  {D.}~\bibnamefont {Goldhaber-Gordon}},\ }\href@noop {} {\bibfield  {journal}
  {\bibinfo  {journal} {arXiv preprint arXiv:1412.3189}\ } (\bibinfo {year}
  {2014})}\BibitemShut {NoStop}%
\bibitem [{\citenamefont {Kou}\ \emph {et~al.}(2015)\citenamefont {Kou},
  \citenamefont {Pan}, \citenamefont {Wang}, \citenamefont {Fan}, \citenamefont
  {Choi}, \citenamefont {Lee}, \citenamefont {Nie}, \citenamefont {Murata},
  \citenamefont {Shao}, \citenamefont {Zhang} \emph {et~al.}}]{kou2015}%
  \BibitemOpen
  \bibfield  {author} {\bibinfo {author} {\bibfnamefont {X.}~\bibnamefont
  {Kou}}, \bibinfo {author} {\bibfnamefont {L.}~\bibnamefont {Pan}}, \bibinfo
  {author} {\bibfnamefont {J.}~\bibnamefont {Wang}}, \bibinfo {author}
  {\bibfnamefont {Y.}~\bibnamefont {Fan}}, \bibinfo {author} {\bibfnamefont
  {E.~S.}\ \bibnamefont {Choi}}, \bibinfo {author} {\bibfnamefont {W.-L.}\
  \bibnamefont {Lee}}, \bibinfo {author} {\bibfnamefont {T.}~\bibnamefont
  {Nie}}, \bibinfo {author} {\bibfnamefont {K.}~\bibnamefont {Murata}},
  \bibinfo {author} {\bibfnamefont {Q.}~\bibnamefont {Shao}}, \bibinfo {author}
  {\bibfnamefont {S.-C.}\ \bibnamefont {Zhang}},  \emph {et~al.},\ }\href@noop
  {} {\bibfield  {journal} {\bibinfo  {journal} {arXiv preprint
  arXiv:1503.04150}\ } (\bibinfo {year} {2015})}\BibitemShut {NoStop}%
\bibitem [{\citenamefont {Feng}\ \emph {et~al.}(2015)\citenamefont {Feng},
  \citenamefont {Feng}, \citenamefont {Ou}, \citenamefont {Wang}, \citenamefont
  {Liu}, \citenamefont {Zhang}, \citenamefont {Zhao}, \citenamefont {Jiang},
  \citenamefont {Zhang}, \citenamefont {He} \emph {et~al.}}]{feng2015}%
  \BibitemOpen
  \bibfield  {author} {\bibinfo {author} {\bibfnamefont {Y.}~\bibnamefont
  {Feng}}, \bibinfo {author} {\bibfnamefont {X.}~\bibnamefont {Feng}}, \bibinfo
  {author} {\bibfnamefont {Y.}~\bibnamefont {Ou}}, \bibinfo {author}
  {\bibfnamefont {J.}~\bibnamefont {Wang}}, \bibinfo {author} {\bibfnamefont
  {C.}~\bibnamefont {Liu}}, \bibinfo {author} {\bibfnamefont {L.}~\bibnamefont
  {Zhang}}, \bibinfo {author} {\bibfnamefont {D.}~\bibnamefont {Zhao}},
  \bibinfo {author} {\bibfnamefont {G.}~\bibnamefont {Jiang}}, \bibinfo
  {author} {\bibfnamefont {S.-C.}\ \bibnamefont {Zhang}}, \bibinfo {author}
  {\bibfnamefont {K.}~\bibnamefont {He}},  \emph {et~al.},\ }\href@noop {}
  {\bibfield  {journal} {\bibinfo  {journal} {arXiv preprint arXiv:1503.04569}\
  } (\bibinfo {year} {2015})}\BibitemShut {NoStop}%
\bibitem [{\citenamefont {Chang}\ \emph {et~al.}(2014)\citenamefont {Chang},
  \citenamefont {Zhao}, \citenamefont {Kim}, \citenamefont {Zhang},
  \citenamefont {Assaf}, \citenamefont {Heiman}, \citenamefont {Zhang},
  \citenamefont {Liu}, \citenamefont {Chan},\ and\ \citenamefont
  {Moodera}}]{chang2014}%
  \BibitemOpen
  \bibfield  {author} {\bibinfo {author} {\bibfnamefont {C.-Z.}\ \bibnamefont
  {Chang}}, \bibinfo {author} {\bibfnamefont {W.}~\bibnamefont {Zhao}},
  \bibinfo {author} {\bibfnamefont {D.~Y.}\ \bibnamefont {Kim}}, \bibinfo
  {author} {\bibfnamefont {H.}~\bibnamefont {Zhang}}, \bibinfo {author}
  {\bibfnamefont {B.~A.}\ \bibnamefont {Assaf}}, \bibinfo {author}
  {\bibfnamefont {D.}~\bibnamefont {Heiman}}, \bibinfo {author} {\bibfnamefont
  {S.-C.}\ \bibnamefont {Zhang}}, \bibinfo {author} {\bibfnamefont
  {C.}~\bibnamefont {Liu}}, \bibinfo {author} {\bibfnamefont {M.~H.}\
  \bibnamefont {Chan}}, \ and\ \bibinfo {author} {\bibfnamefont {J.~S.}\
  \bibnamefont {Moodera}},\ }\href@noop {} {\bibfield  {journal} {\bibinfo
  {journal} {arXiv preprint arXiv:1412.3758}\ } (\bibinfo {year}
  {2014})}\BibitemShut {NoStop}%
\bibitem [{\citenamefont {Kandala}\ \emph {et~al.}(2015)\citenamefont
  {Kandala}, \citenamefont {Richardella}, \citenamefont {Kempinger},
  \citenamefont {Liu},\ and\ \citenamefont {Samarth}}]{kandala2015}%
  \BibitemOpen
  \bibfield  {author} {\bibinfo {author} {\bibfnamefont {A.}~\bibnamefont
  {Kandala}}, \bibinfo {author} {\bibfnamefont {A.}~\bibnamefont
  {Richardella}}, \bibinfo {author} {\bibfnamefont {S.}~\bibnamefont
  {Kempinger}}, \bibinfo {author} {\bibfnamefont {C.}~\bibnamefont {Liu}}, \
  and\ \bibinfo {author} {\bibfnamefont {N.}~\bibnamefont {Samarth}},\
  }\href@noop {} {\bibfield  {journal} {\bibinfo  {journal} {arXiv preprint
  arXiv:1503.03556}\ } (\bibinfo {year} {2015})}\BibitemShut {NoStop}%
\bibitem [{\citenamefont {Wang}\ \emph
  {et~al.}(2014{\natexlab{a}})\citenamefont {Wang}, \citenamefont {Lian},\ and\
  \citenamefont {Zhang}}]{wang2014c}%
  \BibitemOpen
  \bibfield  {author} {\bibinfo {author} {\bibfnamefont {J.}~\bibnamefont
  {Wang}}, \bibinfo {author} {\bibfnamefont {B.}~\bibnamefont {Lian}}, \ and\
  \bibinfo {author} {\bibfnamefont {S.-C.}\ \bibnamefont {Zhang}},\ }\href@noop
  {} {\bibfield  {journal} {\bibinfo  {journal} {arXiv preprint
  arXiv:1409.6715}\ } (\bibinfo {year} {2014}{\natexlab{a}})}\BibitemShut
  {NoStop}%
\bibitem [{\citenamefont {Semenoff}(1984)}]{semenoff1984}%
  \BibitemOpen
  \bibfield  {author} {\bibinfo {author} {\bibfnamefont {G.~W.}\ \bibnamefont
  {Semenoff}},\ }\href {\doibase 10.1103/PhysRevLett.53.2449} {\bibfield
  {journal} {\bibinfo  {journal} {Phys. Rev. Lett.}\ }\textbf {\bibinfo
  {volume} {53}},\ \bibinfo {pages} {2449} (\bibinfo {year}
  {1984})}\BibitemShut {NoStop}%
\bibitem [{\citenamefont {Khveshchenko}\ and\ \citenamefont
  {Wiegmann}(1989)}]{khveshchenko1989}%
  \BibitemOpen
  \bibfield  {author} {\bibinfo {author} {\bibfnamefont {D.}~\bibnamefont
  {Khveshchenko}}\ and\ \bibinfo {author} {\bibfnamefont {P.~B.}\ \bibnamefont
  {Wiegmann}},\ }\href@noop {} {\bibfield  {journal} {\bibinfo  {journal}
  {Physics Letters B}\ }\textbf {\bibinfo {volume} {225}},\ \bibinfo {pages}
  {279} (\bibinfo {year} {1989})}\BibitemShut {NoStop}%
\bibitem [{\citenamefont {Volovik}(2003)}]{volovik2003}%
  \BibitemOpen
  \bibfield  {author} {\bibinfo {author} {\bibfnamefont {G.~E.}\ \bibnamefont
  {Volovik}},\ }\href@noop {} {\emph {\bibinfo {title} {The Universe in a
  Helium Droplet}}}\ (\bibinfo  {publisher} {Oxford University Press},\
  \bibinfo {address} {Oxford},\ \bibinfo {year} {2003})\BibitemShut {NoStop}%
\bibitem [{\citenamefont {Jackiw}(1984)}]{jackiw1984}%
  \BibitemOpen
  \bibfield  {author} {\bibinfo {author} {\bibfnamefont {R.}~\bibnamefont
  {Jackiw}},\ }\href {\doibase 10.1103/PhysRevD.29.2375} {\bibfield  {journal}
  {\bibinfo  {journal} {Phys. Rev. D}\ }\textbf {\bibinfo {volume} {29}},\
  \bibinfo {pages} {2375} (\bibinfo {year} {1984})}\BibitemShut {NoStop}%
\bibitem [{\citenamefont {Hsu}\ \emph {et~al.}(2013)\citenamefont {Hsu},
  \citenamefont {Liu},\ and\ \citenamefont {Liu}}]{hsu2013}%
  \BibitemOpen
  \bibfield  {author} {\bibinfo {author} {\bibfnamefont {H.-C.}\ \bibnamefont
  {Hsu}}, \bibinfo {author} {\bibfnamefont {X.}~\bibnamefont {Liu}}, \ and\
  \bibinfo {author} {\bibfnamefont {C.-X.}\ \bibnamefont {Liu}},\ }\href@noop
  {} {\bibfield  {journal} {\bibinfo  {journal} {Physical Review B}\ }\textbf
  {\bibinfo {volume} {88}},\ \bibinfo {pages} {085315} (\bibinfo {year}
  {2013})}\BibitemShut {NoStop}%
\bibitem [{\citenamefont {Wang}\ \emph
  {et~al.}(2014{\natexlab{b}})\citenamefont {Wang}, \citenamefont {Liu},
  \citenamefont {Zhang}, \citenamefont {Samarth}, \citenamefont {Zhang},\ and\
  \citenamefont {Liu}}]{wang2014}%
  \BibitemOpen
  \bibfield  {author} {\bibinfo {author} {\bibfnamefont {Q.-Z.}\ \bibnamefont
  {Wang}}, \bibinfo {author} {\bibfnamefont {X.}~\bibnamefont {Liu}}, \bibinfo
  {author} {\bibfnamefont {H.-J.}\ \bibnamefont {Zhang}}, \bibinfo {author}
  {\bibfnamefont {N.}~\bibnamefont {Samarth}}, \bibinfo {author} {\bibfnamefont
  {S.-C.}\ \bibnamefont {Zhang}}, \ and\ \bibinfo {author} {\bibfnamefont
  {C.-X.}\ \bibnamefont {Liu}},\ }\href@noop {} {\bibfield  {journal} {\bibinfo
   {journal} {Physical review letters}\ }\textbf {\bibinfo {volume} {113}},\
  \bibinfo {pages} {147201} (\bibinfo {year} {2014}{\natexlab{b}})}\BibitemShut
  {NoStop}%
\bibitem [{\citenamefont {Liu}\ \emph {et~al.}(2008{\natexlab{b}})\citenamefont
  {Liu}, \citenamefont {Hughes}, \citenamefont {Qi}, \citenamefont {Wang},\
  and\ \citenamefont {Zhang}}]{liu2008}%
  \BibitemOpen
  \bibfield  {author} {\bibinfo {author} {\bibfnamefont {C.-X.}\ \bibnamefont
  {Liu}}, \bibinfo {author} {\bibfnamefont {T.~L.}\ \bibnamefont {Hughes}},
  \bibinfo {author} {\bibfnamefont {X.-L.}\ \bibnamefont {Qi}}, \bibinfo
  {author} {\bibfnamefont {K.}~\bibnamefont {Wang}}, \ and\ \bibinfo {author}
  {\bibfnamefont {S.-C.}\ \bibnamefont {Zhang}},\ }\href@noop {} {\bibfield
  {journal} {\bibinfo  {journal} {Phys. Rev. Lett.}\ }\textbf {\bibinfo
  {volume} {100}},\ \bibinfo {pages} {236601} (\bibinfo {year}
  {2008}{\natexlab{b}})}\BibitemShut {NoStop}%
\bibitem [{\citenamefont {Dietl}\ \emph {et~al.}(2000)\citenamefont {Dietl},
  \citenamefont {Ohno}, \citenamefont {Matsukura}, \citenamefont {Cibert},\
  and\ \citenamefont {Ferrand}}]{dietl2000}%
  \BibitemOpen
  \bibfield  {author} {\bibinfo {author} {\bibfnamefont {T.}~\bibnamefont
  {Dietl}}, \bibinfo {author} {\bibfnamefont {H.}~\bibnamefont {Ohno}},
  \bibinfo {author} {\bibfnamefont {F.}~\bibnamefont {Matsukura}}, \bibinfo
  {author} {\bibfnamefont {J.}~\bibnamefont {Cibert}}, \ and\ \bibinfo {author}
  {\bibfnamefont {D.}~\bibnamefont {Ferrand}},\ }\href@noop {} {\bibfield
  {journal} {\bibinfo  {journal} {Science}\ }\textbf {\bibinfo {volume}
  {287}},\ \bibinfo {pages} {1019} (\bibinfo {year} {2000})}\BibitemShut
  {NoStop}%
\bibitem [{\citenamefont {Jungwirth}\ \emph {et~al.}(2006)\citenamefont
  {Jungwirth}, \citenamefont {Sinova}, \citenamefont {Ma\ifmmode~\check{s}\else
  \v{s}\fi{}ek}, \citenamefont {Ku\ifmmode~\check{c}\else \v{c}\fi{}era},\ and\
  \citenamefont {MacDonald}}]{jungwirth2006}%
  \BibitemOpen
  \bibfield  {author} {\bibinfo {author} {\bibfnamefont {T.}~\bibnamefont
  {Jungwirth}}, \bibinfo {author} {\bibfnamefont {J.}~\bibnamefont {Sinova}},
  \bibinfo {author} {\bibfnamefont {J.}~\bibnamefont {Ma\ifmmode~\check{s}\else
  \v{s}\fi{}ek}}, \bibinfo {author} {\bibfnamefont {J.}~\bibnamefont
  {Ku\ifmmode~\check{c}\else \v{c}\fi{}era}}, \ and\ \bibinfo {author}
  {\bibfnamefont {A.~H.}\ \bibnamefont {MacDonald}},\ }\href {\doibase
  10.1103/RevModPhys.78.809} {\bibfield  {journal} {\bibinfo  {journal} {Rev.
  Mod. Phys.}\ }\textbf {\bibinfo {volume} {78}},\ \bibinfo {pages} {809}
  (\bibinfo {year} {2006})}\BibitemShut {NoStop}%
\bibitem [{\citenamefont {Sato}\ \emph {et~al.}(2010)\citenamefont {Sato},
  \citenamefont {Bergqvist}, \citenamefont {Kudrnovsk\'y}, \citenamefont
  {Dederichs}, \citenamefont {Eriksson}, \citenamefont {Turek}, \citenamefont
  {Sanyal}, \citenamefont {Bouzerar}, \citenamefont {Katayama-Yoshida},
  \citenamefont {Dinh}, \citenamefont {Fukushima}, \citenamefont {Kizaki},\
  and\ \citenamefont {Zeller}}]{sato2010a}%
  \BibitemOpen
  \bibfield  {author} {\bibinfo {author} {\bibfnamefont {K.}~\bibnamefont
  {Sato}}, \bibinfo {author} {\bibfnamefont {L.}~\bibnamefont {Bergqvist}},
  \bibinfo {author} {\bibfnamefont {J.}~\bibnamefont {Kudrnovsk\'y}}, \bibinfo
  {author} {\bibfnamefont {P.~H.}\ \bibnamefont {Dederichs}}, \bibinfo {author}
  {\bibfnamefont {O.}~\bibnamefont {Eriksson}}, \bibinfo {author}
  {\bibfnamefont {I.}~\bibnamefont {Turek}}, \bibinfo {author} {\bibfnamefont
  {B.}~\bibnamefont {Sanyal}}, \bibinfo {author} {\bibfnamefont
  {G.}~\bibnamefont {Bouzerar}}, \bibinfo {author} {\bibfnamefont
  {H.}~\bibnamefont {Katayama-Yoshida}}, \bibinfo {author} {\bibfnamefont
  {V.~A.}\ \bibnamefont {Dinh}}, \bibinfo {author} {\bibfnamefont
  {T.}~\bibnamefont {Fukushima}}, \bibinfo {author} {\bibfnamefont
  {H.}~\bibnamefont {Kizaki}}, \ and\ \bibinfo {author} {\bibfnamefont
  {R.}~\bibnamefont {Zeller}},\ }\href@noop {} {\bibfield  {journal} {\bibinfo
  {journal} {Rev. Mod. Phys.}\ }\textbf {\bibinfo {volume} {82}},\ \bibinfo
  {pages} {1633} (\bibinfo {year} {2010})}\BibitemShut {NoStop}%
\bibitem [{\citenamefont {Dietl}(2010)}]{dietl2010}%
  \BibitemOpen
  \bibfield  {author} {\bibinfo {author} {\bibfnamefont {T.}~\bibnamefont
  {Dietl}},\ }\href@noop {} {\bibfield  {journal} {\bibinfo  {journal} {Nature
  materials}\ }\textbf {\bibinfo {volume} {9}},\ \bibinfo {pages} {965}
  (\bibinfo {year} {2010})}\BibitemShut {NoStop}%
\bibitem [{\citenamefont {Zhang}\ \emph
  {et~al.}(2014{\natexlab{a}})\citenamefont {Zhang}, \citenamefont {Xu},
  \citenamefont {Wang}, \citenamefont {Chang},\ and\ \citenamefont
  {Zhang}}]{zhang2014}%
  \BibitemOpen
  \bibfield  {author} {\bibinfo {author} {\bibfnamefont {H.}~\bibnamefont
  {Zhang}}, \bibinfo {author} {\bibfnamefont {Y.}~\bibnamefont {Xu}}, \bibinfo
  {author} {\bibfnamefont {J.}~\bibnamefont {Wang}}, \bibinfo {author}
  {\bibfnamefont {K.}~\bibnamefont {Chang}}, \ and\ \bibinfo {author}
  {\bibfnamefont {S.-C.}\ \bibnamefont {Zhang}},\ }\href@noop {} {\bibfield
  {journal} {\bibinfo  {journal} {Physical Review Letters}\ }\textbf {\bibinfo
  {volume} {112}},\ \bibinfo {pages} {216803} (\bibinfo {year}
  {2014}{\natexlab{a}})}\BibitemShut {NoStop}%
\bibitem [{\citenamefont {Liu}\ \emph {et~al.}(2010{\natexlab{a}})\citenamefont
  {Liu}, \citenamefont {Zhang}, \citenamefont {Yan}, \citenamefont {Qi},
  \citenamefont {Frauenheim}, \citenamefont {Dai}, \citenamefont {Fang},\ and\
  \citenamefont {Zhang}}]{liu2010a}%
  \BibitemOpen
  \bibfield  {author} {\bibinfo {author} {\bibfnamefont {C.-X.}\ \bibnamefont
  {Liu}}, \bibinfo {author} {\bibfnamefont {H.}~\bibnamefont {Zhang}}, \bibinfo
  {author} {\bibfnamefont {B.}~\bibnamefont {Yan}}, \bibinfo {author}
  {\bibfnamefont {X.-L.}\ \bibnamefont {Qi}}, \bibinfo {author} {\bibfnamefont
  {T.}~\bibnamefont {Frauenheim}}, \bibinfo {author} {\bibfnamefont
  {X.}~\bibnamefont {Dai}}, \bibinfo {author} {\bibfnamefont {Z.}~\bibnamefont
  {Fang}}, \ and\ \bibinfo {author} {\bibfnamefont {S.-C.}\ \bibnamefont
  {Zhang}},\ }\href@noop {} {\bibfield  {journal} {\bibinfo  {journal} {Phys.
  Rev. B}\ }\textbf {\bibinfo {volume} {81}},\ \bibinfo {pages} {041307}
  (\bibinfo {year} {2010}{\natexlab{a}})}\BibitemShut {NoStop}%
\bibitem [{\citenamefont {Lu}\ \emph {et~al.}(2010)\citenamefont {Lu},
  \citenamefont {Shan}, \citenamefont {Yao}, \citenamefont {Niu},\ and\
  \citenamefont {Shen}}]{lu2010}%
  \BibitemOpen
  \bibfield  {author} {\bibinfo {author} {\bibfnamefont {H.~Z.}\ \bibnamefont
  {Lu}}, \bibinfo {author} {\bibfnamefont {W.~Y.}\ \bibnamefont {Shan}},
  \bibinfo {author} {\bibfnamefont {W.}~\bibnamefont {Yao}}, \bibinfo {author}
  {\bibfnamefont {Q.}~\bibnamefont {Niu}}, \ and\ \bibinfo {author}
  {\bibfnamefont {S.~Q.}\ \bibnamefont {Shen}},\ }\href@noop {} {\bibfield
  {journal} {\bibinfo  {journal} {Phys. Rev. B}\ }\textbf {\bibinfo {volume}
  {81}},\ \bibinfo {pages} {115407} (\bibinfo {year} {2010})}\BibitemShut
  {NoStop}%
\bibitem [{\citenamefont {Zhang}\ \emph
  {et~al.}(2013{\natexlab{a}})\citenamefont {Zhang}, \citenamefont {Chang},
  \citenamefont {Tang}, \citenamefont {Zhang}, \citenamefont {Feng},
  \citenamefont {Li}, \citenamefont {Wang}, \citenamefont {Chen}, \citenamefont
  {Liu}, \citenamefont {Duan}, \citenamefont {He}, \citenamefont {Xue},
  \citenamefont {Ma},\ and\ \citenamefont {Wang}}]{zhang2013a}%
  \BibitemOpen
  \bibfield  {author} {\bibinfo {author} {\bibfnamefont {J.}~\bibnamefont
  {Zhang}}, \bibinfo {author} {\bibfnamefont {C.-Z.}\ \bibnamefont {Chang}},
  \bibinfo {author} {\bibfnamefont {P.}~\bibnamefont {Tang}}, \bibinfo {author}
  {\bibfnamefont {Z.}~\bibnamefont {Zhang}}, \bibinfo {author} {\bibfnamefont
  {X.}~\bibnamefont {Feng}}, \bibinfo {author} {\bibfnamefont {K.}~\bibnamefont
  {Li}}, \bibinfo {author} {\bibfnamefont {L.-l.}\ \bibnamefont {Wang}},
  \bibinfo {author} {\bibfnamefont {X.}~\bibnamefont {Chen}}, \bibinfo {author}
  {\bibfnamefont {C.}~\bibnamefont {Liu}}, \bibinfo {author} {\bibfnamefont
  {W.}~\bibnamefont {Duan}}, \bibinfo {author} {\bibfnamefont {K.}~\bibnamefont
  {He}}, \bibinfo {author} {\bibfnamefont {Q.-K.}\ \bibnamefont {Xue}},
  \bibinfo {author} {\bibfnamefont {X.}~\bibnamefont {Ma}}, \ and\ \bibinfo
  {author} {\bibfnamefont {Y.}~\bibnamefont {Wang}},\ }\href@noop {} {\bibfield
   {journal} {\bibinfo  {journal} {Science}\ }\textbf {\bibinfo {volume}
  {339}},\ \bibinfo {pages} {1582} (\bibinfo {year}
  {2013}{\natexlab{a}})}\BibitemShut {NoStop}%
\bibitem [{\citenamefont {Redlich}(1984)}]{redlich1984}%
  \BibitemOpen
  \bibfield  {author} {\bibinfo {author} {\bibfnamefont {A.~N.}\ \bibnamefont
  {Redlich}},\ }\href@noop {} {\bibfield  {journal} {\bibinfo  {journal}
  {Physical Review D}\ }\textbf {\bibinfo {volume} {29}},\ \bibinfo {pages}
  {2366} (\bibinfo {year} {1984})}\BibitemShut {NoStop}%
\bibitem [{\citenamefont {Nielsen}\ and\ \citenamefont
  {Ninomiya}(1981{\natexlab{a}})}]{nielsen1981}%
  \BibitemOpen
  \bibfield  {author} {\bibinfo {author} {\bibfnamefont {H.~B.}\ \bibnamefont
  {Nielsen}}\ and\ \bibinfo {author} {\bibfnamefont {M.}~\bibnamefont
  {Ninomiya}},\ }\href@noop {} {\bibfield  {journal} {\bibinfo  {journal}
  {Nucl. Phys. B}\ }\textbf {\bibinfo {volume} {185}},\ \bibinfo {pages} {20}
  (\bibinfo {year} {1981}{\natexlab{a}})}\BibitemShut {NoStop}%
\bibitem [{\citenamefont {Nielsen}\ and\ \citenamefont
  {Ninomiya}(1981{\natexlab{b}})}]{nielsen1981a}%
  \BibitemOpen
  \bibfield  {author} {\bibinfo {author} {\bibfnamefont {H.~B.}\ \bibnamefont
  {Nielsen}}\ and\ \bibinfo {author} {\bibfnamefont {M.}~\bibnamefont
  {Ninomiya}},\ }\href@noop {} {\bibfield  {journal} {\bibinfo  {journal}
  {Nucl. Phys. B}\ }\textbf {\bibinfo {volume} {193}},\ \bibinfo {pages} {173}
  (\bibinfo {year} {1981}{\natexlab{b}})}\BibitemShut {NoStop}%
\bibitem [{\citenamefont {Chang}\ \emph
  {et~al.}(2013{\natexlab{b}})\citenamefont {Chang}, \citenamefont {Zhang},
  \citenamefont {Liu}, \citenamefont {Zhang}, \citenamefont {Feng},
  \citenamefont {Li}, \citenamefont {Wang}, \citenamefont {Chen}, \citenamefont
  {Dai}, \citenamefont {Fang}, \citenamefont {Qi}, \citenamefont {Zhang},
  \citenamefont {Wang}, \citenamefont {He}, \citenamefont {Ma},\ and\
  \citenamefont {Xue}}]{chang2013b}%
  \BibitemOpen
  \bibfield  {author} {\bibinfo {author} {\bibfnamefont {C.}~\bibnamefont
  {Chang}}, \bibinfo {author} {\bibfnamefont {J.}~\bibnamefont {Zhang}},
  \bibinfo {author} {\bibfnamefont {M.}~\bibnamefont {Liu}}, \bibinfo {author}
  {\bibfnamefont {Z.}~\bibnamefont {Zhang}}, \bibinfo {author} {\bibfnamefont
  {X.}~\bibnamefont {Feng}}, \bibinfo {author} {\bibfnamefont {K.}~\bibnamefont
  {Li}}, \bibinfo {author} {\bibfnamefont {L.}~\bibnamefont {Wang}}, \bibinfo
  {author} {\bibfnamefont {X.}~\bibnamefont {Chen}}, \bibinfo {author}
  {\bibfnamefont {X.}~\bibnamefont {Dai}}, \bibinfo {author} {\bibfnamefont
  {Z.}~\bibnamefont {Fang}}, \bibinfo {author} {\bibfnamefont {X.-L.}\
  \bibnamefont {Qi}}, \bibinfo {author} {\bibfnamefont {S.-C.}\ \bibnamefont
  {Zhang}}, \bibinfo {author} {\bibfnamefont {Y.}~\bibnamefont {Wang}},
  \bibinfo {author} {\bibfnamefont {K.}~\bibnamefont {He}}, \bibinfo {author}
  {\bibfnamefont {X.-C.}\ \bibnamefont {Ma}}, \ and\ \bibinfo {author}
  {\bibfnamefont {Q.-K.}\ \bibnamefont {Xue}},\ }\href@noop {} {\bibfield
  {journal} {\bibinfo  {journal} {Advanced Materials}\ }\textbf {\bibinfo
  {volume} {25}},\ \bibinfo {pages} {1065} (\bibinfo {year}
  {2013}{\natexlab{b}})}\BibitemShut {NoStop}%
\bibitem [{\citenamefont {Hor}\ \emph {et~al.}(2010)\citenamefont {Hor},
  \citenamefont {Roushan}, \citenamefont {Beidenkopf}, \citenamefont {Seo},
  \citenamefont {Qu}, \citenamefont {Checkelsky}, \citenamefont {Wray},
  \citenamefont {Hsieh}, \citenamefont {Xia}, \citenamefont {Xu}, \citenamefont
  {Qian}, \citenamefont {Hasan}, \citenamefont {Ong}, \citenamefont {Yazdani},\
  and\ \citenamefont {Cava}}]{hor2010b}%
  \BibitemOpen
  \bibfield  {author} {\bibinfo {author} {\bibfnamefont {Y.~S.}\ \bibnamefont
  {Hor}}, \bibinfo {author} {\bibfnamefont {P.}~\bibnamefont {Roushan}},
  \bibinfo {author} {\bibfnamefont {H.}~\bibnamefont {Beidenkopf}}, \bibinfo
  {author} {\bibfnamefont {J.}~\bibnamefont {Seo}}, \bibinfo {author}
  {\bibfnamefont {D.}~\bibnamefont {Qu}}, \bibinfo {author} {\bibfnamefont
  {J.~G.}\ \bibnamefont {Checkelsky}}, \bibinfo {author} {\bibfnamefont
  {L.~A.}\ \bibnamefont {Wray}}, \bibinfo {author} {\bibfnamefont
  {D.}~\bibnamefont {Hsieh}}, \bibinfo {author} {\bibfnamefont
  {Y.}~\bibnamefont {Xia}}, \bibinfo {author} {\bibfnamefont {S.}~\bibnamefont
  {Xu}}, \bibinfo {author} {\bibfnamefont {D.}~\bibnamefont {Qian}}, \bibinfo
  {author} {\bibfnamefont {M.~Z.}\ \bibnamefont {Hasan}}, \bibinfo {author}
  {\bibfnamefont {N.~P.}\ \bibnamefont {Ong}}, \bibinfo {author} {\bibfnamefont
  {A.}~\bibnamefont {Yazdani}}, \ and\ \bibinfo {author} {\bibfnamefont
  {R.~J.}\ \bibnamefont {Cava}},\ }\href@noop {} {\bibfield  {journal}
  {\bibinfo  {journal} {Phys. Rev. B}\ }\textbf {\bibinfo {volume} {81}},\
  \bibinfo {pages} {195203} (\bibinfo {year} {2010})}\BibitemShut {NoStop}%
\bibitem [{\citenamefont {Kou}\ \emph {et~al.}(2013{\natexlab{a}})\citenamefont
  {Kou}, \citenamefont {Lang}, \citenamefont {Fan}, \citenamefont {Jiang},
  \citenamefont {Nie}, \citenamefont {Zhang}, \citenamefont {Jiang},
  \citenamefont {Wang}, \citenamefont {Yao}, \citenamefont {He} \emph
  {et~al.}}]{kou2013b}%
  \BibitemOpen
  \bibfield  {author} {\bibinfo {author} {\bibfnamefont {X.}~\bibnamefont
  {Kou}}, \bibinfo {author} {\bibfnamefont {M.}~\bibnamefont {Lang}}, \bibinfo
  {author} {\bibfnamefont {Y.}~\bibnamefont {Fan}}, \bibinfo {author}
  {\bibfnamefont {Y.}~\bibnamefont {Jiang}}, \bibinfo {author} {\bibfnamefont
  {T.}~\bibnamefont {Nie}}, \bibinfo {author} {\bibfnamefont {J.}~\bibnamefont
  {Zhang}}, \bibinfo {author} {\bibfnamefont {W.}~\bibnamefont {Jiang}},
  \bibinfo {author} {\bibfnamefont {Y.}~\bibnamefont {Wang}}, \bibinfo {author}
  {\bibfnamefont {Y.}~\bibnamefont {Yao}}, \bibinfo {author} {\bibfnamefont
  {L.}~\bibnamefont {He}},  \emph {et~al.},\ }\href@noop {} {\bibfield
  {journal} {\bibinfo  {journal} {ACS nano}\ }\textbf {\bibinfo {volume} {7}},\
  \bibinfo {pages} {9205} (\bibinfo {year} {2013}{\natexlab{a}})}\BibitemShut
  {NoStop}%
\bibitem [{\citenamefont {Kou}\ \emph {et~al.}(2013{\natexlab{b}})\citenamefont
  {Kou}, \citenamefont {He}, \citenamefont {Lang}, \citenamefont {Fan},
  \citenamefont {Wong}, \citenamefont {Jiang}, \citenamefont {Nie},
  \citenamefont {Jiang}, \citenamefont {Upadhyaya}, \citenamefont {Xing} \emph
  {et~al.}}]{kou2013a}%
  \BibitemOpen
  \bibfield  {author} {\bibinfo {author} {\bibfnamefont {X.}~\bibnamefont
  {Kou}}, \bibinfo {author} {\bibfnamefont {L.}~\bibnamefont {He}}, \bibinfo
  {author} {\bibfnamefont {M.}~\bibnamefont {Lang}}, \bibinfo {author}
  {\bibfnamefont {Y.}~\bibnamefont {Fan}}, \bibinfo {author} {\bibfnamefont
  {K.}~\bibnamefont {Wong}}, \bibinfo {author} {\bibfnamefont {Y.}~\bibnamefont
  {Jiang}}, \bibinfo {author} {\bibfnamefont {T.}~\bibnamefont {Nie}}, \bibinfo
  {author} {\bibfnamefont {W.}~\bibnamefont {Jiang}}, \bibinfo {author}
  {\bibfnamefont {P.}~\bibnamefont {Upadhyaya}}, \bibinfo {author}
  {\bibfnamefont {Z.}~\bibnamefont {Xing}},  \emph {et~al.},\ }\href@noop {}
  {\bibfield  {journal} {\bibinfo  {journal} {Nano letters}\ }\textbf {\bibinfo
  {volume} {13}},\ \bibinfo {pages} {4587} (\bibinfo {year}
  {2013}{\natexlab{b}})}\BibitemShut {NoStop}%
\bibitem [{\citenamefont {Li}\ \emph {et~al.}(2015)\citenamefont {Li},
  \citenamefont {Chang}, \citenamefont {Wu}, \citenamefont {Tao}, \citenamefont
  {Zhao}, \citenamefont {Chan}, \citenamefont {Moodera}, \citenamefont {Li},\
  and\ \citenamefont {Zhu}}]{li2015}%
  \BibitemOpen
  \bibfield  {author} {\bibinfo {author} {\bibfnamefont {M.}~\bibnamefont
  {Li}}, \bibinfo {author} {\bibfnamefont {C.-Z.}\ \bibnamefont {Chang}},
  \bibinfo {author} {\bibfnamefont {L.}~\bibnamefont {Wu}}, \bibinfo {author}
  {\bibfnamefont {J.}~\bibnamefont {Tao}}, \bibinfo {author} {\bibfnamefont
  {W.}~\bibnamefont {Zhao}}, \bibinfo {author} {\bibfnamefont {W.}~\bibnamefont
  {Chan}, \bibfnamefont {Moses~H.}}, \bibinfo {author} {\bibfnamefont {J.~S.}\
  \bibnamefont {Moodera}}, \bibinfo {author} {\bibfnamefont {J.}~\bibnamefont
  {Li}}, \ and\ \bibinfo {author} {\bibfnamefont {Y.}~\bibnamefont {Zhu}},\
  }\href@noop {} {\bibfield  {journal} {\bibinfo  {journal} {Phys. Rev. Lett.}\
  }\textbf {\bibinfo {volume} {114}},\ \bibinfo {pages} {146802} (\bibinfo
  {year} {2015})}\BibitemShut {NoStop}%
\bibitem [{\citenamefont {Dyck}\ \emph {et~al.}(2005)\citenamefont {Dyck},
  \citenamefont {Dra{\v{s}}ar}, \citenamefont {Lo{\v{s}}t¡¯{\'a}k},\ and\
  \citenamefont {Uher}}]{dyck2005}%
  \BibitemOpen
  \bibfield  {author} {\bibinfo {author} {\bibfnamefont {J.}~\bibnamefont
  {Dyck}}, \bibinfo {author} {\bibfnamefont {{\v{C}}.}~\bibnamefont
  {Dra{\v{s}}ar}}, \bibinfo {author} {\bibfnamefont {P.}~\bibnamefont
  {Lo{\v{s}}t¡¯{\'a}k}}, \ and\ \bibinfo {author} {\bibfnamefont
  {C.}~\bibnamefont {Uher}},\ }\href@noop {} {\bibfield  {journal} {\bibinfo
  {journal} {Physical Review B}\ }\textbf {\bibinfo {volume} {71}},\ \bibinfo
  {pages} {115214} (\bibinfo {year} {2005})}\BibitemShut {NoStop}%
\bibitem [{\citenamefont {Kim}\ \emph {et~al.}(2013)\citenamefont {Kim},
  \citenamefont {Kim}, \citenamefont {Wang}, \citenamefont {Kulbachinskii},
  \citenamefont {Ogawa}, \citenamefont {Sasaki}, \citenamefont {Ohnishi},
  \citenamefont {Kitaura}, \citenamefont {Wu}, \citenamefont {Li},
  \citenamefont {Yamamoto}, \citenamefont {Azuma}, \citenamefont {Kamada},\
  and\ \citenamefont {Dobrosavljevi\ifmmode~\acute{c}\else
  \'{c}\fi{}}}]{kim2013a}%
  \BibitemOpen
  \bibfield  {author} {\bibinfo {author} {\bibfnamefont {H.-J.}\ \bibnamefont
  {Kim}}, \bibinfo {author} {\bibfnamefont {K.-S.}\ \bibnamefont {Kim}},
  \bibinfo {author} {\bibfnamefont {J.-F.}\ \bibnamefont {Wang}}, \bibinfo
  {author} {\bibfnamefont {V.~A.}\ \bibnamefont {Kulbachinskii}}, \bibinfo
  {author} {\bibfnamefont {K.}~\bibnamefont {Ogawa}}, \bibinfo {author}
  {\bibfnamefont {M.}~\bibnamefont {Sasaki}}, \bibinfo {author} {\bibfnamefont
  {A.}~\bibnamefont {Ohnishi}}, \bibinfo {author} {\bibfnamefont
  {M.}~\bibnamefont {Kitaura}}, \bibinfo {author} {\bibfnamefont {Y.-Y.}\
  \bibnamefont {Wu}}, \bibinfo {author} {\bibfnamefont {L.}~\bibnamefont {Li}},
  \bibinfo {author} {\bibfnamefont {I.}~\bibnamefont {Yamamoto}}, \bibinfo
  {author} {\bibfnamefont {J.}~\bibnamefont {Azuma}}, \bibinfo {author}
  {\bibfnamefont {M.}~\bibnamefont {Kamada}}, \ and\ \bibinfo {author}
  {\bibfnamefont {V.}~\bibnamefont {Dobrosavljevi\ifmmode~\acute{c}\else
  \'{c}\fi{}}},\ }\href@noop {} {\bibfield  {journal} {\bibinfo  {journal}
  {Phys. Rev. Lett.}\ }\textbf {\bibinfo {volume} {110}},\ \bibinfo {pages}
  {136601} (\bibinfo {year} {2013})}\BibitemShut {NoStop}%
\bibitem [{\citenamefont {von Bardeleben}\ \emph {et~al.}(2013)\citenamefont
  {von Bardeleben}, \citenamefont {Cantin}, \citenamefont {Zhang},
  \citenamefont {Richardella}, \citenamefont {Rench}, \citenamefont {Samarth},\
  and\ \citenamefont {Borchers}}]{bardeleben2013}%
  \BibitemOpen
  \bibfield  {author} {\bibinfo {author} {\bibfnamefont {H.~J.}\ \bibnamefont
  {von Bardeleben}}, \bibinfo {author} {\bibfnamefont {J.~L.}\ \bibnamefont
  {Cantin}}, \bibinfo {author} {\bibfnamefont {D.~M.}\ \bibnamefont {Zhang}},
  \bibinfo {author} {\bibfnamefont {A.}~\bibnamefont {Richardella}}, \bibinfo
  {author} {\bibfnamefont {D.~W.}\ \bibnamefont {Rench}}, \bibinfo {author}
  {\bibfnamefont {N.}~\bibnamefont {Samarth}}, \ and\ \bibinfo {author}
  {\bibfnamefont {J.~A.}\ \bibnamefont {Borchers}},\ }\href@noop {} {\bibfield
  {journal} {\bibinfo  {journal} {Phys. Rev. B}\ }\textbf {\bibinfo {volume}
  {88}},\ \bibinfo {pages} {075149} (\bibinfo {year} {2013})}\BibitemShut
  {NoStop}%
\bibitem [{\citenamefont {Zhang}\ \emph
  {et~al.}(2012{\natexlab{a}})\citenamefont {Zhang}, \citenamefont
  {Richardella}, \citenamefont {Rench}, \citenamefont {Xu}, \citenamefont
  {Kandala}, \citenamefont {Flanagan}, \citenamefont {Beidenkopf},
  \citenamefont {Yeats}, \citenamefont {Buckley}, \citenamefont {Klimov},
  \citenamefont {Awschalom}, \citenamefont {Yazdani}, \citenamefont {Schiffer},
  \citenamefont {Hasan},\ and\ \citenamefont {Samarth}}]{zhang2012a}%
  \BibitemOpen
  \bibfield  {author} {\bibinfo {author} {\bibfnamefont {D.}~\bibnamefont
  {Zhang}}, \bibinfo {author} {\bibfnamefont {A.}~\bibnamefont {Richardella}},
  \bibinfo {author} {\bibfnamefont {D.~W.}\ \bibnamefont {Rench}}, \bibinfo
  {author} {\bibfnamefont {S.-Y.}\ \bibnamefont {Xu}}, \bibinfo {author}
  {\bibfnamefont {A.}~\bibnamefont {Kandala}}, \bibinfo {author} {\bibfnamefont
  {T.~C.}\ \bibnamefont {Flanagan}}, \bibinfo {author} {\bibfnamefont
  {H.}~\bibnamefont {Beidenkopf}}, \bibinfo {author} {\bibfnamefont {A.~L.}\
  \bibnamefont {Yeats}}, \bibinfo {author} {\bibfnamefont {B.~B.}\ \bibnamefont
  {Buckley}}, \bibinfo {author} {\bibfnamefont {P.~V.}\ \bibnamefont {Klimov}},
  \bibinfo {author} {\bibfnamefont {D.~D.}\ \bibnamefont {Awschalom}}, \bibinfo
  {author} {\bibfnamefont {A.}~\bibnamefont {Yazdani}}, \bibinfo {author}
  {\bibfnamefont {P.}~\bibnamefont {Schiffer}}, \bibinfo {author}
  {\bibfnamefont {M.~Z.}\ \bibnamefont {Hasan}}, \ and\ \bibinfo {author}
  {\bibfnamefont {N.}~\bibnamefont {Samarth}},\ }\href@noop {} {\bibfield
  {journal} {\bibinfo  {journal} {Phys. Rev. B}\ }\textbf {\bibinfo {volume}
  {86}},\ \bibinfo {pages} {205127} (\bibinfo {year}
  {2012}{\natexlab{a}})}\BibitemShut {NoStop}%
\bibitem [{\citenamefont {Wang}\ \emph
  {et~al.}(2013{\natexlab{a}})\citenamefont {Wang}, \citenamefont {Lian},
  \citenamefont {Zhang},\ and\ \citenamefont {Zhang}}]{wang2013b}%
  \BibitemOpen
  \bibfield  {author} {\bibinfo {author} {\bibfnamefont {J.}~\bibnamefont
  {Wang}}, \bibinfo {author} {\bibfnamefont {B.}~\bibnamefont {Lian}}, \bibinfo
  {author} {\bibfnamefont {H.}~\bibnamefont {Zhang}}, \ and\ \bibinfo {author}
  {\bibfnamefont {S.-C.}\ \bibnamefont {Zhang}},\ }\href@noop {} {\bibfield
  {journal} {\bibinfo  {journal} {Physical review letters}\ }\textbf {\bibinfo
  {volume} {111}},\ \bibinfo {pages} {086803} (\bibinfo {year}
  {2013}{\natexlab{a}})}\BibitemShut {NoStop}%
\bibitem [{\citenamefont {Wang}\ \emph
  {et~al.}(2014{\natexlab{c}})\citenamefont {Wang}, \citenamefont {Lian},\ and\
  \citenamefont {Zhang}}]{wang2014a}%
  \BibitemOpen
  \bibfield  {author} {\bibinfo {author} {\bibfnamefont {J.}~\bibnamefont
  {Wang}}, \bibinfo {author} {\bibfnamefont {B.}~\bibnamefont {Lian}}, \ and\
  \bibinfo {author} {\bibfnamefont {S.-C.}\ \bibnamefont {Zhang}},\ }\href@noop
  {} {\bibfield  {journal} {\bibinfo  {journal} {Physical Review B}\ }\textbf
  {\bibinfo {volume} {89}},\ \bibinfo {pages} {085106} (\bibinfo {year}
  {2014}{\natexlab{c}})}\BibitemShut {NoStop}%
\bibitem [{\citenamefont {Kivelson}\ \emph {et~al.}(1992)\citenamefont
  {Kivelson}, \citenamefont {Lee},\ and\ \citenamefont {Zhang}}]{kivelson1992}%
  \BibitemOpen
  \bibfield  {author} {\bibinfo {author} {\bibfnamefont {S.}~\bibnamefont
  {Kivelson}}, \bibinfo {author} {\bibfnamefont {D.-H.}\ \bibnamefont {Lee}}, \
  and\ \bibinfo {author} {\bibfnamefont {S.-C.}\ \bibnamefont {Zhang}},\
  }\href@noop {} {\bibfield  {journal} {\bibinfo  {journal} {Physical Review
  B}\ }\textbf {\bibinfo {volume} {46}},\ \bibinfo {pages} {2223} (\bibinfo
  {year} {1992})}\BibitemShut {NoStop}%
\bibitem [{\citenamefont {Khmel¡¯nitskii}(1983)}]{khmel1983}%
  \BibitemOpen
  \bibfield  {author} {\bibinfo {author} {\bibfnamefont {D.}~\bibnamefont
  {Khmel¡¯nitskii}},\ }\href@noop {} {\bibfield  {journal} {\bibinfo  {journal}
  {JETP lett}\ }\textbf {\bibinfo {volume} {38}} (\bibinfo {year}
  {1983})}\BibitemShut {NoStop}%
\bibitem [{\citenamefont {Qiao}\ \emph {et~al.}(2010)\citenamefont {Qiao},
  \citenamefont {Yang}, \citenamefont {Feng}, \citenamefont {Tse},
  \citenamefont {Ding}, \citenamefont {Yao}, \citenamefont {Wang},\ and\
  \citenamefont {Niu}}]{qiao2010}%
  \BibitemOpen
  \bibfield  {author} {\bibinfo {author} {\bibfnamefont {Z.}~\bibnamefont
  {Qiao}}, \bibinfo {author} {\bibfnamefont {S.~A.}\ \bibnamefont {Yang}},
  \bibinfo {author} {\bibfnamefont {W.}~\bibnamefont {Feng}}, \bibinfo {author}
  {\bibfnamefont {W.-K.}\ \bibnamefont {Tse}}, \bibinfo {author} {\bibfnamefont
  {J.}~\bibnamefont {Ding}}, \bibinfo {author} {\bibfnamefont {Y.}~\bibnamefont
  {Yao}}, \bibinfo {author} {\bibfnamefont {J.}~\bibnamefont {Wang}}, \ and\
  \bibinfo {author} {\bibfnamefont {Q.}~\bibnamefont {Niu}},\ }\href {\doibase
  10.1103/PhysRevB.82.161414} {\bibfield  {journal} {\bibinfo  {journal} {Phys.
  Rev. B}\ }\textbf {\bibinfo {volume} {82}},\ \bibinfo {pages} {161414}
  (\bibinfo {year} {2010})}\BibitemShut {NoStop}%
\bibitem [{\citenamefont {Tse}\ \emph {et~al.}(2011)\citenamefont {Tse},
  \citenamefont {Qiao}, \citenamefont {Yao}, \citenamefont {MacDonald},\ and\
  \citenamefont {Niu}}]{tse2011}%
  \BibitemOpen
  \bibfield  {author} {\bibinfo {author} {\bibfnamefont {W.-K.}\ \bibnamefont
  {Tse}}, \bibinfo {author} {\bibfnamefont {Z.}~\bibnamefont {Qiao}}, \bibinfo
  {author} {\bibfnamefont {Y.}~\bibnamefont {Yao}}, \bibinfo {author}
  {\bibfnamefont {A.}~\bibnamefont {MacDonald}}, \ and\ \bibinfo {author}
  {\bibfnamefont {Q.}~\bibnamefont {Niu}},\ }\href@noop {} {\bibfield
  {journal} {\bibinfo  {journal} {Physical Review B}\ }\textbf {\bibinfo
  {volume} {83}},\ \bibinfo {pages} {155447} (\bibinfo {year}
  {2011})}\BibitemShut {NoStop}%
\bibitem [{\citenamefont {Qiao}\ \emph {et~al.}(2014)\citenamefont {Qiao},
  \citenamefont {Ren}, \citenamefont {Chen}, \citenamefont {Bellaiche},
  \citenamefont {Zhang}, \citenamefont {MacDonald},\ and\ \citenamefont
  {Niu}}]{qiao2014}%
  \BibitemOpen
  \bibfield  {author} {\bibinfo {author} {\bibfnamefont {Z.}~\bibnamefont
  {Qiao}}, \bibinfo {author} {\bibfnamefont {W.}~\bibnamefont {Ren}}, \bibinfo
  {author} {\bibfnamefont {H.}~\bibnamefont {Chen}}, \bibinfo {author}
  {\bibfnamefont {L.}~\bibnamefont {Bellaiche}}, \bibinfo {author}
  {\bibfnamefont {Z.}~\bibnamefont {Zhang}}, \bibinfo {author} {\bibfnamefont
  {A.}~\bibnamefont {MacDonald}}, \ and\ \bibinfo {author} {\bibfnamefont
  {Q.}~\bibnamefont {Niu}},\ }\href@noop {} {\bibfield  {journal} {\bibinfo
  {journal} {Physical review letters}\ }\textbf {\bibinfo {volume} {112}},\
  \bibinfo {pages} {116404} (\bibinfo {year} {2014})}\BibitemShut {NoStop}%
\bibitem [{\citenamefont {Zhang}\ \emph
  {et~al.}(2012{\natexlab{b}})\citenamefont {Zhang}, \citenamefont {Lazo},
  \citenamefont {Bl\"ugel}, \citenamefont {Heinze},\ and\ \citenamefont
  {Mokrousov}}]{zhang2012b}%
  \BibitemOpen
  \bibfield  {author} {\bibinfo {author} {\bibfnamefont {H.}~\bibnamefont
  {Zhang}}, \bibinfo {author} {\bibfnamefont {C.}~\bibnamefont {Lazo}},
  \bibinfo {author} {\bibfnamefont {S.}~\bibnamefont {Bl\"ugel}}, \bibinfo
  {author} {\bibfnamefont {S.}~\bibnamefont {Heinze}}, \ and\ \bibinfo {author}
  {\bibfnamefont {Y.}~\bibnamefont {Mokrousov}},\ }\href@noop {} {\bibfield
  {journal} {\bibinfo  {journal} {Phys. Rev. Lett.}\ }\textbf {\bibinfo
  {volume} {108}},\ \bibinfo {pages} {056802} (\bibinfo {year}
  {2012}{\natexlab{b}})}\BibitemShut {NoStop}%
\bibitem [{\citenamefont {Ezawa}(2012)}]{ezawa2012}%
  \BibitemOpen
  \bibfield  {author} {\bibinfo {author} {\bibfnamefont {M.}~\bibnamefont
  {Ezawa}},\ }\href@noop {} {\bibfield  {journal} {\bibinfo  {journal}
  {Physical review letters}\ }\textbf {\bibinfo {volume} {109}},\ \bibinfo
  {pages} {055502} (\bibinfo {year} {2012})}\BibitemShut {NoStop}%
\bibitem [{\citenamefont {Pan}\ \emph {et~al.}(2014)\citenamefont {Pan},
  \citenamefont {Li}, \citenamefont {Liu}, \citenamefont {Zhu}, \citenamefont
  {Qiao},\ and\ \citenamefont {Yao}}]{pan2014}%
  \BibitemOpen
  \bibfield  {author} {\bibinfo {author} {\bibfnamefont {H.}~\bibnamefont
  {Pan}}, \bibinfo {author} {\bibfnamefont {Z.}~\bibnamefont {Li}}, \bibinfo
  {author} {\bibfnamefont {C.-C.}\ \bibnamefont {Liu}}, \bibinfo {author}
  {\bibfnamefont {G.}~\bibnamefont {Zhu}}, \bibinfo {author} {\bibfnamefont
  {Z.}~\bibnamefont {Qiao}}, \ and\ \bibinfo {author} {\bibfnamefont
  {Y.}~\bibnamefont {Yao}},\ }\href@noop {} {\bibfield  {journal} {\bibinfo
  {journal} {Physical review letters}\ }\textbf {\bibinfo {volume} {112}},\
  \bibinfo {pages} {106802} (\bibinfo {year} {2014})}\BibitemShut {NoStop}%
\bibitem [{\citenamefont {Wu}\ \emph {et~al.}(2014{\natexlab{a}})\citenamefont
  {Wu}, \citenamefont {Shan},\ and\ \citenamefont {Yan}}]{wu2014}%
  \BibitemOpen
  \bibfield  {author} {\bibinfo {author} {\bibfnamefont {S.-C.}\ \bibnamefont
  {Wu}}, \bibinfo {author} {\bibfnamefont {G.}~\bibnamefont {Shan}}, \ and\
  \bibinfo {author} {\bibfnamefont {B.}~\bibnamefont {Yan}},\ }\href@noop {}
  {\bibfield  {journal} {\bibinfo  {journal} {Physical review letters}\
  }\textbf {\bibinfo {volume} {113}},\ \bibinfo {pages} {256401} (\bibinfo
  {year} {2014}{\natexlab{a}})}\BibitemShut {NoStop}%
\bibitem [{\citenamefont {Xu}\ \emph {et~al.}(2013)\citenamefont {Xu},
  \citenamefont {Yan}, \citenamefont {Zhang}, \citenamefont {Wang},
  \citenamefont {Xu}, \citenamefont {Tang}, \citenamefont {Duan},\ and\
  \citenamefont {Zhang}}]{xu2013}%
  \BibitemOpen
  \bibfield  {author} {\bibinfo {author} {\bibfnamefont {Y.}~\bibnamefont
  {Xu}}, \bibinfo {author} {\bibfnamefont {B.}~\bibnamefont {Yan}}, \bibinfo
  {author} {\bibfnamefont {H.-J.}\ \bibnamefont {Zhang}}, \bibinfo {author}
  {\bibfnamefont {J.}~\bibnamefont {Wang}}, \bibinfo {author} {\bibfnamefont
  {G.}~\bibnamefont {Xu}}, \bibinfo {author} {\bibfnamefont {P.}~\bibnamefont
  {Tang}}, \bibinfo {author} {\bibfnamefont {W.}~\bibnamefont {Duan}}, \ and\
  \bibinfo {author} {\bibfnamefont {S.-C.}\ \bibnamefont {Zhang}},\ }\href@noop
  {} {\bibfield  {journal} {\bibinfo  {journal} {Phys. Rev. Lett.}\ }\textbf
  {\bibinfo {volume} {111}},\ \bibinfo {pages} {136804} (\bibinfo {year}
  {2013})}\BibitemShut {NoStop}%
\bibitem [{\citenamefont {Xiao}\ \emph {et~al.}(2011)\citenamefont {Xiao},
  \citenamefont {Zhu}, \citenamefont {Ran}, \citenamefont {Nagaosa},\ and\
  \citenamefont {Okamoto}}]{xiao2011}%
  \BibitemOpen
  \bibfield  {author} {\bibinfo {author} {\bibfnamefont {D.}~\bibnamefont
  {Xiao}}, \bibinfo {author} {\bibfnamefont {W.}~\bibnamefont {Zhu}}, \bibinfo
  {author} {\bibfnamefont {Y.}~\bibnamefont {Ran}}, \bibinfo {author}
  {\bibfnamefont {N.}~\bibnamefont {Nagaosa}}, \ and\ \bibinfo {author}
  {\bibfnamefont {S.}~\bibnamefont {Okamoto}},\ }\href@noop {} {\bibfield
  {journal} {\bibinfo  {journal} {Nature Communications}\ }\textbf {\bibinfo
  {volume} {2}},\ \bibinfo {pages} {596} (\bibinfo {year} {2011})}\BibitemShut
  {NoStop}%
\bibitem [{\citenamefont {Wang}\ \emph
  {et~al.}(2014{\natexlab{d}})\citenamefont {Wang}, \citenamefont {Wang},
  \citenamefont {Fang},\ and\ \citenamefont {Dai}}]{wang2014b}%
  \BibitemOpen
  \bibfield  {author} {\bibinfo {author} {\bibfnamefont {Y.}~\bibnamefont
  {Wang}}, \bibinfo {author} {\bibfnamefont {Z.}~\bibnamefont {Wang}}, \bibinfo
  {author} {\bibfnamefont {Z.}~\bibnamefont {Fang}}, \ and\ \bibinfo {author}
  {\bibfnamefont {X.}~\bibnamefont {Dai}},\ }\href@noop {} {\bibfield
  {journal} {\bibinfo  {journal} {arXiv preprint arXiv:1409.6797}\ } (\bibinfo
  {year} {2014}{\natexlab{d}})}\BibitemShut {NoStop}%
\bibitem [{\citenamefont {Cook}\ and\ \citenamefont
  {Paramekanti}(2014)}]{cook2014}%
  \BibitemOpen
  \bibfield  {author} {\bibinfo {author} {\bibfnamefont {A.~M.}\ \bibnamefont
  {Cook}}\ and\ \bibinfo {author} {\bibfnamefont {A.}~\bibnamefont
  {Paramekanti}},\ }\href {\doibase 10.1103/PhysRevLett.113.077203} {\bibfield
  {journal} {\bibinfo  {journal} {Phys. Rev. Lett.}\ }\textbf {\bibinfo
  {volume} {113}},\ \bibinfo {pages} {077203} (\bibinfo {year}
  {2014})}\BibitemShut {NoStop}%
\bibitem [{\citenamefont {Dong}\ \emph {et~al.}(2014)\citenamefont {Dong},
  \citenamefont {Wang}, \citenamefont {Zhang}, \citenamefont {Duan},
  \citenamefont {Zhu}, \citenamefont {Sofo},\ and\ \citenamefont
  {Liu}}]{dong2014}%
  \BibitemOpen
  \bibfield  {author} {\bibinfo {author} {\bibfnamefont {X.-Y.}\ \bibnamefont
  {Dong}}, \bibinfo {author} {\bibfnamefont {J.-F.}\ \bibnamefont {Wang}},
  \bibinfo {author} {\bibfnamefont {R.-X.}\ \bibnamefont {Zhang}}, \bibinfo
  {author} {\bibfnamefont {W.-H.}\ \bibnamefont {Duan}}, \bibinfo {author}
  {\bibfnamefont {B.-F.}\ \bibnamefont {Zhu}}, \bibinfo {author} {\bibfnamefont
  {J.}~\bibnamefont {Sofo}}, \ and\ \bibinfo {author} {\bibfnamefont {C.-X.}\
  \bibnamefont {Liu}},\ }\href@noop {} {\bibfield  {journal} {\bibinfo
  {journal} {arXiv preprint arXiv:1409.3641}\ } (\bibinfo {year}
  {2014})}\BibitemShut {NoStop}%
\bibitem [{\citenamefont {Garrity}\ and\ \citenamefont
  {Vanderbilt}(2013)}]{garrity2013}%
  \BibitemOpen
  \bibfield  {author} {\bibinfo {author} {\bibfnamefont {K.~F.}\ \bibnamefont
  {Garrity}}\ and\ \bibinfo {author} {\bibfnamefont {D.}~\bibnamefont
  {Vanderbilt}},\ }\href@noop {} {\bibfield  {journal} {\bibinfo  {journal}
  {Physical review letters}\ }\textbf {\bibinfo {volume} {110}},\ \bibinfo
  {pages} {116802} (\bibinfo {year} {2013})}\BibitemShut {NoStop}%
\bibitem [{\citenamefont {Raghu}\ \emph {et~al.}(2008)\citenamefont {Raghu},
  \citenamefont {Qi}, \citenamefont {Honerkamp},\ and\ \citenamefont
  {Zhang}}]{raghu2008}%
  \BibitemOpen
  \bibfield  {author} {\bibinfo {author} {\bibfnamefont {S.}~\bibnamefont
  {Raghu}}, \bibinfo {author} {\bibfnamefont {X.-L.}\ \bibnamefont {Qi}},
  \bibinfo {author} {\bibfnamefont {C.}~\bibnamefont {Honerkamp}}, \ and\
  \bibinfo {author} {\bibfnamefont {S.-C.}\ \bibnamefont {Zhang}},\ }\href@noop
  {} {\bibfield  {journal} {\bibinfo  {journal} {Phys. Rev. Lett.}\ }\textbf
  {\bibinfo {volume} {100}},\ \bibinfo {pages} {156401} (\bibinfo {year}
  {2008})}\BibitemShut {NoStop}%
\bibitem [{\citenamefont {Zhang}\ \emph
  {et~al.}(2014{\natexlab{b}})\citenamefont {Zhang}, \citenamefont {Huang},
  \citenamefont {Haule},\ and\ \citenamefont {Vanderbilt}}]{zhang2014b}%
  \BibitemOpen
  \bibfield  {author} {\bibinfo {author} {\bibfnamefont {H.}~\bibnamefont
  {Zhang}}, \bibinfo {author} {\bibfnamefont {H.}~\bibnamefont {Huang}},
  \bibinfo {author} {\bibfnamefont {K.}~\bibnamefont {Haule}}, \ and\ \bibinfo
  {author} {\bibfnamefont {D.}~\bibnamefont {Vanderbilt}},\ }\href {\doibase
  10.1103/PhysRevB.90.165143} {\bibfield  {journal} {\bibinfo  {journal} {Phys.
  Rev. B}\ }\textbf {\bibinfo {volume} {90}},\ \bibinfo {pages} {165143}
  (\bibinfo {year} {2014}{\natexlab{b}})}\BibitemShut {NoStop}%
\bibitem [{\citenamefont {Zhang}\ \emph
  {et~al.}(2014{\natexlab{c}})\citenamefont {Zhang}, \citenamefont {Wang},
  \citenamefont {Xu}, \citenamefont {Xu},\ and\ \citenamefont
  {Zhang}}]{zhang2014a}%
  \BibitemOpen
  \bibfield  {author} {\bibinfo {author} {\bibfnamefont {H.}~\bibnamefont
  {Zhang}}, \bibinfo {author} {\bibfnamefont {J.}~\bibnamefont {Wang}},
  \bibinfo {author} {\bibfnamefont {G.}~\bibnamefont {Xu}}, \bibinfo {author}
  {\bibfnamefont {Y.}~\bibnamefont {Xu}}, \ and\ \bibinfo {author}
  {\bibfnamefont {S.-C.}\ \bibnamefont {Zhang}},\ }\href@noop {} {\bibfield
  {journal} {\bibinfo  {journal} {Physical review letters}\ }\textbf {\bibinfo
  {volume} {112}},\ \bibinfo {pages} {096804} (\bibinfo {year}
  {2014}{\natexlab{c}})}\BibitemShut {NoStop}%
\bibitem [{\citenamefont {Wang}\ \emph
  {et~al.}(2013{\natexlab{b}})\citenamefont {Wang}, \citenamefont {Lian},
  \citenamefont {Zhang}, \citenamefont {Xu},\ and\ \citenamefont
  {Zhang}}]{wang2013a}%
  \BibitemOpen
  \bibfield  {author} {\bibinfo {author} {\bibfnamefont {J.}~\bibnamefont
  {Wang}}, \bibinfo {author} {\bibfnamefont {B.}~\bibnamefont {Lian}}, \bibinfo
  {author} {\bibfnamefont {H.}~\bibnamefont {Zhang}}, \bibinfo {author}
  {\bibfnamefont {Y.}~\bibnamefont {Xu}}, \ and\ \bibinfo {author}
  {\bibfnamefont {S.-C.}\ \bibnamefont {Zhang}},\ }\href@noop {} {\bibfield
  {journal} {\bibinfo  {journal} {Physical review letters}\ }\textbf {\bibinfo
  {volume} {111}},\ \bibinfo {pages} {136801} (\bibinfo {year}
  {2013}{\natexlab{b}})}\BibitemShut {NoStop}%
\bibitem [{\citenamefont {Zhang}\ \emph
  {et~al.}(2013{\natexlab{b}})\citenamefont {Zhang}, \citenamefont {Li},
  \citenamefont {Feng}, \citenamefont {Kane},\ and\ \citenamefont
  {Mele}}]{zhang2013b}%
  \BibitemOpen
  \bibfield  {author} {\bibinfo {author} {\bibfnamefont {F.}~\bibnamefont
  {Zhang}}, \bibinfo {author} {\bibfnamefont {X.}~\bibnamefont {Li}}, \bibinfo
  {author} {\bibfnamefont {J.}~\bibnamefont {Feng}}, \bibinfo {author}
  {\bibfnamefont {C.}~\bibnamefont {Kane}}, \ and\ \bibinfo {author}
  {\bibfnamefont {E.}~\bibnamefont {Mele}},\ }\href@noop {} {\bibfield
  {journal} {\bibinfo  {journal} {arXiv preprint arXiv:1309.7682}\ } (\bibinfo
  {year} {2013}{\natexlab{b}})}\BibitemShut {NoStop}%
\bibitem [{\citenamefont {Fang}\ \emph {et~al.}(2014)\citenamefont {Fang},
  \citenamefont {Gilbert},\ and\ \citenamefont {Bernevig}}]{fang2014}%
  \BibitemOpen
  \bibfield  {author} {\bibinfo {author} {\bibfnamefont {C.}~\bibnamefont
  {Fang}}, \bibinfo {author} {\bibfnamefont {M.~J.}\ \bibnamefont {Gilbert}}, \
  and\ \bibinfo {author} {\bibfnamefont {B.~A.}\ \bibnamefont {Bernevig}},\
  }\href {\doibase 10.1103/PhysRevLett.112.046801} {\bibfield  {journal}
  {\bibinfo  {journal} {Phys. Rev. Lett.}\ }\textbf {\bibinfo {volume} {112}},\
  \bibinfo {pages} {046801} (\bibinfo {year} {2014})}\BibitemShut {NoStop}%
\bibitem [{\citenamefont {Liu}\ \emph {et~al.}(2013)\citenamefont {Liu},
  \citenamefont {Hsu},\ and\ \citenamefont {Liu}}]{liu2013a}%
  \BibitemOpen
  \bibfield  {author} {\bibinfo {author} {\bibfnamefont {X.}~\bibnamefont
  {Liu}}, \bibinfo {author} {\bibfnamefont {H.-C.}\ \bibnamefont {Hsu}}, \ and\
  \bibinfo {author} {\bibfnamefont {C.-X.}\ \bibnamefont {Liu}},\ }\href@noop
  {} {\bibfield  {journal} {\bibinfo  {journal} {Physical review letters}\
  }\textbf {\bibinfo {volume} {111}},\ \bibinfo {pages} {086802} (\bibinfo
  {year} {2013})}\BibitemShut {NoStop}%
\bibitem [{\citenamefont {Liang}\ \emph {et~al.}(2013)\citenamefont {Liang},
  \citenamefont {Wu},\ and\ \citenamefont {Hu}}]{liang2013}%
  \BibitemOpen
  \bibfield  {author} {\bibinfo {author} {\bibfnamefont {Q.-F.}\ \bibnamefont
  {Liang}}, \bibinfo {author} {\bibfnamefont {L.-H.}\ \bibnamefont {Wu}}, \
  and\ \bibinfo {author} {\bibfnamefont {X.}~\bibnamefont {Hu}},\ }\href@noop
  {} {\bibfield  {journal} {\bibinfo  {journal} {New Journal of Physics}\
  }\textbf {\bibinfo {volume} {15}},\ \bibinfo {pages} {063031} (\bibinfo
  {year} {2013})}\BibitemShut {NoStop}%
\bibitem [{\citenamefont {Qi}\ \emph {et~al.}(2009)\citenamefont {Qi},
  \citenamefont {Li}, \citenamefont {Zang},\ and\ \citenamefont
  {Zhang}}]{qi2009}%
  \BibitemOpen
  \bibfield  {author} {\bibinfo {author} {\bibfnamefont {X.-L.}\ \bibnamefont
  {Qi}}, \bibinfo {author} {\bibfnamefont {R.}~\bibnamefont {Li}}, \bibinfo
  {author} {\bibfnamefont {J.}~\bibnamefont {Zang}}, \ and\ \bibinfo {author}
  {\bibfnamefont {S.-C.}\ \bibnamefont {Zhang}},\ }\href@noop {} {\bibfield
  {journal} {\bibinfo  {journal} {Science}\ }\textbf {\bibinfo {volume}
  {323}},\ \bibinfo {pages} {1184} (\bibinfo {year} {2009})}\BibitemShut
  {NoStop}%
\bibitem [{\citenamefont {Maciejko}\ \emph {et~al.}(2010)\citenamefont
  {Maciejko}, \citenamefont {Qi}, \citenamefont {Karch},\ and\ \citenamefont
  {Zhang}}]{maciejko2010b}%
  \BibitemOpen
  \bibfield  {author} {\bibinfo {author} {\bibfnamefont {J.}~\bibnamefont
  {Maciejko}}, \bibinfo {author} {\bibfnamefont {X.~L.}\ \bibnamefont {Qi}},
  \bibinfo {author} {\bibfnamefont {A.}~\bibnamefont {Karch}}, \ and\ \bibinfo
  {author} {\bibfnamefont {S.~C.}\ \bibnamefont {Zhang}},\ }\href@noop {}
  {\bibfield  {journal} {\bibinfo  {journal} {Phys. Rev. Lett.}\ }\textbf
  {\bibinfo {volume} {105}},\ \bibinfo {pages} {246809} (\bibinfo {year}
  {2010})}\BibitemShut {NoStop}%
\bibitem [{\citenamefont {Tang}\ \emph {et~al.}(2011)\citenamefont {Tang},
  \citenamefont {Mei},\ and\ \citenamefont {Wen}}]{tang2011}%
  \BibitemOpen
  \bibfield  {author} {\bibinfo {author} {\bibfnamefont {E.}~\bibnamefont
  {Tang}}, \bibinfo {author} {\bibfnamefont {J.-W.}\ \bibnamefont {Mei}}, \
  and\ \bibinfo {author} {\bibfnamefont {X.-G.}\ \bibnamefont {Wen}},\ }\href
  {\doibase 10.1103/PhysRevLett.106.236802} {\bibfield  {journal} {\bibinfo
  {journal} {Phys. Rev. Lett.}\ }\textbf {\bibinfo {volume} {106}},\ \bibinfo
  {pages} {236802} (\bibinfo {year} {2011})}\BibitemShut {NoStop}%
\bibitem [{\citenamefont {Neupert}\ \emph {et~al.}(2011)\citenamefont
  {Neupert}, \citenamefont {Santos}, \citenamefont {Chamon},\ and\
  \citenamefont {Mudry}}]{neupert2011a}%
  \BibitemOpen
  \bibfield  {author} {\bibinfo {author} {\bibfnamefont {T.}~\bibnamefont
  {Neupert}}, \bibinfo {author} {\bibfnamefont {L.}~\bibnamefont {Santos}},
  \bibinfo {author} {\bibfnamefont {C.}~\bibnamefont {Chamon}}, \ and\ \bibinfo
  {author} {\bibfnamefont {C.}~\bibnamefont {Mudry}},\ }\href@noop {}
  {\bibfield  {journal} {\bibinfo  {journal} {Physical review letters}\
  }\textbf {\bibinfo {volume} {106}},\ \bibinfo {pages} {236804} (\bibinfo
  {year} {2011})}\BibitemShut {NoStop}%
\bibitem [{\citenamefont {Sun}\ \emph {et~al.}(2011)\citenamefont {Sun},
  \citenamefont {Gu}, \citenamefont {Katsura},\ and\ \citenamefont
  {Sarma}}]{sun2011}%
  \BibitemOpen
  \bibfield  {author} {\bibinfo {author} {\bibfnamefont {K.}~\bibnamefont
  {Sun}}, \bibinfo {author} {\bibfnamefont {Z.}~\bibnamefont {Gu}}, \bibinfo
  {author} {\bibfnamefont {H.}~\bibnamefont {Katsura}}, \ and\ \bibinfo
  {author} {\bibfnamefont {S.~D.}\ \bibnamefont {Sarma}},\ }\href@noop {}
  {\bibfield  {journal} {\bibinfo  {journal} {Physical review letters}\
  }\textbf {\bibinfo {volume} {106}},\ \bibinfo {pages} {236803} (\bibinfo
  {year} {2011})}\BibitemShut {NoStop}%
\bibitem [{\citenamefont {Sheng}\ \emph {et~al.}(2011)\citenamefont {Sheng},
  \citenamefont {Gu}, \citenamefont {Sun},\ and\ \citenamefont
  {Sheng}}]{sheng2011}%
  \BibitemOpen
  \bibfield  {author} {\bibinfo {author} {\bibfnamefont {D.}~\bibnamefont
  {Sheng}}, \bibinfo {author} {\bibfnamefont {Z.-C.}\ \bibnamefont {Gu}},
  \bibinfo {author} {\bibfnamefont {K.}~\bibnamefont {Sun}}, \ and\ \bibinfo
  {author} {\bibfnamefont {L.}~\bibnamefont {Sheng}},\ }\href@noop {}
  {\bibfield  {journal} {\bibinfo  {journal} {Nature communications}\ }\textbf
  {\bibinfo {volume} {2}},\ \bibinfo {pages} {389} (\bibinfo {year}
  {2011})}\BibitemShut {NoStop}%
\bibitem [{\citenamefont {Qi}(2011)}]{qi2011a}%
  \BibitemOpen
  \bibfield  {author} {\bibinfo {author} {\bibfnamefont {X.-L.}\ \bibnamefont
  {Qi}},\ }\href@noop {} {\bibfield  {journal} {\bibinfo  {journal} {Physical
  review letters}\ }\textbf {\bibinfo {volume} {107}},\ \bibinfo {pages}
  {126803} (\bibinfo {year} {2011})}\BibitemShut {NoStop}%
\bibitem [{\citenamefont {Regnault}\ and\ \citenamefont
  {Bernevig}(2011)}]{regnault2011}%
  \BibitemOpen
  \bibfield  {author} {\bibinfo {author} {\bibfnamefont {N.}~\bibnamefont
  {Regnault}}\ and\ \bibinfo {author} {\bibfnamefont {B.~A.}\ \bibnamefont
  {Bernevig}},\ }\href@noop {} {\enquote {\bibinfo {title} {Fractional chern
  insulator},}\ }\bibinfo {howpublished} {e-print arXiv:1105.4867} (\bibinfo
  {year} {2011})\BibitemShut {NoStop}%
\bibitem [{\citenamefont {Parameswaran}\ \emph {et~al.}(2012)\citenamefont
  {Parameswaran}, \citenamefont {Roy},\ and\ \citenamefont
  {Sondhi}}]{parameswaran2012}%
  \BibitemOpen
  \bibfield  {author} {\bibinfo {author} {\bibfnamefont {S.}~\bibnamefont
  {Parameswaran}}, \bibinfo {author} {\bibfnamefont {R.}~\bibnamefont {Roy}}, \
  and\ \bibinfo {author} {\bibfnamefont {S.}~\bibnamefont {Sondhi}},\
  }\href@noop {} {\bibfield  {journal} {\bibinfo  {journal} {Physical Review
  B}\ }\textbf {\bibinfo {volume} {85}},\ \bibinfo {pages} {241308} (\bibinfo
  {year} {2012})}\BibitemShut {NoStop}%
\bibitem [{\citenamefont {Barkeshli}\ and\ \citenamefont
  {Qi}(2012)}]{barkeshli2012}%
  \BibitemOpen
  \bibfield  {author} {\bibinfo {author} {\bibfnamefont {M.}~\bibnamefont
  {Barkeshli}}\ and\ \bibinfo {author} {\bibfnamefont {X.-L.}\ \bibnamefont
  {Qi}},\ }\href@noop {} {\bibfield  {journal} {\bibinfo  {journal} {Physical
  Review X}\ }\textbf {\bibinfo {volume} {2}},\ \bibinfo {pages} {031013}
  (\bibinfo {year} {2012})}\BibitemShut {NoStop}%
\bibitem [{\citenamefont {Wu}\ \emph {et~al.}(2013)\citenamefont {Wu},
  \citenamefont {Regnault},\ and\ \citenamefont {Bernevig}}]{wu2013}%
  \BibitemOpen
  \bibfield  {author} {\bibinfo {author} {\bibfnamefont {Y.-L.}\ \bibnamefont
  {Wu}}, \bibinfo {author} {\bibfnamefont {N.}~\bibnamefont {Regnault}}, \ and\
  \bibinfo {author} {\bibfnamefont {B.~A.}\ \bibnamefont {Bernevig}},\
  }\href@noop {} {\bibfield  {journal} {\bibinfo  {journal} {Physical review
  letters}\ }\textbf {\bibinfo {volume} {110}},\ \bibinfo {pages} {106802}
  (\bibinfo {year} {2013})}\BibitemShut {NoStop}%
\bibitem [{\citenamefont {Bergholtz}\ and\ \citenamefont
  {Liu}(2013)}]{bergholtz2013}%
  \BibitemOpen
  \bibfield  {author} {\bibinfo {author} {\bibfnamefont {E.~J.}\ \bibnamefont
  {Bergholtz}}\ and\ \bibinfo {author} {\bibfnamefont {Z.}~\bibnamefont
  {Liu}},\ }\href@noop {} {\bibfield  {journal} {\bibinfo  {journal}
  {International Journal of Modern Physics B}\ }\textbf {\bibinfo {volume}
  {27}},\ \bibinfo {pages} {1330017} (\bibinfo {year} {2013})}\BibitemShut
  {NoStop}%
\bibitem [{\citenamefont {Lindner}\ \emph {et~al.}(2011)\citenamefont
  {Lindner}, \citenamefont {Refael},\ and\ \citenamefont
  {Galitski}}]{lindner2011}%
  \BibitemOpen
  \bibfield  {author} {\bibinfo {author} {\bibfnamefont {N.~H.}\ \bibnamefont
  {Lindner}}, \bibinfo {author} {\bibfnamefont {G.}~\bibnamefont {Refael}}, \
  and\ \bibinfo {author} {\bibfnamefont {V.}~\bibnamefont {Galitski}},\
  }\href@noop {} {\bibfield  {journal} {\bibinfo  {journal} {Nature Physics}\
  }\textbf {\bibinfo {volume} {7}},\ \bibinfo {pages} {490} (\bibinfo {year}
  {2011})}\BibitemShut {NoStop}%
\bibitem [{\citenamefont {Kitagawa}\ \emph {et~al.}(2011)\citenamefont
  {Kitagawa}, \citenamefont {Oka}, \citenamefont {Brataas}, \citenamefont
  {Fu},\ and\ \citenamefont {Demler}}]{kitagawa2011}%
  \BibitemOpen
  \bibfield  {author} {\bibinfo {author} {\bibfnamefont {T.}~\bibnamefont
  {Kitagawa}}, \bibinfo {author} {\bibfnamefont {T.}~\bibnamefont {Oka}},
  \bibinfo {author} {\bibfnamefont {A.}~\bibnamefont {Brataas}}, \bibinfo
  {author} {\bibfnamefont {L.}~\bibnamefont {Fu}}, \ and\ \bibinfo {author}
  {\bibfnamefont {E.}~\bibnamefont {Demler}},\ }\href@noop {} {\bibfield
  {journal} {\bibinfo  {journal} {Physical Review B}\ }\textbf {\bibinfo
  {volume} {84}},\ \bibinfo {pages} {235108} (\bibinfo {year}
  {2011})}\BibitemShut {NoStop}%
\bibitem [{\citenamefont {Gu}\ \emph {et~al.}(2011)\citenamefont {Gu},
  \citenamefont {Fertig}, \citenamefont {Arovas},\ and\ \citenamefont
  {Auerbach}}]{gu2011}%
  \BibitemOpen
  \bibfield  {author} {\bibinfo {author} {\bibfnamefont {Z.}~\bibnamefont
  {Gu}}, \bibinfo {author} {\bibfnamefont {H.}~\bibnamefont {Fertig}}, \bibinfo
  {author} {\bibfnamefont {D.~P.}\ \bibnamefont {Arovas}}, \ and\ \bibinfo
  {author} {\bibfnamefont {A.}~\bibnamefont {Auerbach}},\ }\href@noop {}
  {\bibfield  {journal} {\bibinfo  {journal} {Physical review letters}\
  }\textbf {\bibinfo {volume} {107}},\ \bibinfo {pages} {216601} (\bibinfo
  {year} {2011})}\BibitemShut {NoStop}%
\bibitem [{\citenamefont {Wang}\ \emph
  {et~al.}(2013{\natexlab{c}})\citenamefont {Wang}, \citenamefont {Steinberg},
  \citenamefont {Jarillo-Herrero},\ and\ \citenamefont {Gedik}}]{wang2013d}%
  \BibitemOpen
  \bibfield  {author} {\bibinfo {author} {\bibfnamefont {Y.}~\bibnamefont
  {Wang}}, \bibinfo {author} {\bibfnamefont {H.}~\bibnamefont {Steinberg}},
  \bibinfo {author} {\bibfnamefont {P.}~\bibnamefont {Jarillo-Herrero}}, \ and\
  \bibinfo {author} {\bibfnamefont {N.}~\bibnamefont {Gedik}},\ }\href@noop {}
  {\bibfield  {journal} {\bibinfo  {journal} {Science}\ }\textbf {\bibinfo
  {volume} {342}},\ \bibinfo {pages} {453} (\bibinfo {year}
  {2013}{\natexlab{c}})}\BibitemShut {NoStop}%
\bibitem [{\citenamefont {Wu}(2008)}]{wu2008}%
  \BibitemOpen
  \bibfield  {author} {\bibinfo {author} {\bibfnamefont {C.}~\bibnamefont
  {Wu}},\ }\href@noop {} {\bibfield  {journal} {\bibinfo  {journal} {Physical
  review letters}\ }\textbf {\bibinfo {volume} {101}},\ \bibinfo {pages}
  {186807} (\bibinfo {year} {2008})}\BibitemShut {NoStop}%
\bibitem [{\citenamefont {Zheng}\ and\ \citenamefont {Zhai}(2014)}]{zhai2014}%
  \BibitemOpen
  \bibfield  {author} {\bibinfo {author} {\bibfnamefont {W.}~\bibnamefont
  {Zheng}}\ and\ \bibinfo {author} {\bibfnamefont {H.}~\bibnamefont {Zhai}},\
  }\href {\doibase 10.1103/PhysRevA.89.061603} {\bibfield  {journal} {\bibinfo
  {journal} {Phys. Rev. A}\ }\textbf {\bibinfo {volume} {89}},\ \bibinfo
  {pages} {061603} (\bibinfo {year} {2014})}\BibitemShut {NoStop}%
\bibitem [{\citenamefont {Liu}\ \emph {et~al.}(2010{\natexlab{b}})\citenamefont
  {Liu}, \citenamefont {Liu}, \citenamefont {Wu},\ and\ \citenamefont
  {Sinova}}]{liu2010b}%
  \BibitemOpen
  \bibfield  {author} {\bibinfo {author} {\bibfnamefont {X.-J.}\ \bibnamefont
  {Liu}}, \bibinfo {author} {\bibfnamefont {X.}~\bibnamefont {Liu}}, \bibinfo
  {author} {\bibfnamefont {C.}~\bibnamefont {Wu}}, \ and\ \bibinfo {author}
  {\bibfnamefont {J.}~\bibnamefont {Sinova}},\ }\href {\doibase
  10.1103/PhysRevA.81.033622} {\bibfield  {journal} {\bibinfo  {journal} {Phys.
  Rev. A}\ }\textbf {\bibinfo {volume} {81}},\ \bibinfo {pages} {033622}
  (\bibinfo {year} {2010}{\natexlab{b}})}\BibitemShut {NoStop}%
\bibitem [{\citenamefont {Duca}\ \emph {et~al.}(2015)\citenamefont {Duca},
  \citenamefont {Li}, \citenamefont {Reitter}, \citenamefont {Bloch},
  \citenamefont {Schleier-Smith},\ and\ \citenamefont {Schneider}}]{duca2015}%
  \BibitemOpen
  \bibfield  {author} {\bibinfo {author} {\bibfnamefont {L.}~\bibnamefont
  {Duca}}, \bibinfo {author} {\bibfnamefont {T.}~\bibnamefont {Li}}, \bibinfo
  {author} {\bibfnamefont {M.}~\bibnamefont {Reitter}}, \bibinfo {author}
  {\bibfnamefont {I.}~\bibnamefont {Bloch}}, \bibinfo {author} {\bibfnamefont
  {M.}~\bibnamefont {Schleier-Smith}}, \ and\ \bibinfo {author} {\bibfnamefont
  {U.}~\bibnamefont {Schneider}},\ }\href@noop {} {\bibfield  {journal}
  {\bibinfo  {journal} {Science}\ }\textbf {\bibinfo {volume} {347}},\ \bibinfo
  {pages} {288} (\bibinfo {year} {2015})}\BibitemShut {NoStop}%
\bibitem [{\citenamefont {Goldman}\ \emph {et~al.}(2013)\citenamefont
  {Goldman}, \citenamefont {Dalibard}, \citenamefont {Dauphin}, \citenamefont
  {Gerbier}, \citenamefont {Lewenstein}, \citenamefont {Zoller},\ and\
  \citenamefont {Spielman}}]{goldman2013}%
  \BibitemOpen
  \bibfield  {author} {\bibinfo {author} {\bibfnamefont {N.}~\bibnamefont
  {Goldman}}, \bibinfo {author} {\bibfnamefont {J.}~\bibnamefont {Dalibard}},
  \bibinfo {author} {\bibfnamefont {A.}~\bibnamefont {Dauphin}}, \bibinfo
  {author} {\bibfnamefont {F.}~\bibnamefont {Gerbier}}, \bibinfo {author}
  {\bibfnamefont {M.}~\bibnamefont {Lewenstein}}, \bibinfo {author}
  {\bibfnamefont {P.}~\bibnamefont {Zoller}}, \ and\ \bibinfo {author}
  {\bibfnamefont {I.~B.}\ \bibnamefont {Spielman}},\ }\href@noop {} {\bibfield
  {journal} {\bibinfo  {journal} {Proceedings of the National Academy of
  Sciences}\ }\textbf {\bibinfo {volume} {110}},\ \bibinfo {pages} {6736}
  (\bibinfo {year} {2013})}\BibitemShut {NoStop}%
\bibitem [{\citenamefont {Jotzu}\ \emph {et~al.}(2014)\citenamefont {Jotzu},
  \citenamefont {Messer}, \citenamefont {Desbuquois}, \citenamefont {Lebrat},
  \citenamefont {Uehlinger}, \citenamefont {Greif},\ and\ \citenamefont
  {Esslinger}}]{jotzu2014}%
  \BibitemOpen
  \bibfield  {author} {\bibinfo {author} {\bibfnamefont {G.}~\bibnamefont
  {Jotzu}}, \bibinfo {author} {\bibfnamefont {M.}~\bibnamefont {Messer}},
  \bibinfo {author} {\bibfnamefont {R.}~\bibnamefont {Desbuquois}}, \bibinfo
  {author} {\bibfnamefont {M.}~\bibnamefont {Lebrat}}, \bibinfo {author}
  {\bibfnamefont {T.}~\bibnamefont {Uehlinger}}, \bibinfo {author}
  {\bibfnamefont {D.}~\bibnamefont {Greif}}, \ and\ \bibinfo {author}
  {\bibfnamefont {T.}~\bibnamefont {Esslinger}},\ }\href@noop {} {\bibfield
  {journal} {\bibinfo  {journal} {Nature}\ }\textbf {\bibinfo {volume} {515}},\
  \bibinfo {pages} {237} (\bibinfo {year} {2014})}\BibitemShut {NoStop}%
\bibitem [{\citenamefont {Haldane}\ and\ \citenamefont
  {Raghu}(2008)}]{haldane2008}%
  \BibitemOpen
  \bibfield  {author} {\bibinfo {author} {\bibfnamefont {F.}~\bibnamefont
  {Haldane}}\ and\ \bibinfo {author} {\bibfnamefont {S.}~\bibnamefont
  {Raghu}},\ }\href@noop {} {\bibfield  {journal} {\bibinfo  {journal}
  {Physical review letters}\ }\textbf {\bibinfo {volume} {100}},\ \bibinfo
  {pages} {013904} (\bibinfo {year} {2008})}\BibitemShut {NoStop}%
\bibitem [{\citenamefont {Wang}\ \emph {et~al.}(2008)\citenamefont {Wang},
  \citenamefont {Chong}, \citenamefont {Joannopoulos},\ and\ \citenamefont
  {Solja\ifmmode \check{c}\else \v{c}\fi{}i\ifmmode~\acute{c}\else
  \'{c}\fi{}}}]{wang2008}%
  \BibitemOpen
  \bibfield  {author} {\bibinfo {author} {\bibfnamefont {Z.}~\bibnamefont
  {Wang}}, \bibinfo {author} {\bibfnamefont {Y.~D.}\ \bibnamefont {Chong}},
  \bibinfo {author} {\bibfnamefont {J.~D.}\ \bibnamefont {Joannopoulos}}, \
  and\ \bibinfo {author} {\bibfnamefont {M.}~\bibnamefont {Solja\ifmmode
  \check{c}\else \v{c}\fi{}i\ifmmode~\acute{c}\else \'{c}\fi{}}},\ }\href
  {\doibase 10.1103/PhysRevLett.100.013905} {\bibfield  {journal} {\bibinfo
  {journal} {Phys. Rev. Lett.}\ }\textbf {\bibinfo {volume} {100}},\ \bibinfo
  {pages} {013905} (\bibinfo {year} {2008})}\BibitemShut {NoStop}%
\bibitem [{\citenamefont {Wang}\ \emph {et~al.}(2009)\citenamefont {Wang},
  \citenamefont {Chong}, \citenamefont {Joannopoulos},\ and\ \citenamefont
  {Solja{\v{c}}i{\'c}}}]{wang2009}%
  \BibitemOpen
  \bibfield  {author} {\bibinfo {author} {\bibfnamefont {Z.}~\bibnamefont
  {Wang}}, \bibinfo {author} {\bibfnamefont {Y.}~\bibnamefont {Chong}},
  \bibinfo {author} {\bibfnamefont {J.}~\bibnamefont {Joannopoulos}}, \ and\
  \bibinfo {author} {\bibfnamefont {M.}~\bibnamefont {Solja{\v{c}}i{\'c}}},\
  }\href@noop {} {\bibfield  {journal} {\bibinfo  {journal} {Nature}\ }\textbf
  {\bibinfo {volume} {461}},\ \bibinfo {pages} {772} (\bibinfo {year}
  {2009})}\BibitemShut {NoStop}%
\bibitem [{\citenamefont {Rechtsman}\ \emph {et~al.}(2013)\citenamefont
  {Rechtsman}, \citenamefont {Zeuner}, \citenamefont {Plotnik}, \citenamefont
  {Lumer}, \citenamefont {Podolsky}, \citenamefont {Dreisow}, \citenamefont
  {Nolte}, \citenamefont {Segev},\ and\ \citenamefont
  {Szameit}}]{rechtsman2013}%
  \BibitemOpen
  \bibfield  {author} {\bibinfo {author} {\bibfnamefont {M.~C.}\ \bibnamefont
  {Rechtsman}}, \bibinfo {author} {\bibfnamefont {J.~M.}\ \bibnamefont
  {Zeuner}}, \bibinfo {author} {\bibfnamefont {Y.}~\bibnamefont {Plotnik}},
  \bibinfo {author} {\bibfnamefont {Y.}~\bibnamefont {Lumer}}, \bibinfo
  {author} {\bibfnamefont {D.}~\bibnamefont {Podolsky}}, \bibinfo {author}
  {\bibfnamefont {F.}~\bibnamefont {Dreisow}}, \bibinfo {author} {\bibfnamefont
  {S.}~\bibnamefont {Nolte}}, \bibinfo {author} {\bibfnamefont
  {M.}~\bibnamefont {Segev}}, \ and\ \bibinfo {author} {\bibfnamefont
  {A.}~\bibnamefont {Szameit}},\ }\href@noop {} {\bibfield  {journal} {\bibinfo
   {journal} {Nature}\ }\textbf {\bibinfo {volume} {496}},\ \bibinfo {pages}
  {196} (\bibinfo {year} {2013})}\BibitemShut {NoStop}%
\bibitem [{\citenamefont {Zhang}\ \emph
  {et~al.}(2013{\natexlab{c}})\citenamefont {Zhang}, \citenamefont {Ren},
  \citenamefont {Wang},\ and\ \citenamefont {Li}}]{zhang2013c}%
  \BibitemOpen
  \bibfield  {author} {\bibinfo {author} {\bibfnamefont {L.}~\bibnamefont
  {Zhang}}, \bibinfo {author} {\bibfnamefont {J.}~\bibnamefont {Ren}}, \bibinfo
  {author} {\bibfnamefont {J.-S.}\ \bibnamefont {Wang}}, \ and\ \bibinfo
  {author} {\bibfnamefont {B.}~\bibnamefont {Li}},\ }\href@noop {} {\bibfield
  {journal} {\bibinfo  {journal} {Physical Review B}\ }\textbf {\bibinfo
  {volume} {87}},\ \bibinfo {pages} {144101} (\bibinfo {year}
  {2013}{\natexlab{c}})}\BibitemShut {NoStop}%
\bibitem [{\citenamefont {Shindou}\ \emph {et~al.}(2013)\citenamefont
  {Shindou}, \citenamefont {Matsumoto}, \citenamefont {Murakami},\ and\
  \citenamefont {Ohe}}]{shindou2013}%
  \BibitemOpen
  \bibfield  {author} {\bibinfo {author} {\bibfnamefont {R.}~\bibnamefont
  {Shindou}}, \bibinfo {author} {\bibfnamefont {R.}~\bibnamefont {Matsumoto}},
  \bibinfo {author} {\bibfnamefont {S.}~\bibnamefont {Murakami}}, \ and\
  \bibinfo {author} {\bibfnamefont {J.-i.}\ \bibnamefont {Ohe}},\ }\href@noop
  {} {\bibfield  {journal} {\bibinfo  {journal} {Physical Review B}\ }\textbf
  {\bibinfo {volume} {87}},\ \bibinfo {pages} {174427} (\bibinfo {year}
  {2013})}\BibitemShut {NoStop}%
\bibitem [{\citenamefont {Mook}\ \emph {et~al.}(2014)\citenamefont {Mook},
  \citenamefont {Henk},\ and\ \citenamefont {Mertig}}]{mook2014}%
  \BibitemOpen
  \bibfield  {author} {\bibinfo {author} {\bibfnamefont {A.}~\bibnamefont
  {Mook}}, \bibinfo {author} {\bibfnamefont {J.}~\bibnamefont {Henk}}, \ and\
  \bibinfo {author} {\bibfnamefont {I.}~\bibnamefont {Mertig}},\ }\href@noop {}
  {\bibfield  {journal} {\bibinfo  {journal} {Physical Review B}\ }\textbf
  {\bibinfo {volume} {90}},\ \bibinfo {pages} {024412} (\bibinfo {year}
  {2014})}\BibitemShut {NoStop}%
\bibitem [{\citenamefont {Kim}\ \emph {et~al.}(2003)\citenamefont {Kim},
  \citenamefont {Austin}, \citenamefont {Baauw}, \citenamefont {Mudge},
  \citenamefont {Flautner}, \citenamefont {Hu}, \citenamefont {Irwin},
  \citenamefont {Kandemir},\ and\ \citenamefont {Narayanan}}]{kim2003}%
  \BibitemOpen
  \bibfield  {author} {\bibinfo {author} {\bibfnamefont {N.~S.}\ \bibnamefont
  {Kim}}, \bibinfo {author} {\bibfnamefont {T.}~\bibnamefont {Austin}},
  \bibinfo {author} {\bibfnamefont {D.}~\bibnamefont {Baauw}}, \bibinfo
  {author} {\bibfnamefont {T.}~\bibnamefont {Mudge}}, \bibinfo {author}
  {\bibfnamefont {K.}~\bibnamefont {Flautner}}, \bibinfo {author}
  {\bibfnamefont {J.~S.}\ \bibnamefont {Hu}}, \bibinfo {author} {\bibfnamefont
  {M.~J.}\ \bibnamefont {Irwin}}, \bibinfo {author} {\bibfnamefont
  {M.}~\bibnamefont {Kandemir}}, \ and\ \bibinfo {author} {\bibfnamefont
  {V.}~\bibnamefont {Narayanan}},\ }\href@noop {} {\bibfield  {journal}
  {\bibinfo  {journal} {Computer}\ }\textbf {\bibinfo {volume} {36}},\ \bibinfo
  {pages} {68} (\bibinfo {year} {2003})}\BibitemShut {NoStop}%
\bibitem [{\citenamefont {Zhang}\ and\ \citenamefont
  {Zhang}(2012)}]{zhang2012c}%
  \BibitemOpen
  \bibfield  {author} {\bibinfo {author} {\bibfnamefont {X.}~\bibnamefont
  {Zhang}}\ and\ \bibinfo {author} {\bibfnamefont {S.-C.}\ \bibnamefont
  {Zhang}},\ }in\ \href@noop {} {\emph {\bibinfo {booktitle} {SPIE Defense,
  Security, and Sensing}}}\ (\bibinfo {organization} {International Society for
  Optics and Photonics},\ \bibinfo {year} {2012})\ pp.\ \bibinfo {pages}
  {837309--837309}\BibitemShut {NoStop}%
\bibitem [{\citenamefont {Wu}\ \emph {et~al.}(2014{\natexlab{b}})\citenamefont
  {Wu}, \citenamefont {Liu},\ and\ \citenamefont {Liu}}]{wu2014a}%
  \BibitemOpen
  \bibfield  {author} {\bibinfo {author} {\bibfnamefont {J.}~\bibnamefont
  {Wu}}, \bibinfo {author} {\bibfnamefont {J.}~\bibnamefont {Liu}}, \ and\
  \bibinfo {author} {\bibfnamefont {X.-J.}\ \bibnamefont {Liu}},\ }\href@noop
  {} {\bibfield  {journal} {\bibinfo  {journal} {Physical review letters}\
  }\textbf {\bibinfo {volume} {113}},\ \bibinfo {pages} {136403} (\bibinfo
  {year} {2014}{\natexlab{b}})}\BibitemShut {NoStop}%
\end{thebibliography}%
\end{document}